\documentclass[arps,prd,onecolumn,twoside,floatfix,noshowpacs,nofootinbib,hyperref]{revtex4}
\pdfoutput=1
 
\usepackage{amsmath}
\usepackage{bm}

\newcommand\beq{\begin{equation}}
\newcommand\eeq{\end{equation}}
\newcommand\beqn{\begin{eqnarray}}
\newcommand\eeqn{\end{eqnarray}}

\newcommand\fsky{f_{\mathrm{sky}}}

\newcommand{\ba}{\begin{eqnarray}}
\newcommand{\ea}{\end{eqnarray}}
\newcommand{\be}{\begin{equation}}
\newcommand{\ee}{\end{equation}}

\newcommand\lsim{\mathrel{\rlap{\lower4pt\hbox{\hskip1pt$\sim$}}
        \raise1pt\hbox{$<$}}}
\newcommand\gsim{\mathrel{\rlap{\lower4pt\hbox{\hskip1pt$\sim$}}
        \raise1pt\hbox{$>$}}}

\newcommand{\jcap}{{J.~Cosm.~Astrop.~Phys.}}
\newcommand{\araa}{{Annu.~Rev.~Astron.~Astrophys.}}
\newcommand{\aap}{{Astron.~Astrophys.}}
\newcommand{\apjl}{{Astrophys.~J.~Lett.}}
\newcommand{\apjs}{{Astrophys.~J.~Supp.}}

\newcommand{\mnras}{{Mon.~Not.~R.~Astron.~Soc.}}
\usepackage{graphicx}
\usepackage[large]{subfigure}
\usepackage{amssymb, amsmath}
\usepackage[amssymb]{SIunits}
\usepackage{epstopdf}
\usepackage{hyperref}
\usepackage{url}
\usepackage{aas_macros}
\usepackage{natbib}

\newcommand{\DetSig}{6.2$\sigma$}

\begin{document}
\title{Detection of Thermal SZ -- CMB Lensing Cross-Correlation in Planck Nominal Mission Data}
\author{J.~Colin Hill\footnote{jch@astro.princeton.edu}}\affiliation{Dept.~of Astrophysical Sciences, Peyton Hall, Princeton University, 4 Ivy Lane, Princeton, NJ USA 08544}
\author{David~N.~Spergel\footnote{dns@astro.princeton.edu}}\affiliation{Dept.~of Astrophysical Sciences, Peyton Hall, Princeton University, 4 Ivy Lane, Princeton, NJ USA 08544}
\begin{abstract}
The nominal mission maps from the Planck satellite contain a wealth of information about
secondary anisotropies in the cosmic microwave background (CMB),
including those induced by the thermal Sunyaev-Zel'dovich (tSZ) effect
and gravitational lensing.  As both the tSZ and CMB lensing signals
trace the large-scale matter density field, the anisotropies sourced
by these processes are expected to be correlated.  We report the first
detection of this cross-correlation signal, which we measure at
\DetSig~significance using the Planck data.  We take advantage of
Planck's multifrequency coverage to construct a tSZ map using
internal linear combination techniques, which we subsequently
cross-correlate with the publicly-released Planck CMB lensing potential map.
The cross-correlation is subject to contamination from the cosmic
infrared background (CIB), which is known to correlate strongly with
CMB lensing.  We correct for this contamination via cross-correlating
our tSZ map with the Planck 857 GHz map and confirm the robustness of our measurement using several
null tests.  We interpret the signal using halo model calculations,
which indicate that the tSZ -- CMB lensing cross-correlation
is a unique probe of the physics of intracluster gas in high-redshift, low-mass
groups and clusters.  Our results are consistent with extrapolations of existing
gas physics models to this previously unexplored regime and show clear evidence for contributions from
both the one- and two-halo terms, but no statistically significant evidence for contributions from diffuse, unbound
gas outside of collapsed halos.  We also show that the amplitude of
the signal depends rather sensitively on the amplitude of fluctuations
($\sigma_8$) and the matter density ($\Omega_m$), scaling as
$\sigma_8^{6.1} \Omega_m^{1.5}$ at $\ell=1000$.  We constrain the
degenerate combination $\sigma_8 (\Omega_m/0.282)^{0.26} = 0.824 \pm 0.029$, a result that is in less tension with primordial CMB
constraints than some recent tSZ analyses.  We also combine our
measurement with the Planck measurement of the tSZ auto-power spectrum to
demonstrate a technique that can in principle constrain both cosmology
and the physics of intracluster gas simultaneously.  Our detection
is a direct confirmation that hot, ionized gas traces the dark matter distribution over a wide
range of scales in the universe ($\sim 0.1$--$50 \, {\rm Mpc}/h$).
\end{abstract}
\maketitle

\section{Introduction}
The primordial anisotropies in the cosmic microwave background radiation (CMB) have been a powerful source
of cosmological information over the past two decades, establishing the $\Lambda$CDM standard model and
constraining its parameters to nearly percent-level precision~\cite{Hinshawetal2013,Bennettetal2013,Planck2013params,Sieversetal2013,Storyetal2013}.  However, as CMB photons propagate from the surface of last scattering, they are affected by a number of physical processes that produce secondary anisotropies.  These processes include gravitational lensing of CMB photons by intervening large-scale structures along the line-of-sight (LOS)~\cite{Blanchard-Schneider1987} and the Sunyaev-Zel'dovich (SZ) effect due to inverse Compton scattering of CMB photons off free electrons along the LOS~\cite{Zeldovich-Sunyaev1969,Sunyaev-Zeldovich1970}.  The SZ effect contains two distinct contributions: one due to the thermal motion of hot electrons, primarily located in the intracluster medium (ICM) of galaxy clusters (the thermal SZ effect), and one due to the bulk motion of electrons along the LOS (the kinetic SZ effect).  The thermal SZ (tSZ) effect is roughly an order of magnitude larger than the kinetic SZ (kSZ) effect for typical massive galaxy clusters~\cite{Carlstrometal2002} and is now recognized as a robust method with which to find and characterize clusters in blind surveys of the microwave sky~\cite{Marriageetal2011,Vanderlindeetal2010,Planck2013counts}.  The first detection of the kSZ effect was only achieved recently due to its much smaller amplitude~\cite{Handetal2012}.  We will consider only the tSZ effect in this work.

Measurements of CMB lensing have improved dramatically in recent years, from first detections using cross-correlation techniques~\cite{Smithetal2007,Hirataetal2008} to precision measurements using CMB data alone~\cite{Dasetal2011,vanEngelenetal2012,Planck2013lensing}.  The lensing effect leads to $\approx 2$--$3$ arcminute coherent distortions of $\sim$degree-sized regions in the CMB temperature field.  This distortion produces statistical anisotropy in the small-scale CMB fluctuations, which allows the lensing potential (or convergence) to be reconstructed~(e.g.,~\cite{Okamoto-Hu2003}).  The CMB lensing signal is a probe of the integrated mass distribution out to the surface of last scattering at $z \approx 1100$ --- the most distant source plane possible.  The redshift kernel for the CMB lensing signal peaks around $z \sim 2$, although it receives significant contributions over a wide redshift range ($0.1 \lsim z \lsim 10$).  The CMB lensing power spectrum is a robust cosmological probe as it is primarily sourced by Fourier modes that are still in the linear regime, allowing accurate theoretical predictions to be computed, at least for multipoles $\ell \lsim 1000$~\cite{Lewis-Challinor2006}.  In short, the CMB lensing field is an excellent tracer of the large-scale matter density field over a broad redshift range.

The tSZ signal is predominantly sourced by hot, ionized electrons located in the deep potential wells of massive galaxy clusters.  These halos trace the large-scale matter density field --- in fact, they are highly biased tracers ($b \sim 3$--$4$).  Thus, the tSZ and CMB lensing signals must be correlated.  We report the first detection of this cross-correlation in this paper.  The level of correlation is sensitive to the particular details of how ICM gas traces the dark matter overdensity field, on both large scales (the ``two-halo'' term) and small scales (the ``one-halo term'').  In addition, due to the rare nature of the objects responsible for the tSZ signal, the amplitude of the tSZ -- CMB lensing cross-correlation is quite sensitive to the amplitude of fluctuations ($\sigma_8$) and the matter density ($\Omega_m$).  We show below that the normalized cross-correlation coefficient between the tSZ and CMB lensing fields is $\approx 30$--$40$\% at $\ell \lsim 1000$ for our fiducial model, with a stronger correlation on smaller angular scales.  This value is somewhat less than the normalized correlation between the cosmic infrared background (CIB) and the CMB lensing field, which reaches values as high as $\approx 80$\%~\cite{Planck2013CIBxlens}, because the redshift kernels of the tSZ and CMB lensing fields are not as well-matched as those of the CIB and CMB lensing signals.  Physically, this arises because the clusters responsible for the tSZ signal have not formed until recent epochs ($z \lsim 1.5$), while the dusty star-forming galaxies responsible for most of the CIB emission formed much earlier ($z \gtrsim 2$--$3$), providing a stronger overlap with the CMB lensing redshift kernel.

The tSZ -- CMB lensing cross-correlation is a bispectrum of the CMB temperature field, as the CMB lensing potential (or convergence) is reconstructed from quadratic combinations of the temperature fluctuations in multipole space~\cite{Okamoto-Hu2003}, while the tSZ field is reconstructed from linear combinations of the temperature maps (see Section~\ref{sec:ILC}).  To our knowledge, this bispectrum was first estimated in~\cite{Goldberg-Spergel1999}, where its signal-to-noise ratio (SNR) was forecasted for the (then-forthcoming) WMAP experiment, and compared to the signal from the integrated Sachs-Wolfe (ISW) -- CMB lensing bispectrum.  The authors estimated a SNR $\approx 3$ for the total tSZ+ISW -- CMB lensing bispectrum in WMAP, with most of the signal arising from the tSZ -- CMB lensing term.  These estimates were later updated using more detailed halo model calculations in~\cite{Cooray-Hu2000}, who considered contamination to the tSZ -- CMB lensing bispectrum from reionization-induced bispectra.  Shortly thereafter,~\cite{Coorayetal2000} investigated the use of multifrequency subtraction techniques to reconstruct the tSZ signal using forecasted WMAP and Planck data, and estimated the SNR with which various tSZ statistics could be detected.  Using a simplified model for the ICM gas physics (which they stated was likely only valid at best to the order-of-magnitude level), they estimated a SNR $\sim \mathcal{O}(10)$ for the tSZ -- CMB lensing cross-spectrum in Planck data, depending on the maximum multipole used in the analysis.  Finally,~\cite{Cooray2000} computed predictions for the cross-correlation between the tSZ signal and weak lensing maps constructed from forecasted Sloan Digital Sky Survey (SDSS) data, although no SNR estimates were calculated.  Predictions for the cross-correlation of SDSS galaxies (not lensing) with the tSZ signal from WMAP were computed in~\cite{Peiris-Spergel2000}, and a first detection of this signal was made in~\cite{Afshordietal2004} at SNR $\approx 3.1$.  In the intervening decade since these early papers, our knowledge of the cosmological parameters, ICM gas physics, and contaminating signals from other sources (e.g., the CIB) has improved immensely, making the initial tSZ -- CMB lensing cross-correlation SNR estimates somewhat obsolete.  However, they are roughly accurate at the order-of-magnitude level, and it is indeed the case that the SNR of the tSZ -- CMB lensing cross-spectrum is higher than that of the ISW -- CMB lensing cross-spectrum: we detect the former at \DetSig~significance in this paper, while only $2.5$--$2.6\sigma$ evidence has been found for the latter in Planck collaboration analyses~\cite{Planck2013NG,Planck2013ISW}.  

We measure the tSZ -- CMB lensing cross-correlation by computing the cross-power spectrum of the publicly released Planck CMB lensing potential map~\cite{Planck2013lensing} and a tSZ (or ``Compton-$y$'') map that we construct from a subset of the Planck channel maps.  As shown for more general CMB lensing-induced bispectra in Appendix B of~\cite{Smithetal2007}, the optimal estimator for the tSZ -- CMB lensing temperature bispectrum factors into the individual steps of performing lens reconstruction and correlating with a $y$-map that utilizes the spectral signature of the tSZ effect.  Thus, although we do not claim optimality in this analysis, our approach should be close to the optimal estimator (technically this statement holds only in the limit of weak non-Gaussianity; future high-SNR measurements of this cross-correlation may require an improved estimator).  A possible goal for future analyses may be the simultaneous measurement of the tSZ -- CMB lensing and CIB -- CMB lensing bispectra, rather than measuring each individually while treating the other as a systematic (as done for the CIB in~\cite{Planck2013CIBxlens}, and in this paper for the tSZ).  Also, it is worth noting that the tSZ -- CMB lensing bispectrum (and CIB -- CMB lensing as well) can contaminate measurements of primordial non-Gaussianity; these bispectra have thus far been neglected in such constraints (e.g.,~\cite{Planck2013NG}).  However, the smooth shapes of these late-time bispectra are rather distinct from the acoustic oscillations in the primordial bispectrum arising from the transfer function, and thus the contamination to standard ``shapes'' considered in non-Gaussianity analyses may not be significant.

In this paper, we demonstrate that the tSZ -- CMB lensing cross-correlation signal is a unique probe of the ICM physics in high-redshift ($z \sim 1$), relatively low-mass ($M \sim 10^{13}$--$10^{14} \, \, M_{\odot}/h$) groups and clusters.  In fact, this signal receives contributions from objects at even higher redshifts and lower masses than the tSZ auto-power spectrum, which is well-known for its dependence on the ICM physics in high-$z$, low-mass groups and clusters that have been unobserved thus far with direct X-ray or optical observations~(e.g.,~\cite{Battagliaetal2012,Tracetal2011}).  In a broader sense, this cross-correlation presents a new method with which to constrain the pressure--mass relation, which remains the limiting factor in cosmological constraints based on tSZ measurements~\cite{Planck2013counts,Planck2013ymap,Hasselfieldetal2013,Reichardtetal2013}.  Although any individual tSZ statistic, such as the tSZ auto-power spectrum or the tSZ -- CMB lensing cross-power spectrum, is subject to a complete degeneracy between the normalization of the pressure--mass relation and the amplitude of cosmic density fluctuations (i.e., $\sigma_8$), it is possible to break this degeneracy by combining multiple such statistics with different dependences on the ICM physics and the background cosmology.  A simple version of this idea was applied to measurements of the tSZ auto-power spectrum and the tSZ skewness in~\cite{Hill-Sherwin2013} to show that the physics of intracluster gas could be constrained in a manner nearly independent of the background cosmology.  We demonstrate a similar technique in this paper, which will soon become much more powerful with higher SNR detections of tSZ statistics.

The remainder of this paper is organized as follows.  In Section~\ref{sec:theory}, we describe the theory underlying our halo model calculations of the tSZ -- CMB lensing cross-power spectrum, including our modeling of the ICM physics, and compute the mass and redshift contributions to the signal.  Section~\ref{sec:tSZrecon} presents our Compton-$y$ map constructed from the Planck channel maps using internal linear combination (ILC) techniques.  We also briefly discuss the tSZ auto-power spectrum.  In Section~\ref{sec:tSZxlensing}, we describe our measurement of the tSZ -- CMB lensing cross-power spectrum, including a number of null tests and a small correction for leakage of CIB emission into the Compton-$y$ map.  In Section~\ref{sec:interp}, we use our results to constrain the physics of the ICM and the cosmological parameters $\sigma_8$ and $\Omega_m$.  We also demonstrate that by combining multiple tSZ statistics --- in this case, the tSZ -- CMB lensing cross-power spectrum and the tSZ auto-power spectrum --- it is possible to break the ICM--cosmology degeneracy and simultaneously constrain both.  We discuss our results and conclude in Section~\ref{sec:discussion}.

We assume a flat $\Lambda$CDM cosmology throughout, with parameters
set to their WMAP9+eCMB+BAO+$H_0$ maximum-likelihood values \cite{Hinshawetal2013} unless otherwise specified.  In particular, $\Omega_m = 0.282$ and $\sigma_8 = 0.817$ are our fiducial values for the matter density and the rms amplitude of linear density fluctuations on $8$ Mpc$/h$ scales at $z=0$, respectively.  All masses are quoted in units of $M_{\odot}/h$, where $h \equiv H_0/(100 \, \mathrm{km} \, \mathrm{s}^{-1} \, \mathrm{Mpc}^{-1})$ and $H_0$ is the Hubble parameter today.  All distances and wavenumbers are in comoving units of $\mathrm{Mpc}/h$.

\section{Theory}
\label{sec:theory}
\subsection{Thermal SZ Effect}
The tSZ effect is a frequency-dependent change in the observed CMB temperature due to the inverse Compton scattering of CMB photons off of hot electrons along the LOS, e.g., ionized gas in the ICM of a galaxy cluster.  The temperature change $\Delta T$ at angular position $\vec{\theta}$ with respect to the center of a cluster of mass $M$ at redshift $z$ is given by~\cite{Sunyaev-Zeldovich1970}:
\ba
\label{eq.tSZdef}
\frac{\Delta T(\vec{\theta}, M, z)}{T_{\mathrm{CMB}}} & = & g_{\nu} y(\vec{\theta}, M, z) \nonumber\\
 & = & g_{\nu} \frac{\sigma_T}{m_e c^2} \int_{\mathrm{LOS}} P_e \left( \sqrt{l^2 + d_A^2 |\vec{\theta}|^2}, M, z \right) dl \,,
\ea
where $g_{\nu} = x\,\mathrm{coth}(x/2)-4$ is the tSZ spectral function
with $x \equiv h\nu/k_B T_{\mathrm{CMB}}$, $y(\vec{\theta}, M, z)$ is
the Compton-$y$ parameter, $\sigma_T$ is the Thomson scattering
cross-section, $m_e$ is the electron mass, and $P_e(\vec{r},M,z)$ is the
ICM electron pressure at (three-dimensional) location $\vec{r}$ with
respect to the cluster center.  We have neglected relativistic
corrections to the tSZ spectral function in Eq.~(\ref{eq.tSZdef})
(e.g.,~\cite{Nozawaetal2006}), as these effects are only
non-negligible for the most massive clusters in the universe ($\gtrsim
10^{15} \,\, M_{\odot}/h$).  The tSZ -- CMB lensing cross-spectrum is
dominated by contributions from clusters well below this mass scale
(see Figs.~\ref{fig.dCldzell100} and~\ref{fig.dCldzell1000}), and
thus we do not expect relativistic corrections to be important in our
analysis.  Our measurements and theoretical calculations will be given
in terms of the Compton-$y$ parameter, which is the
frequency-independent quantity characterizing the tSZ signal (in the
absence of relativistic corrections).

Our theoretical calculations assume a spherically symmetric pressure
profile, i.e., $P_e(\vec{r},M,z) = P_e(r,M,z)$ in Eq.~(\ref{eq.tSZdef}). Note
that the integral in Eq.~(\ref{eq.tSZdef}) is computed along the LOS
such that $r^2 = l^2 + d_A(z)^2 \theta^2$, where $d_A(z)$ is the
angular diameter distance to redshift $z$ and $\theta \equiv
|\vec{\theta}|$ is the angular distance between $\vec{\theta}$ and the
cluster center in the plane of the sky.  A spherically symmetric
pressure profile implies that the Compton-$y$ profile is azimuthally
symmetric in the plane of the sky, i.e., $\Delta T(\vec{\theta},M,z) =
\Delta T(\theta,M,z)$.  Finally, note that the electron pressure
$P_e(\vec{r},M,z)$ is related to the thermal gas pressure via $P_{th} =
P_e (5 X_H+3)/2(X_H+1) = 1.932 P_e$, where $X_H=0.76$ is the
primordial hydrogen mass fraction.

We define the mass $M$ in Eq.~(\ref{eq.tSZdef}) to be the virial mass, which is the mass
enclosed within a radius $r_{\rm vir}$~\cite{Bryan-Norman1998}:
\be
\label{eq.rvir}
r_{\rm vir} = \left( \frac{3 M}{4 \pi \Delta_{cr}(z) \rho_{cr}(z)} \right)^{1/3} \,,
\ee
where $\Delta_{cr}(z) = 18 \pi^2 + 82(\Omega(z)-1) - 39(\Omega(z)-1)^2$ and $\Omega(z) = \Omega_m(1+z)^3/(\Omega_m(1+z)^3+\Omega_{\Lambda})$.  For some calculations, we require the spherical overdensity (SO) mass
contained within some radius, defined as follows: $M_{\delta c}$ ($M_{\delta d}$) is the mass enclosed
within a sphere of radius $r_{\delta c}$ ($r_{\delta d}$) such that the enclosed density is $\delta$ times
the critical (mean matter) density at redshift $z$.  For clarity, $c$ subscripts refer to masses defined with respect
to the critical density at redshift $z$, $\rho_{cr}(z) = 3H^2(z)/8\pi G$ with $H(z)$ the Hubble parameter
at redshift $z$, whereas $d$ subscripts refer to masses defined with respect to the mean matter density at
redshift $z$, $\bar{\rho}_m(z) \equiv \bar{\rho}_m$ (this quantity is constant in comoving units).   We convert
between the virial mass $M$ and other SO masses (e.g., $M_{200c}$ or $M_{200d}$) using the Navarro-Frenk-White (NFW)
density profile~\cite{NFW1997} and the concentration-mass relation from~\cite{Duffyetal2008}.

Our fiducial model for the ICM pressure profile is the parametrized
fit to the ``AGN feedback'' simulations given
in~\cite{Battagliaetal2012}, which fully describes the pressure
profile as a function of cluster mass ($M_{200c}$) and redshift.  The simulations on which the ``AGN feedback'' model is based
include virial shock heating, radiative cooling,
and sub-grid prescriptions for star formation and feedback from
supernovae and active galactic nuclei (AGN)~\cite{Battagliaetal2010}.
Also, the smoothed particle hydrodynamics method used for the
simulations captures the effects of bulk motions and turbulence, which
provide non-thermal pressure support in the cluster outskirts.
While this non-thermal pressure support suppresses the tSZ signal in
the outer regions of groups and clusters (by allowing the thermal
pressure to decrease while still maintaining balance against the dark
matter-sourced gravitational potential), the AGN feedback heats and
expels gas from the inner regions of the cluster, leading to flatter
pressure profiles in the center and higher temperatures and pressures
in the outskirts compared to calculations with cooling only (no AGN feedback)~\cite{Battagliaetal2010}.  However, the feedback can also
lower the cluster gas fraction by blowing gas out of the cluster (see, e.g.,~\cite{Sijackietal2007,McCarthyetal2010} for
other cosmological simulations incorporating AGN feedback in the ICM).  All
of these effects are accounted for in the phenomenological ``GNFW''
fit provided in~\cite{Battagliaetal2012}.  This profile has been found to
agree with a number of recent X-ray and tSZ
studies~\cite{Arnaudetal2010,Sunetal2011,Planck2012stack,Planck2012Coma}.

We also consider the parametrized fit to the ``adiabatic'' simulations
given in~\cite{Battagliaetal2012}.  These simulations include only
heating from virial shocks; no cooling or feedback prescriptions of
any kind are included.  Thus, this fit predicts much more tSZ signal
from a cluster of a given mass and redshift than our fiducial ``AGN
feedback'' model does.  It is already in tension with many
observations (e.g.,~\cite{Hill-Sherwin2013,Hajianetal2013}), but we include it here as
an extreme example of the ICM physics possibilities.

For comparison purposes, we also show results below computed
using the ``universal pressure profile'' (UPP) of Arnaud et al.~\cite{Arnaudetal2010},
which consists of a similar ``GNFW''-type fitting function that specifies the pressure profile
as a function of mass ($M_{500c}$) and redshift.
Within $r_{500c}$, this profile is derived from observations of
local ($z<0.2$), massive ($10^{14} \, M_{\odot} < M_{500c} < 10^{15}
\, M_{\odot}$) clusters in the X-ray band with XMM-Newton.  Beyond
$r_{500c}$, this profile is a fit to various numerical simulations,
which include radiative cooling, star formation, and supernova
feedback, but do not include AGN feedback~\cite{Borganietal2004,Piffaretti-Valdarnini2008,Nagaietal2007}.  The overall normalization
of this profile is subject to uncertainty due to the so-called
``hydrostatic mass bias'', $(1-b)$, because the cluster masses used in the
analysis are derived from X-ray observations under the assumption of hydrostatic equilibrium (HSE),
which is not expected to be exactly valid in actual clusters.  This bias
is typically expressed via $M^{\rm HSE}_{500c} = (1-b) M_{500c}$.
Typical values are expected to be $(1-b) \approx
0.8$--$0.9$~\cite{Arnaudetal2010,Shawetal2010}, but recent Planck
results require much more extreme values ($(1-b) \approx
0.5$--$0.6$) in order to reconcile observed tSZ cluster counts with
predictions based on cosmological parameters from the Planck observations of the primordial
CMB~\cite{Planck2013counts}.  We note that $(1-b)$ should not be
thought of as a single number, valid for all clusters --- rather, it
is expected to be a function of cluster mass and redshift, and likely
exhibits scatter about any mean relation as well.  Statements
in the literature about this bias thus only reflect comparisons to the
massive, low-redshift population of clusters from which the UPP
of~\cite{Arnaudetal2010} was derived.  As a rough guide, our fiducial ``AGN
feedback'' model corresponds approximately to $(1-b) \approx 15$\% for
this population of clusters, though this varies with radius (see
Fig. 2 of~\cite{Battagliaetal2010}).  Beyond $r_{500c}$, where the
UPP of~\cite{Arnaudetal2010} is not based
on X-ray data, the ``AGN feedback'' fit predicts much larger
pressures than the UPP, likely due to the newer simulations' inclusion of AGN feedback and the earlier
simulations' neglect of this effect.  Finally, we stress that although
we will show the results of calculations using the UPP
of~\cite{Arnaudetal2010}, it is derived from observations of clusters at much lower
redshifts and higher masses than those that dominate the tSZ -- CMB
lensing cross-power spectrum, and hence these calculations are an
extrapolation of this model.  Nonetheless, our results can constrain the mean value of $(1-b)$ averaged
over the cluster population, and can further be combined with the results of~\cite{Planck2013counts} or~\cite{Planck2013ymap} to obtain even tighter constraints.

Finally, we also show results below computed directly from a cosmological simulation described in Sehgal et al.~\cite{Sehgaletal2010}\footnote{\tt http://lambda.gsfc.nasa.gov/toolbox/tb\char`_sim\char`_ov.cfm}.  The simulation is a large dark matter-only $N$-body simulation that is post-processed to include gas and galaxies according to various phenomenological prescriptions.  We use the full-sky CMB lensing and tSZ maps derived from these simulations to compute the tSZ -- CMB lensing cross-power spectrum.  The tSZ auto-power spectrum of this simulation has been studied previously (e.g.,~\cite{Sehgaletal2010,Dunkleyetal2011}) and lies higher than the results from ACT and SPT at $\ell = 3000$, likely due to missing non-thermal physics in the baryonic post-processing applied to the $N$-body simulation.  We find a similar result for the tSZ -- CMB lensing cross-power spectrum in Section~\ref{sec:interp}.  Note that the cosmological parameters used in the simulation ($\sigma_8 = 0.80$ and $\Omega_m = 0.264$) are consistent with the WMAP5 cosmology, but in order to facilitate comparison with our halo model calculations, we rescale the tSZ -- CMB lensing cross-power spectrum derived from the simulation to our fiducial WMAP9 values ($\sigma_8 = 0.817$ and $\Omega_m = 0.282$) using the dependences on these parameters computed in Section~\ref{sec:theoryPS}.  We emphasize that this rescaling is not computed an overall rescaling of the entire cross-spectrum, but rather as an $\ell$-dependent calculation.  As shown later, we find a general agreement in shape between the tSZ -- CMB lensing cross-power spectrum derived from our halo model calculations and from this simulation (see Fig.~\ref{fig.Clyphimodelcomp}); the amplitudes differ as a result of the differing gas physics treatments.

\subsection{CMB Lensing}
Gravitational lensing of the CMB causes a re-mapping of the unlensed
temperature field (e.g.,~\cite{Lewis-Challinor2006}):
\ba
\label{eq.Tlensed}
T(\hat{n}) & = & T^{\rm un}(\hat{n} + \nabla \phi(\hat{n})) \nonumber\\
  & = & T^{\rm un}(\hat{n}) + \nabla^i \phi(\hat{n}) \nabla_i
T^{\rm un}(\hat{n}) + \mathcal{O} (\phi^2) \,,
\ea
where $T^{\rm un}$ is the unlensed primordial
temperature and $\phi(\hat{n})$ is the CMB lensing potential:
\be
\label{eq.phidef}
\phi(\hat{n}) = -2 \int_0^{\chi_*} d\chi \frac{\chi_* - \chi}{\chi_*
  \chi} \Psi (\chi \hat{n}, \chi) \,,
\ee
where we have specialized to the case of a flat universe.  In this
equation, $\chi(z)$ is the comoving distance to redshift $z$, $\chi_*$
is the comoving distance to the surface of last scattering at
$z_* \approx 1100$, and $\Psi(\chi \hat{n}, \chi)$ is the gravitational
potential.  Note that the lensing convergence is given by $\kappa(\hat{n}) =
-\nabla^2 \phi(\hat{n}) / 2$ (where $\nabla^2$ is now the
two-dimensional Laplacian on the sky), or $\kappa_{\ell}
= \ell(\ell+1) \phi_{\ell}/2$ in multipole space.

Analogous to the Compton-$y$ profile for a halo of mass $M$ at
redshift $z$ defined in Eq.~(\ref{eq.tSZdef}), we can define a CMB
lensing convergence profile, $\kappa(\vec{\theta}, M, z)$:
\be
\label{eq.kappaprofile}
\kappa(\vec{\theta}, M, z) = \Sigma_{\rm crit}^{-1}(z) \int_{\rm LOS} \rho \left( \sqrt{l^2 + d_A^2 |\vec{\theta}|^2}, M, z \right) dl \,,
\ee
where $\rho(\vec{r},M,z)$ is the halo density profile and $\Sigma_{\rm crit}(z)$ is the critical surface density for lensing at redshift $z$ assuming a source plane at $z_* \approx 1100$:
\be
\label{eq.Sigmacrit}
\Sigma_{\rm crit}^{-1}(z) = \frac{4 \pi G \chi(z) \left(\chi_{*}-\chi(z) \right)}{c^2 \chi_* (1+z)} \,.
\ee
We adopt the NFW density profile~\cite{NFW1997} (which is spherically symmetric, i.e., $\rho(\vec{r},M,z) = \rho(r,M,z)$) and the concentration-mass relation from~\cite{Duffyetal2008} when calculating Eq.~(\ref{eq.kappaprofile}).  Also, note that Eq.~(\ref{eq.kappaprofile}) describes the lensing convergence profile, but we will work in terms of the lensing potential below, as this is the quantity directly measured in the publicly released Planck lensing map.  The convergence and potential are trivially related in multipole space, as mentioned above: $\phi_{\ell} = 2 \kappa_{\ell}/(\ell(\ell+1) )$.

\subsection{Power Spectra}
\label{sec:theoryPS}
Given models for the Compton-$y$ and lensing potential signals from each halo of mass $M$ and redshift $z$, we compute the tSZ -- CMB lensing cross-spectrum in the halo model framework~(see~\cite{Cooray-Sheth2002} for a review).  The following expressions are directly analogous to those derived in~\cite{Hill-Pajer2013}; the only change is that now one factor of Compton-$y$ will be replaced by the lensing potential $\phi$.  We work in the flat-sky and Limber approximations in this paper, since we only consider multipoles $\ell \gtrsim 100$ in the cross-spectrum analysis (complete full-sky derivations can be found in Appendix A of~\cite{Hill-Pajer2013}).

The tSZ -- CMB lensing cross-power spectrum, $C_{\ell}^{y\phi}$, is given by the sum of the one-halo and two-halo terms:
\be
\label{eq.Cell}
C_{\ell}^{y\phi} = C_{\ell}^{y\phi,1h} + C_{\ell}^{y\phi,2h} \,.
\ee
The one-halo term arises from correlations between the Compton-$y$ and
lensing potential $\phi$ profiles of the same object.  In the flat-sky
limit, the one-halo term is given by~(e.g.,~\cite{Hill-Pajer2013,Komatsu-Seljak2002}):
\be
\label{eq.Cell1h}
C_{\ell}^{y\phi,1h} = \int dz \frac{d^2V}{dz d\Omega} \int dM \frac{dn(M,z)}{dM} \tilde{y}_{\ell}(M,z) \tilde{\phi}_{\ell}(M,z) \,,
\ee
where
\be
\tilde{y}_{\ell}(M,z) = \frac{\sigma_T}{m_e c^2} \frac{4 \pi
  r_{s,y}}{\ell_{s,y}^2} \int dx_y \, x_y^2 \frac{\sin((\ell+1/2)
  x_y/\ell_{s,y})}{(\ell+1/2) x_y/\ell_{s,y}} P_{e}(x_y r_{s,y},M,z)
\label{eq.yelltwid}
\ee
and
\ba
\tilde{\phi}_{\ell}(M,z) & = & \frac{2}{\ell(\ell+1)} \tilde{\kappa}_{\ell}(M,z) \nonumber \\
   & = & \frac{2}{\ell(\ell+1)} \frac{4 \pi
  r_{s,\phi}}{\ell_{s,\phi}^2} \int dx_{\phi} \, x_{\phi}^2
\frac{\sin((\ell+1/2) x_{\phi}/\ell_{s,\phi})}{(\ell+1/2)
  x_{\phi}/\ell_{s,\phi}} \frac{\rho(x_{\phi}
  r_{s,\phi},M,z)}{\Sigma_{\rm crit}(z)} \,.
\label{eq.phielltwid}
\ea
In Eq.~(\ref{eq.Cell1h}), $d^2V/dz d\Omega$ is the comoving volume
element per steradian and $dn(M,z)/dM$ is the halo mass function,
i.e., the comoving number density of halos per unit mass as a function
of redshift.  All calculations in this paper use the mass function
from~\cite{Tinkeretal2008}, but with the updated parameters provided in
Eqs.~(8)--(12) of~\cite{Tinkeretal2010}, which explicitly enforce the
physical constraint that the mean bias of all matter in the universe
at a fixed redshift must equal unity.  (This constraint was not enforced in the original
fits in~\cite{Tinkeretal2008}, and it is relevant when calculating
quantities that receive contributions from very low-mass halos, such
as the CMB lensing power spectrum.)  We work with the mass function
fit given for the SO mass $M_{200d}$.  Further details on the mass
function calculations can be found in Section II.A
of~\cite{Hill-Pajer2013}.

In Eq.~(\ref{eq.yelltwid}), $r_{s,y}$ is a characteristic scale radius
(not the NFW scale radius) of the ICM pressure profile, $\ell_{s,y} =
a(z)\chi(z)/r_{s,y} = d_A(z)/r_{s,y}$ is the associated multipole
moment, and $x_y \equiv r/r_{s,y}$ is a dimensionless radial integration variable.  For
the fiducial ``AGN feedback'' pressure profile
from~\cite{Battagliaetal2012} used in our calculations, the natural
scale radius is $r_{200 c}$.  For the UPP of~\cite{Arnaudetal2010},
the natural scale radius is $r_{500 c}$.

Similarly, in Eq.~(\ref{eq.phielltwid}), $r_{s,\phi}$ is a
characteristic scale radius of the halo density profile, which in this
case is indeed the NFW scale radius.  Likewise, $\ell_{s,\phi} =
a(z)\chi(z)/r_{s,\phi} = d_A(z)/r_{s,\phi}$ is the associated
multipole moment and $x_{\phi} \equiv r/r_{s,\phi}$ is a dimensionless
radial integration variable.

The two-halo term arises from correlations between the Compton-$y$ and lensing potential $\phi$ profiles of two separate objects.  In the Limber approximation~\cite{Limber1954}, the two-halo term is given by~(e.g.,~\cite{Hill-Pajer2013,Komatsu-Kitayama1999}):
\be
\label{eq.Cell2h}
C_{\ell}^{y\phi,2h} = \int dz \frac{d^2V}{dz d\Omega} \left[ \int dM_1 \frac{dn(M_1,z)}{dM_1} b(M_1,z) \tilde{y}_{\ell}(M_1,z) \right] \left[ \int dM_2 \frac{dn(M_2,z)}{dM_2} b(M_2,z) \tilde{\phi}_{\ell}(M_2,z) \right]  P_{\mathrm{lin}}\left(\frac{\ell+1/2}{\chi(z)},z\right) \,,
\ee
where $b(M,z)$ is the linear halo bias and $P_{\rm lin}(k,z)$ is the
linear theory matter power spectrum.  We use the fitting function for
the halo bias from Table 2 of~\cite{Tinkeretal2010}, which matches our mass
function to ensure a mean bias of unity at a given redshift, as
mentioned above.  We compute the matter power spectrum for a given
set of cosmological parameters using {\tt CAMB}\footnote{http://www.camb.info}.

In this work we model the two-halo term using the full expression
in Eq.~(\ref{eq.Cell2h}).  However, in the large-scale limit ($\ell
\rightarrow 0$) this equation can be simplified to match the standard
expression used in analyses of cross-correlation between CMB lensing
and various mass tracers (e.g., Eq.~(48) of~\cite{Planck2013lensing} or
Eq.~(6) of \cite{Sherwinetal2012}).  The simplification rests on the fact that
on very large scales ($k \rightarrow 0$) the 3D Fourier transform of the halo
density profile is simply the total mass $M$ of the halo, i.e., $\tilde{\rho}(k,M,z)
\rightarrow M$ as $k \rightarrow 0$.  Using this fact and the previously noted
result that the mean bias of all matter at a fixed redshift is unity, i.e.,
$\int dM \frac{dn}{dM} b(M,z) \frac{M}{\bar{\rho}_m} = 1$, one obtains
\ba
C_{\ell \rightarrow 0}^{y\phi,2h} & \approx & \frac{2}{\ell(\ell+1)} \int d\chi W^{\kappa}(\chi) P_{\rm lin}\left(\frac{\ell+1/2}{\chi(z)},z\right) \int dM \frac{dn(M,z)}{dM} b(M,z) \tilde{y}_{\ell}(M,z) \nonumber \\
   & \approx & \frac{2}{\ell(\ell+1)} \int d\chi W^{\kappa}(\chi) P_{\rm lin}\left(\frac{\ell+1/2}{\chi(z)},z\right) \int dM \frac{dn(M,z)}{dM} b(M,z) \tilde{y}_{0}(M,z) \,,
\label{eq.Cell2hlowell}
\ea
where $W^{\kappa}(\chi) = \frac{3 \Omega_m H_0^2 (1+z)}{2 c^2} \chi \left( \frac{\chi_{*}-\chi}{\chi_*} \right)$ is the lensing convergence projection kernel (assuming a flat universe) and we have changed integration variables from $z$ to $\chi$ to match the standard CMB lensing literature (e.g., Eq.~(48) of~\cite{Planck2013lensing}).  In obtaining the second line of Eq.~(\ref{eq.Cell2hlowell}), we have noted that the $\ell \rightarrow 0$ limit corresponds to angular scales much larger than the scale of individual clusters, and thus one can further approximate $\tilde{y}_{\ell}(M,z) \approx \tilde{y}_{0}(M,z)$, which is proportional to the mean Compton-$y$ signal from that cluster.  In this limit, one can thus constrain an effective ``Compton-$y$'' bias corresponding to the product of the halo bias and Compton-$y$ signal.  In contrast, when cross-correlating a redshift catalog of objects (e.g., quasars or galaxy clusters) with the CMB lensing field, one directly constrains the typical halo bias (and hence mass) of those objects.  Our measurements are all at multipoles $\ell > 100$, and thus we work with the full expression for the two-halo term in Eq.~(\ref{eq.Cell2h}), in which it is unfortunately not straightforward to disentangle the influence of the bias and the ICM physics.  However, we can still probe the ICM by varying the gas physics model while holding all other ingredients in the calculation fixed; since the gas physics is likely the most uncertain ingredient (especially at these redshifts and masses --- see Figs.~\ref{fig.dCldzell100} and~\ref{fig.dCldzell1000}), this is a reasonable approach.

The fiducial integration limits in our calculations are $0.005 < z < 10$ for all redshift integrals and $10^{5} M_{\odot}/h < M < 5 \times 10^{15} M_{\odot}/h$ for all mass integrals.  We note that this involves an extrapolation of the halo mass and bias functions from~\cite{Tinkeretal2008,Tinkeretal2010} to mass and redshift ranges in which they were not explicitly measured in the simulations.  However, the bulk of the tSZ -- CMB lensing cross-spectrum comes from mass and redshift ranges in which the fitting functions are indeed measured (see Figs.~\ref{fig.dCldzell100} and~\ref{fig.dCldzell1000}), so this extrapolation should not have a huge effect; it is primarily needed in order to compute the CMB lensing auto-spectrum for comparison calculations, as this signal does indeed receive contributions from very high redshifts and low masses.  In addition, we must define a boundary at which to cut off the integrals over the pressure and density profiles in Eqs.~(\ref{eq.yelltwid}) and~(\ref{eq.phielltwid}).  We determine this boundary by requiring our halo model calculation of the CMB lensing auto-spectrum to agree with the standard method in which one simply integrates over the linear theory matter power spectrum:
\be
\label{eq.Cellphiphi2hlowell}
C_{\ell}^{\phi\phi,2h} \approx \frac{4}{\ell^2(\ell+1)^2}  \int d\chi \left( \frac{W^{\kappa}(\chi)}{\chi} \right)^2 P_{\rm lin}\left(\frac{\ell+1/2}{\chi(z)},z\right) \,,
\ee
which is quite accurate for $\ell \lsim 1000$~(see Fig. 2 of ~\cite{Lewis-Challinor2006}).  It is straightforward to derive Eq.~(\ref{eq.Cellphiphi2hlowell}) from the CMB lensing analogue of Eq.~(\ref{eq.Cell2h}) by considering the $\ell \rightarrow 0$ limit and using the procedure described in the previous paragraph.  However, since we are using a variety of fitting functions from the literature, this formal derivation does not hold precisely when applied to the numerical calculations.  The primary obstacle is the logarithmic divergence of the enclosed mass in the NFW profile as $r \rightarrow \infty$~\cite{Binney-Tremaine}.  By testing various values of the outer cut-off in Eq.~(\ref{eq.phielltwid}), we find that $r_{\rm out} = 1.5 r_{\rm vir}$ leads to better than $5$\% agreement between our halo model calculation of the CMB lensing potential power spectrum and the linear theory calculation in Eq.~(\ref{eq.Cellphiphi2hlowell}) for all multipoles $\ell \lsim 900$, above which nonlinear corrections to Eq.~(\ref{eq.Cellphiphi2hlowell}) become important.  Thus, we adopt $r_{\rm out} = 1.5 r_{\rm vir}$ when computing Eqs.~(\ref{eq.yelltwid}) and~(\ref{eq.phielltwid}).  Future work requiring more accurate predictions will necessitate detailed simulations and better understanding of the halo model approximations in this context.

\begin{figure}
\centering
\includegraphics[totalheight=0.4\textheight]{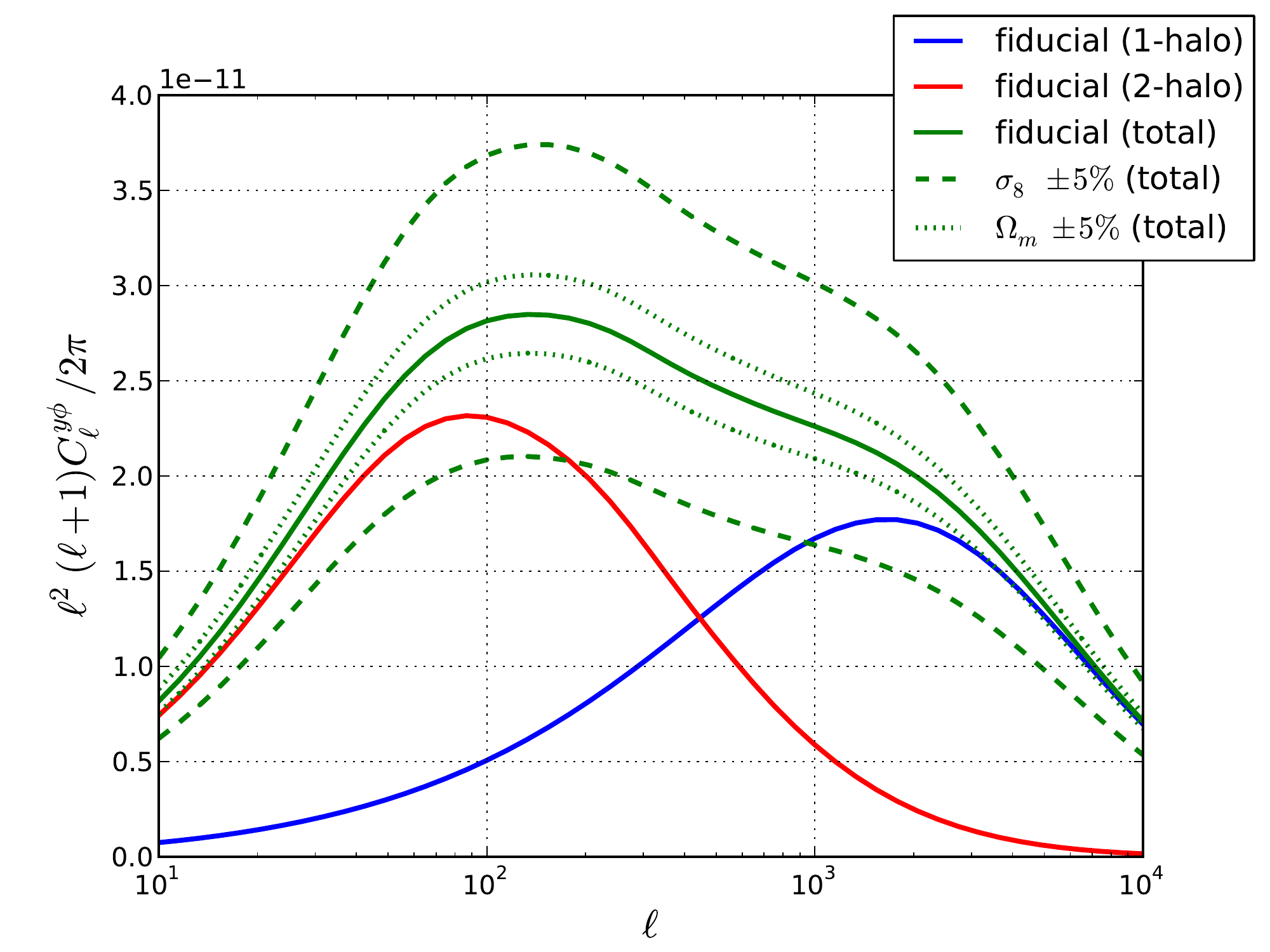}
\caption{The tSZ -- CMB lensing cross-power spectrum computed for our fiducial model (WMAP9 cosmological parameters and the ``AGN feedback'' ICM pressure profile fit from~\cite{Battagliaetal2012}).  The total signal is the green solid curve, while the one-halo and two-halo contributions are the blue and red solid curves, respectively.  The plot also shows the total signal computed for variations around our fiducial model: $\pm 5$\% variations in $\sigma_8$ are the dashed green curves, while $\pm 5$\% variations in $\Omega_m$ are the dotted green curves.  In both cases, an increase (decrease) in the parameter's value yield an increase (decrease) in the amplitude of the cross-spectrum.
\label{fig.Clyphifid}}
\end{figure}

We use the model described thus far to compute the tSZ -- CMB lensing cross-power spectrum for our fiducial cosmology.  The results are shown in Fig.~\ref{fig.Clyphifid}.  As expected, the two-halo term dominates at low $\ell$, while the one-halo term dominates at high $\ell$; the two terms are roughly equal at $\ell \approx 450$.  As one might expect, this behavior lies between the extremes of the CMB lensing power spectrum, which is dominated by the two-halo term for $\ell \lsim 1000$~\cite{Lewis-Challinor2006}, and the tSZ power spectrum, which is dominated by the one-halo term for $\ell \gsim 10$~\cite{Hill-Pajer2013}.  Fig.~\ref{fig.Clyphifid} also shows the cross-power spectrum computed for $\pm 5$\% variations in $\sigma_8$ and $\Omega_m$ around our fiducial cosmology.  Using these variations, we compute approximate power-law scalings of the cross-power spectrum with respect to these parameters.  The scalings vary as a function of $\ell$ --- for example, the scaling with $\sigma_8$ is strongest near $\ell \sim 1000$--$2000$, where the one-halo term dominates but the power is not yet being sourced by the variations within the pressure profile itself, which occurs at very high $\ell$.  The scaling with $\Omega_m$ is strongest near $\ell \sim 800$--$1500$.  Representative scalings are:
\ba
C_{\ell=100}^{y\phi} & \propto & \left( \frac{\sigma_8}{0.817} \right)^{5.7} \left( \frac{\Omega_m}{0.282} \right)^{1.4} \nonumber \\
C_{\ell=1000}^{y\phi} & \propto & \left( \frac{\sigma_8}{0.817} \right)^{6.1} \left( \frac{\Omega_m}{0.282} \right)^{1.5} \,.
\label{eq.scalings}
\ea
Over the $\ell$ range where we measure $C_{\ell}^{y\phi}$ (see Section~\ref{sec:tSZxlensing}) --- $100 < \ell < 1600$ --- the average values of the scalings are 6.0 and 1.5 for $\sigma_8$ and $\Omega_m$, respectively.  These scalings provide the theoretical degeneracy between these parameters, which will be useful in Section~\ref{sec:interp}.  Note that the dependence of the tSZ -- CMB lensing cross-power spectrum on these parameters is not as steep as the dependence of the tSZ auto-power spectrum~\cite{Komatsu-Seljak2002,Hill-Pajer2013} or other higher-order tSZ statistics, such as the skewness~\cite{Wilsonetal2012,Hill-Sherwin2013} or bispectrum~\cite{Bhattacharyaetal2012}.  The dependence is stronger for the tSZ auto-statistics because their signals arise from more massive, rare clusters that lie further in the exponential tail of the mass function than those that source the tSZ -- CMB lensing cross-spectrum (see below).  The cross-spectrum is, however, uncharted territory for measurements of the ICM pressure profile, as its signal comes from groups and clusters at much higher redshifts and lower masses than those that source the tSZ auto-statistics.

\begin{figure}
\begin{minipage}[b]{0.495\linewidth}
\centering
\includegraphics[width=\textwidth]{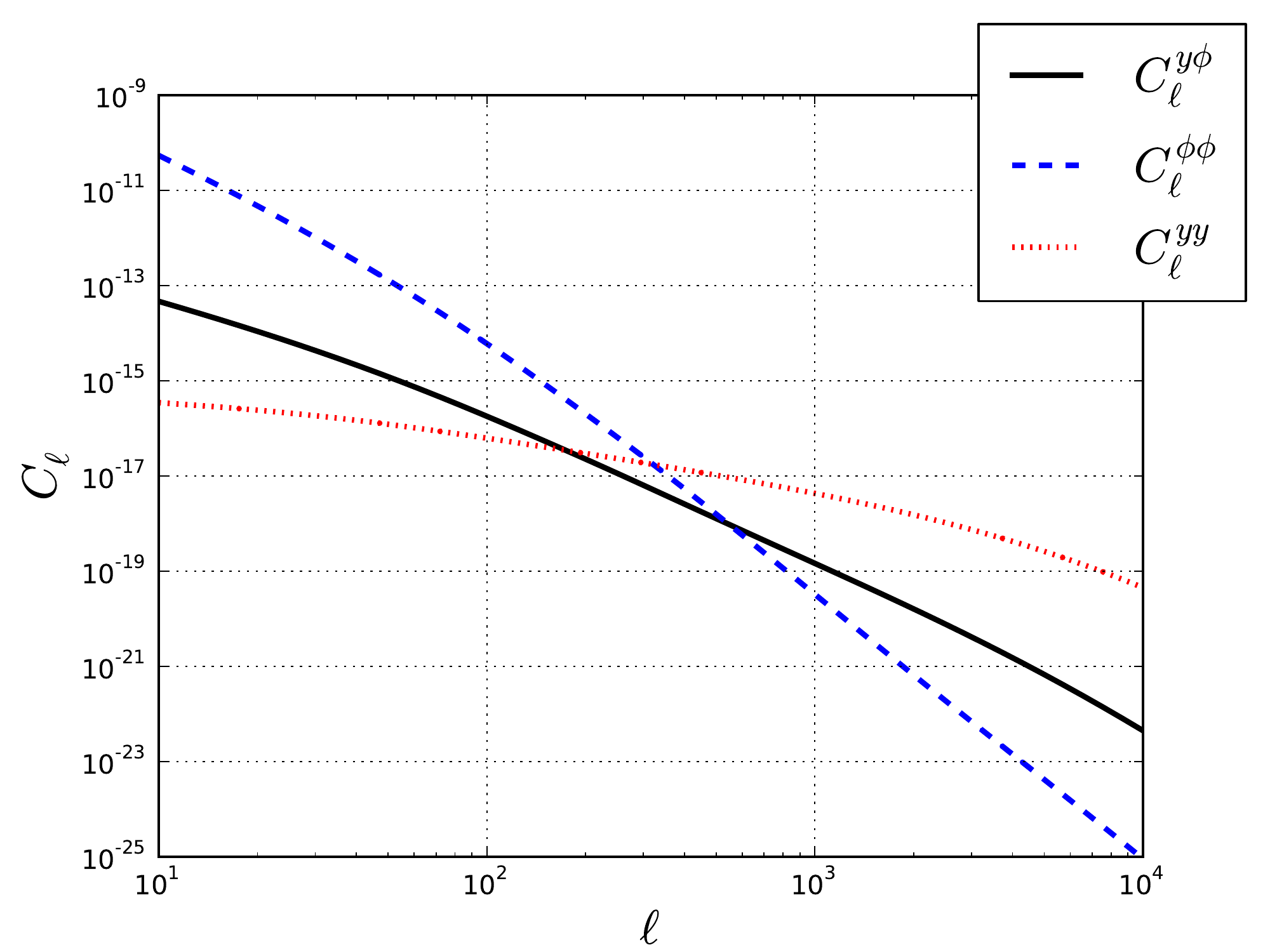}
\end{minipage}
\begin{minipage}[b]{0.495\linewidth}
\centering
\includegraphics[width=\textwidth]{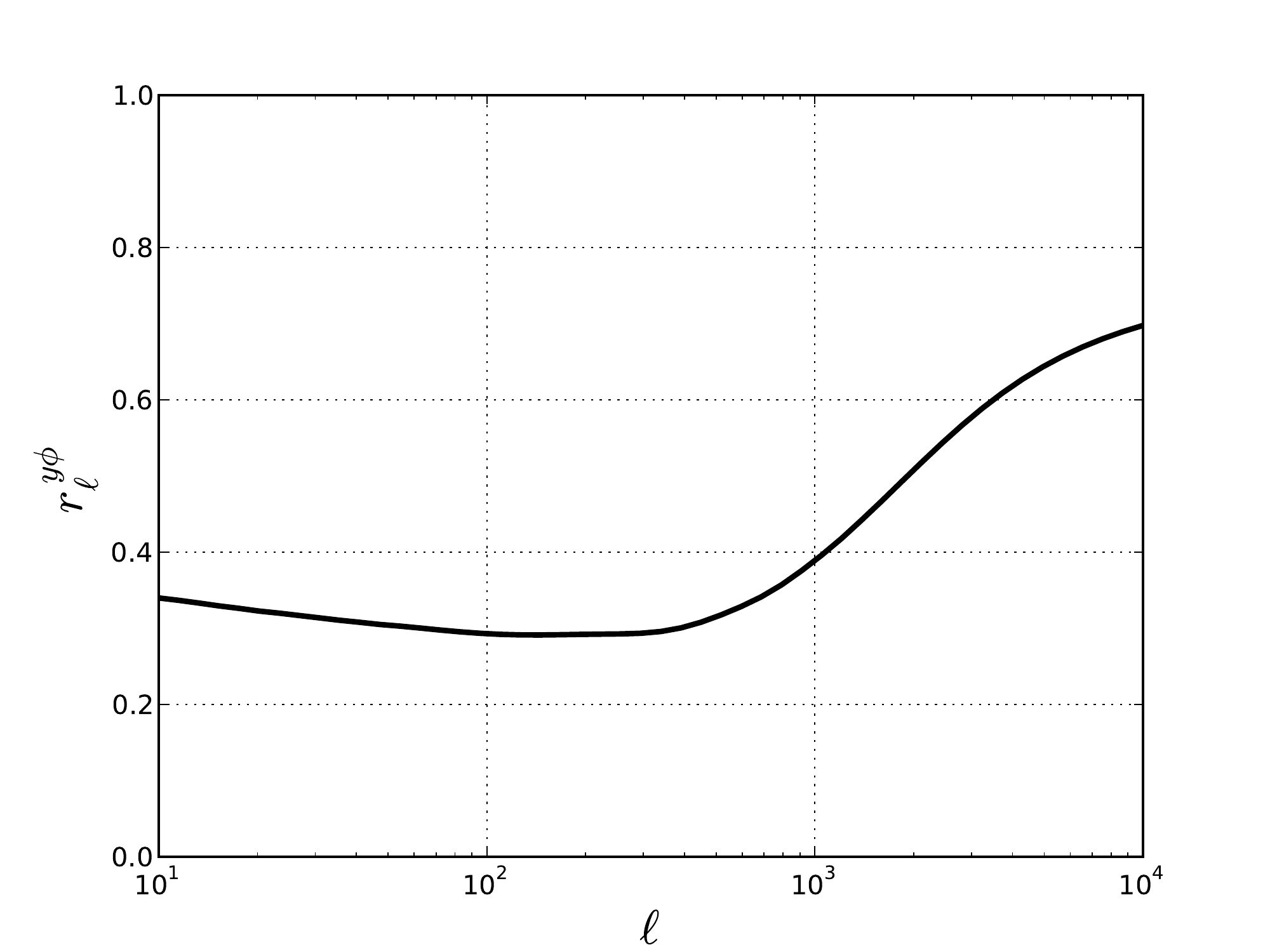}
\end{minipage}
\caption{The left panel shows the power spectra of the tSZ and CMB lensing potential fields for our fiducial model, as labeled in the figure.  The right panel shows the normalized cross-correlation coefficient of the tSZ and CMB lensing potential signals as a function of multipole (see Eq.~(\ref{eq.rellyphi})).  The typical normalized correlation is $30$--$40$\% over the $\ell$ range where we measure $C_{\ell}^{y\phi}$ ($100 < \ell < 1600$).  The strength of the correlation increases at smaller angular scales, where both signals are dominated by the one-halo term.
\label{fig.crosscortheory}}
\end{figure}

Before assessing the origin of the cross-spectrum signal in detail, it is useful to examine the predicted strength of the correlation between the tSZ and CMB lensing potential fields.  A standard method to assess the level of correlation is through the normalized cross-correlation coefficient, $r_{\ell}^{y\phi}$:
\be
\label{eq.rellyphi}
r_{\ell}^{y\phi} = \frac{C_{\ell}^{y\phi}}{\sqrt{C_{\ell}^{yy} C_{\ell}^{\phi\phi}}} \,,
\ee
where $C_{\ell}^{yy}$ and $C_{\ell}^{\phi\phi}$ are the tSZ and CMB lensing auto-power spectra, respectively.  Fig.~\ref {fig.crosscortheory} shows the result of this calculation, as well as a comparison of the auto-spectra to the cross-spectrum.  The behavior of the power spectra is consistent with expectations: $C_{\ell}^{\phi\phi}$ is much larger at low $\ell$ than high $\ell$; $C_{\ell}^{yy}$ is close to Poisson for $\ell < 1000$, above which intracluster structure contributes power; and $C_{\ell}^{y\phi}$ lies between these two extremes.  The normalized cross-correlation $r_{\ell}^{y\phi} \approx 30$--$40$\% over the $\ell$ range where we measure the tSZ -- CMB lensing cross-spectrum ($100 < \ell < 1600$); the average value of the normalized correlation within this range is $33.2$\%.  At smaller angular scales, the correlation reaches values as high as $60$--$70$\%.  This increase occurs because both the tSZ and CMB lensing signals are dominated by the one-halo term from individual objects at these scales, and the gas traces the dark matter quite effectively within halos.  Overall, the tSZ signal is a reasonably strong tracer of the large-scale matter distribution, although not nearly as strong as the CIB~\cite{Planck2013CIBxlens}, at least on large scales.  The tSZ -- CMB lensing cross-correlation is, however, interesting as a probe of a previously unstudied population of high-redshift, low-mass groups and clusters.

\begin{figure}
\begin{minipage}[b]{0.32\linewidth}
\centering
\includegraphics[width=\textwidth]{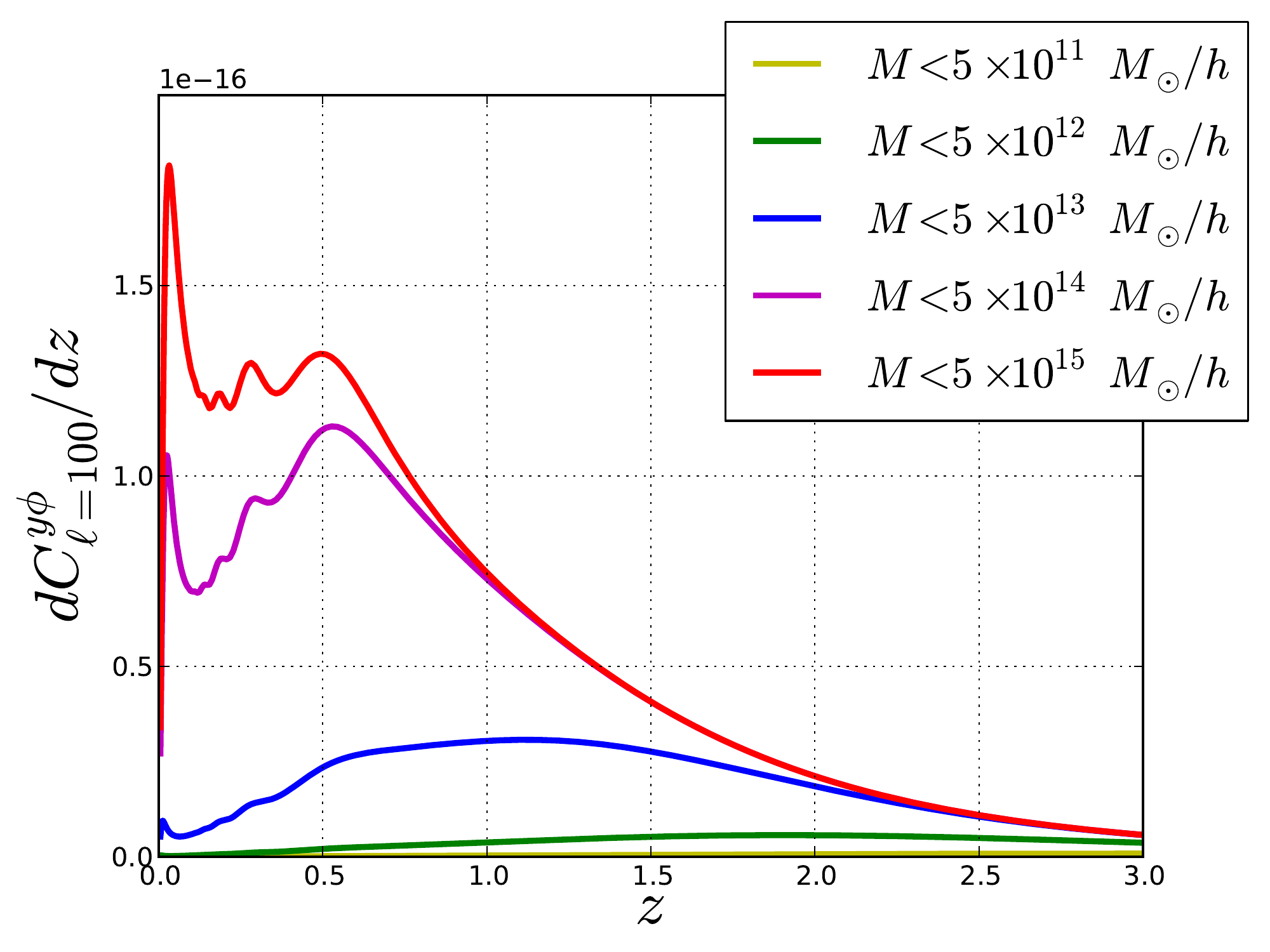}
\end{minipage}
\begin{minipage}[b]{0.32\linewidth}
\centering
\includegraphics[width=\textwidth]{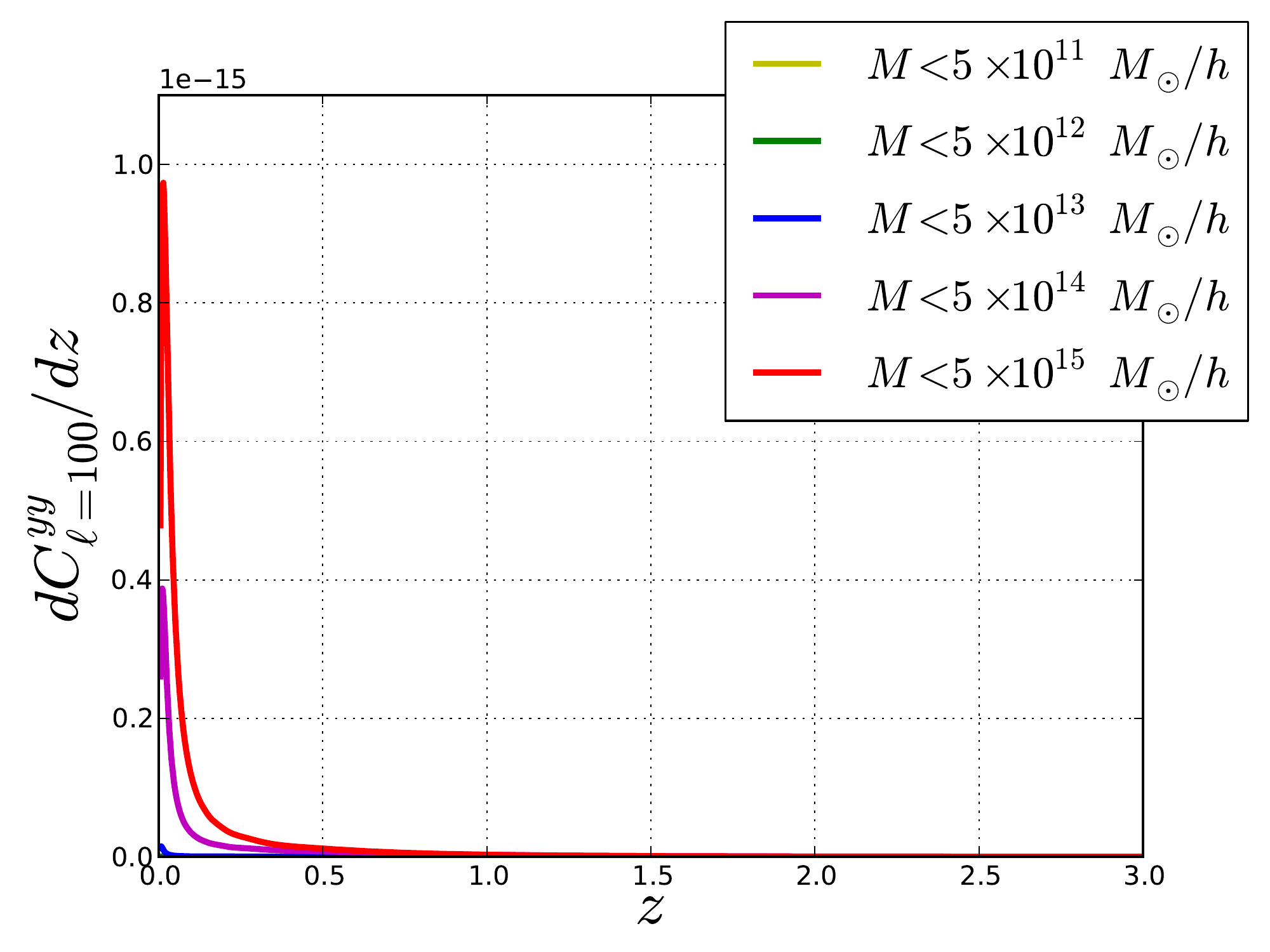}
\end{minipage}
\begin{minipage}[b]{0.32\linewidth}
\centering
\includegraphics[width=\textwidth]{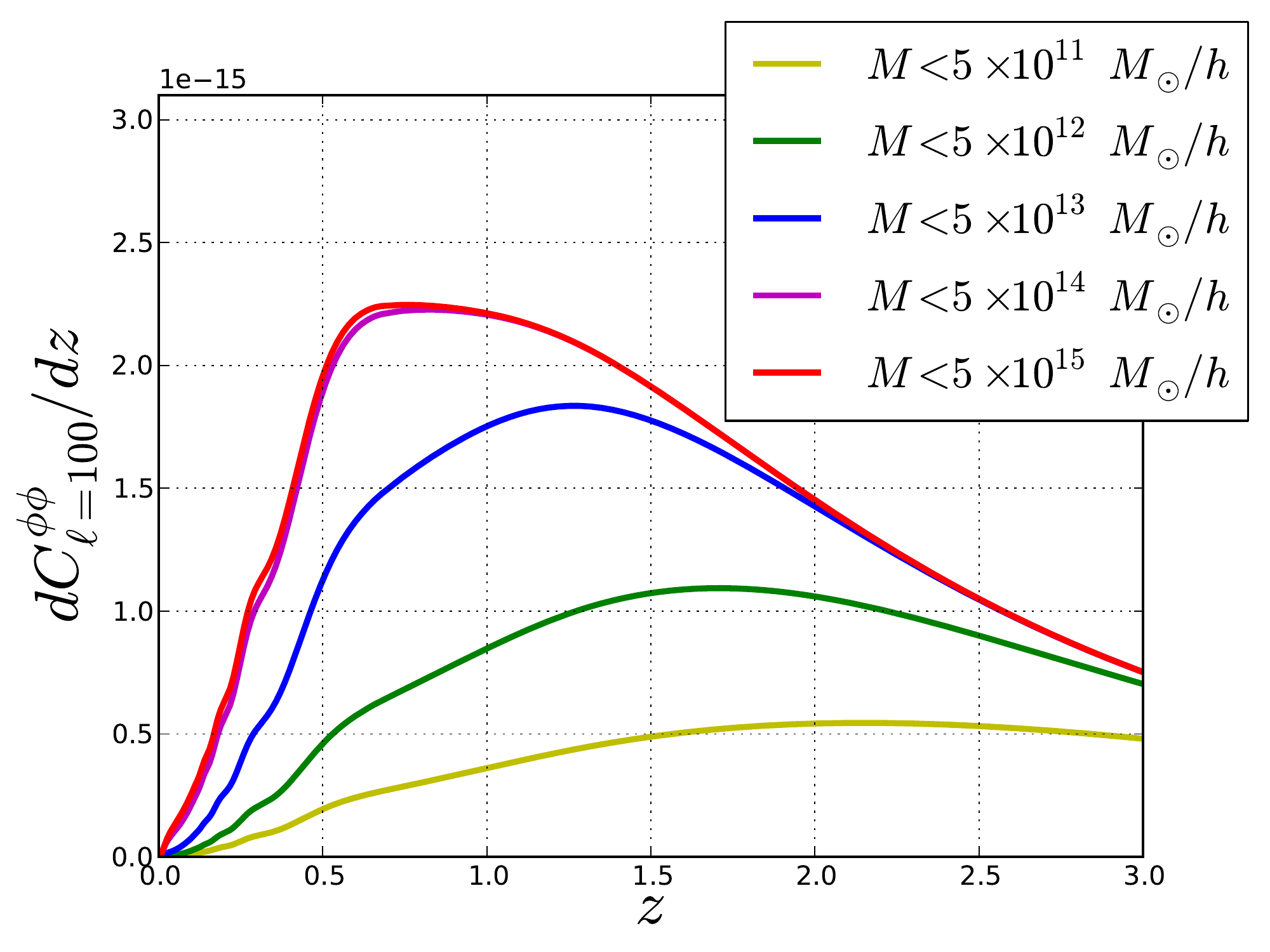}
\end{minipage}
\caption{Mass and redshift contributions to the tSZ -- CMB lensing cross-spectrum (left panel), tSZ auto-spectrum (middle panel), and CMB lensing potential auto-spectrum (right panel), computed at $\ell=100$, where the two-halo term dominates the cross-spectrum.  Each curve includes contributions from progressively higher mass scales, as labelled in the figures, while the vertical axis encodes the differential contribution from each redshift.  At this multipole, the tSZ auto-spectrum is heavily dominated by massive, low-redshift halos, while the CMB lensing auto-spectrum receives contributions from a much wider range of halo masses and redshifts.  The cross-spectrum, as expected, lies between these extremes.  Note that the sharp features in the cross-spectrum curves are an artifact of combining the various fitting functions used in our calculation.  See the text for further discussion.
\label{fig.dCldzell100}}
\end{figure}

\begin{figure}
\begin{minipage}[b]{0.32\linewidth}
\centering
\includegraphics[width=\textwidth]{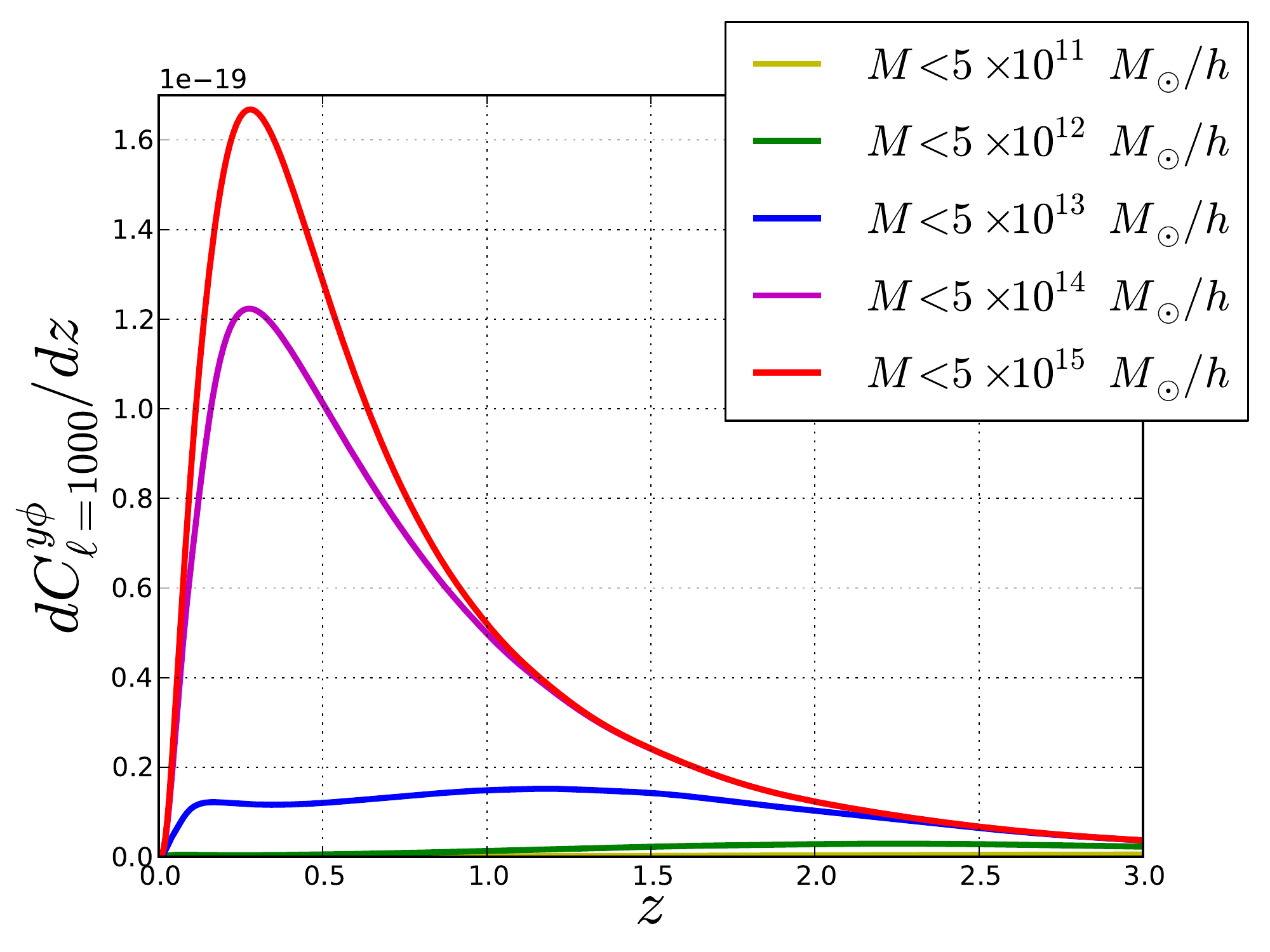}
\end{minipage}
\begin{minipage}[b]{0.32\linewidth}
\centering
\includegraphics[width=\textwidth]{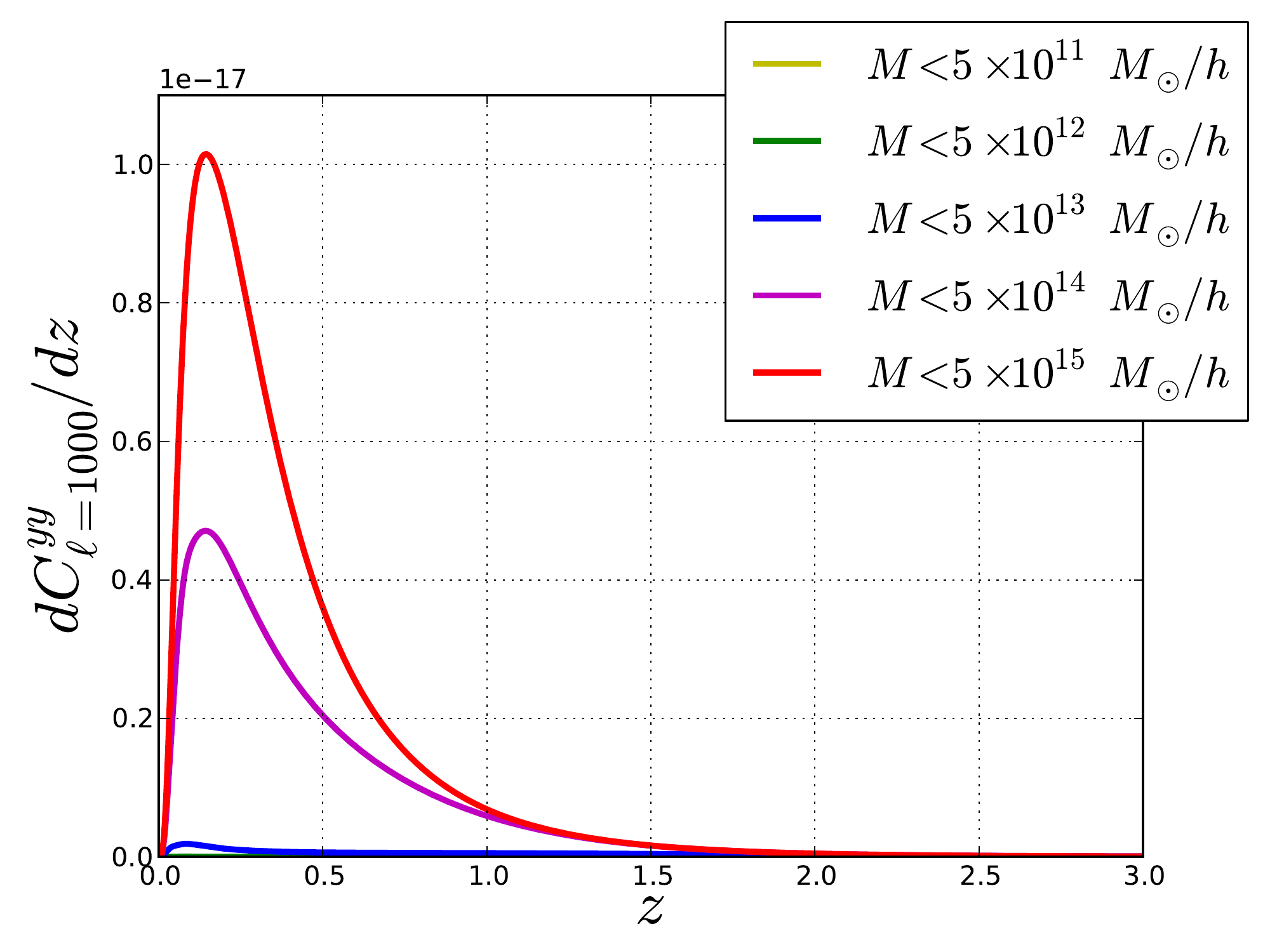}
\end{minipage}
\begin{minipage}[b]{0.32\linewidth}
\centering
\includegraphics[width=\textwidth]{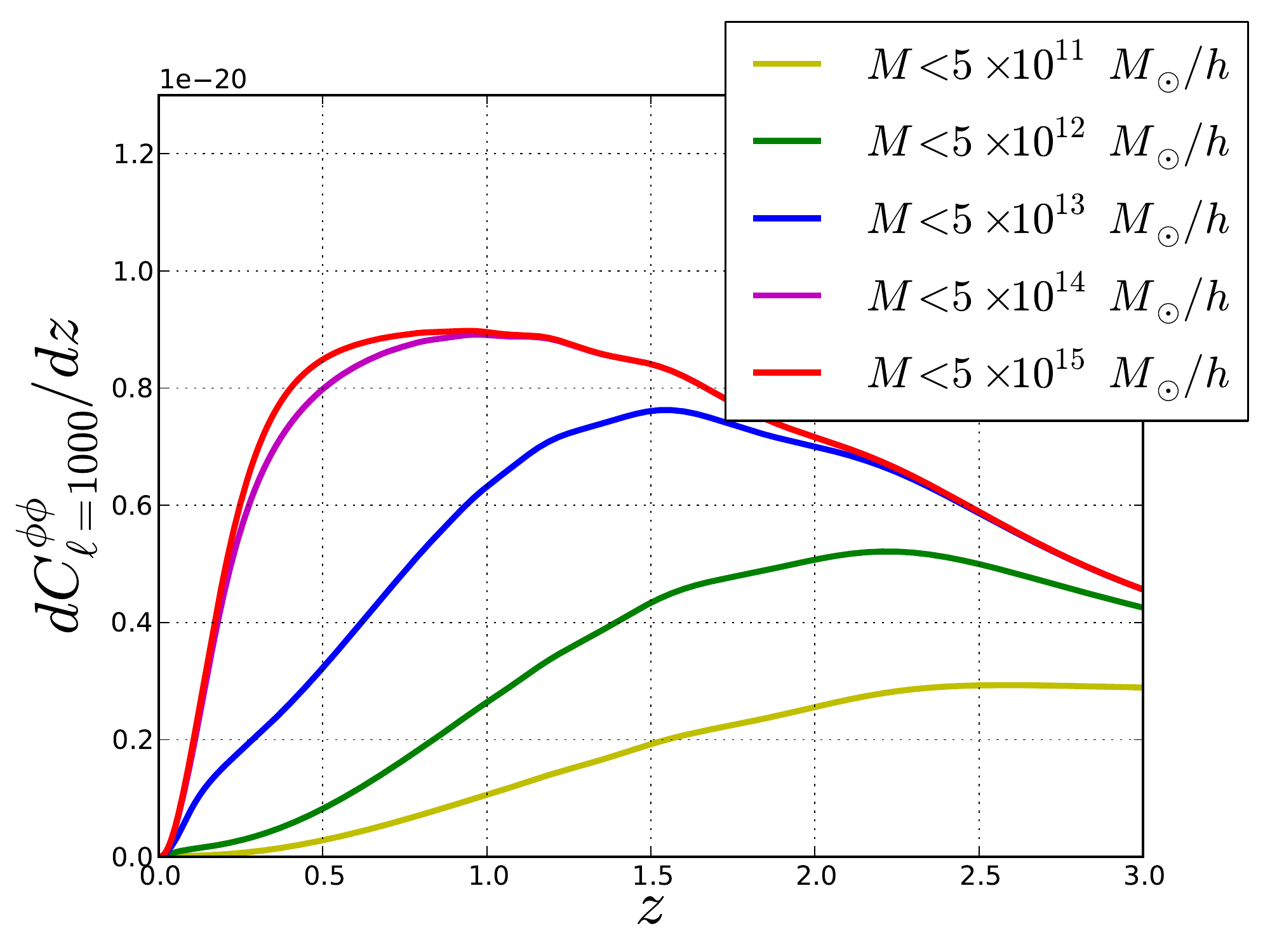}
\end{minipage}
\caption{Identical to Fig.~\ref{fig.dCldzell100}, but computed at $\ell=1000$, where the one-halo term dominates the cross-spectrum.  The tSZ auto-spectrum contributions here extend to lower masses and higher redshifts than at $\ell=100$, leading to a stronger expected cross-correlation with the CMB lensing signal.  See the text for further discussion.
\label{fig.dCldzell1000}}
\end{figure}

We plot the redshift and mass kernels for the tSZ -- CMB lensing cross-spectrum and each auto-spectrum at $\ell = 100$ and $1000$ in Figs.~\ref{fig.dCldzell100} and~\ref{fig.dCldzell1000}, respectively.  These multipoles probe different regimes of the cross-spectrum: at $\ell=100$, the cross-spectrum and CMB lensing auto-spectrum are dominated by the two-halo term, while the tSZ auto-spectrum is dominated by the one-halo term; at $\ell = 1000$, the cross-spectrum and tSZ auto-spectrum are dominated by the one-halo term, while the CMB lensing auto-spectrum receives significant contributions from both terms.  It is clear that at $\ell=100$ the mass and redshift kernels of the tSZ and CMB lensing auto-spectra are not as well-matched as they are at $\ell=1000$, which explains why $r_{\ell}^{y\phi}$ is larger at $\ell=1000$ than at $\ell=100$.

The most striking feature of Fig.~\ref{fig.dCldzell100} is the difference in the mass and redshift kernels for the tSZ auto-spectrum and the tSZ -- CMB lensing cross-spectrum.  At these low multipoles, the auto-spectrum is dominated by massive nearby clusters, while the cross-spectrum receives significant contributions out to very high redshift ($z \lsim 2$) and low masses ($5 \times 10^{12} \, M_{\odot}/h$).  This arises from the fact that the CMB lensing redshift and mass kernels upweight these scales in the cross-spectrum.  Similar behavior is also seen at $\ell=1000$ in Fig.~\ref{fig.dCldzell1000}, though the difference between the tSZ auto-spectrum and tSZ -- CMB lensing cross-spectrum is somewhat less dramatic in this case.

\begin{figure}
\begin{minipage}[b]{0.49\linewidth}
\centering
\includegraphics[width=\textwidth]{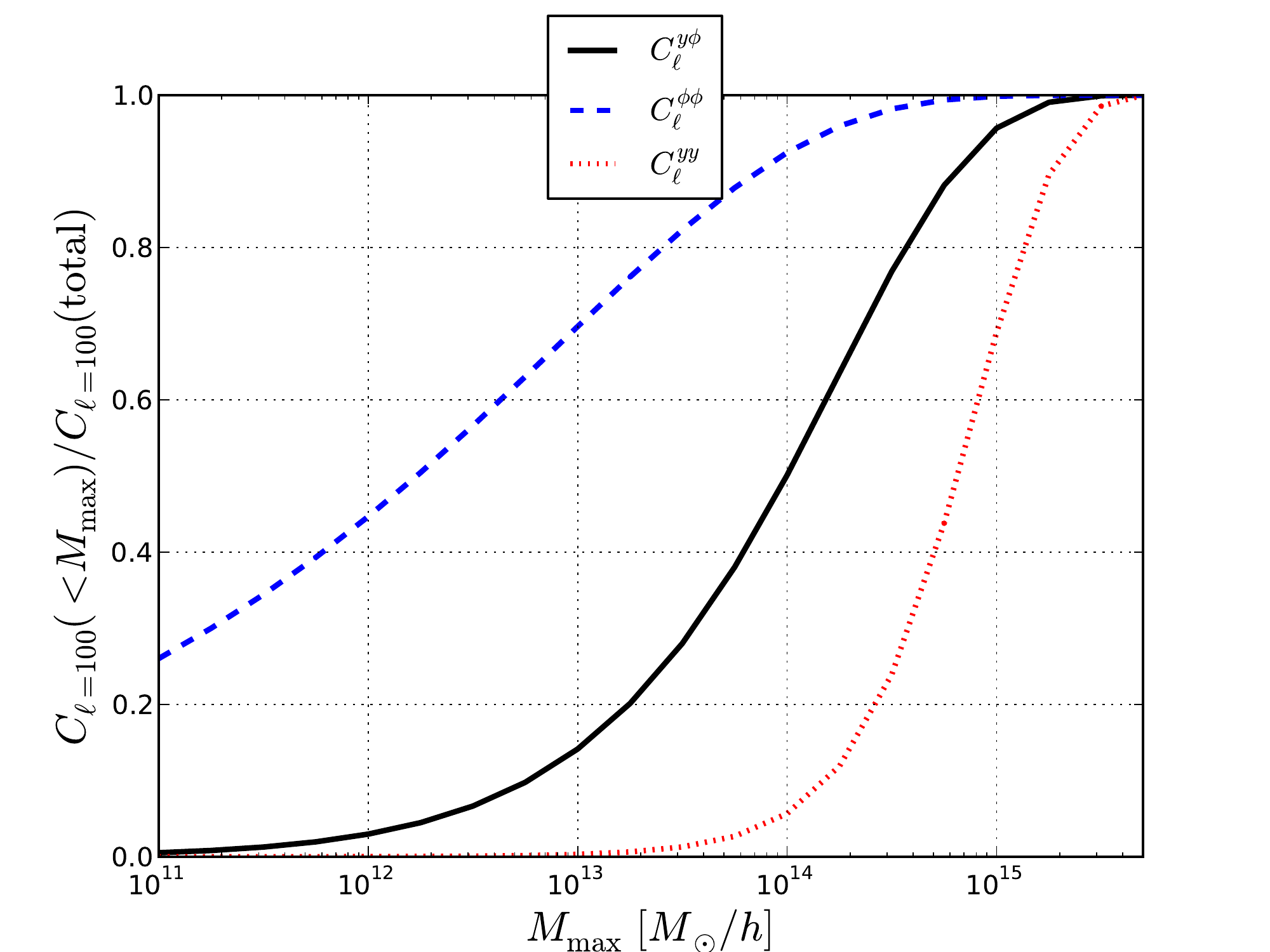}
\end{minipage}
\begin{minipage}[b]{0.49\linewidth}
\centering
\includegraphics[width=\textwidth]{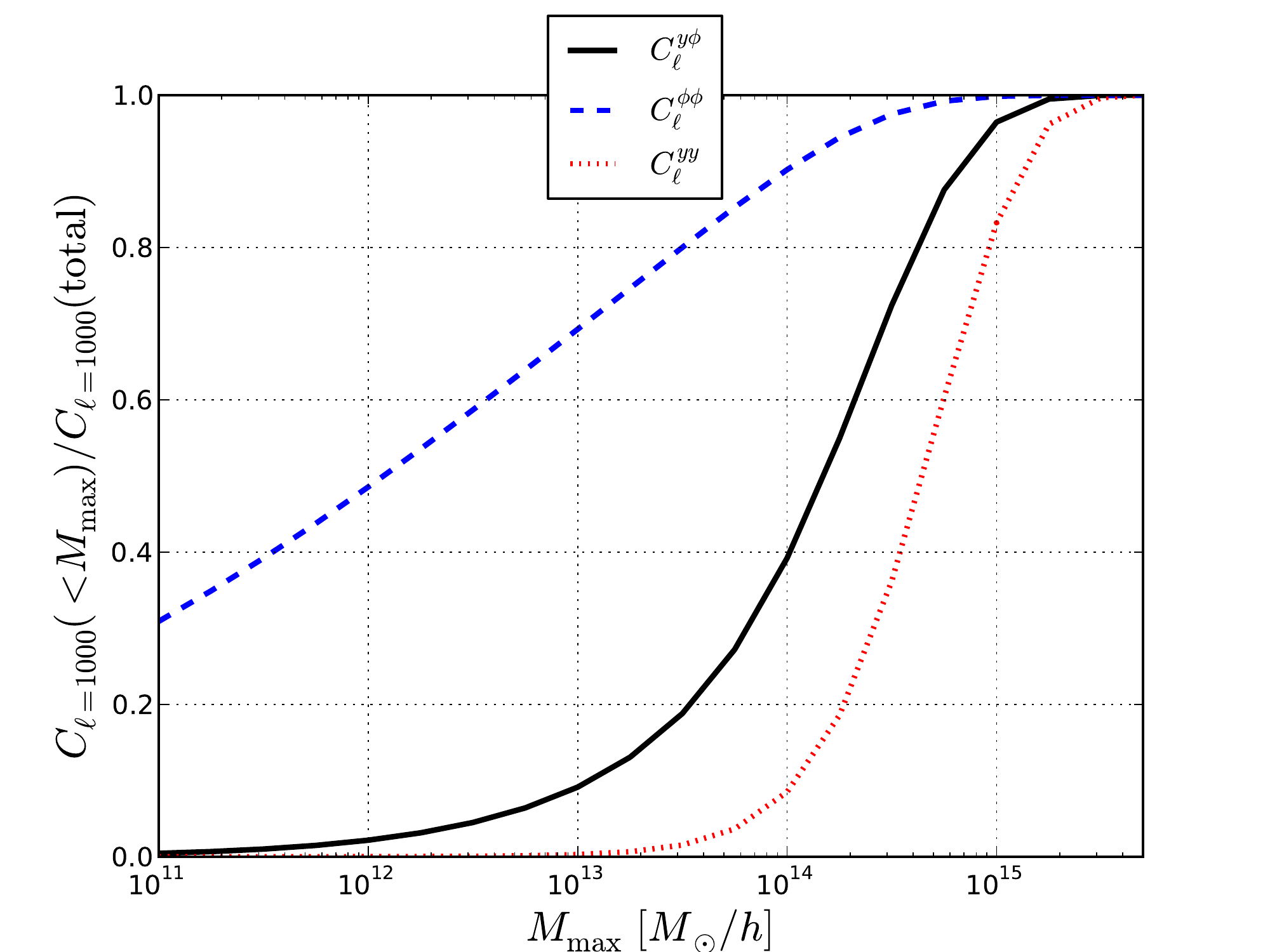}
\end{minipage}
\caption{Cumulative contributions to the tSZ -- CMB lensing cross-spectrum (black), CMB lensing auto-spectrum (blue), and tSZ auto-spectrum (red) as a function of halo mass, at $\ell=100$ (left panel) and $\ell=1000$ (right panel).  These results are integrated over all redshifts.  Note that the cross-spectrum receives significant contributions from much lower mass scales than the tSZ auto-spectrum.  See the text for further discussion.
\label{fig.Clmasscontribs}}
\end{figure}

These results are further illustrated in Fig.~\ref{fig.Clmasscontribs}, which shows the cumulative contribution to each power spectrum as a function of mass (integrated over all redshifts) at $\ell=100$ and $1000$.  The primary takeaway is that the tSZ -- CMB lensing cross-spectrum is indeed sourced by much less massive halos than those that source the tSZ auto-spectrum --- the difference is especially dramatic at $\ell=100$, but also seen at $\ell=1000$.  At $\ell=100$ (1000), $50$\% ($40$\%) of the cross-spectrum signal comes from masses below $10^{14} \, M_{\odot}/h$, while $6$\% ($9$\%) of the tSZ auto-spectrum signal comes from these masses\footnote{Note that masking massive, nearby clusters could change the tSZ auto-spectrum contributions significantly~\cite{Hill-Pajer2013,Shawetal2009}, but we will not consider this possibility here.}.  Our calculations for the tSZ auto-spectrum contributions agree with earlier results~\cite{Battagliaetal2012,Tracetal2011}.  Overall, the tSZ -- CMB lensing cross-spectrum provides a new probe of the ICM at mass and redshift scales that are presently unexplored.

As a final point regarding our theoretical calculations, we note that the cosmological parameter analysis presented in Section~\ref{sec:interp} accounts not only for statistical errors in our measurements, but also sample variance arising from the angular trispectrum of the tSZ -- CMB lensing signals.  Considering only the one-halo term in the flat-sky limit (which dominates the trispectrum at the scales we consider~\cite{Cooray2001}), the trispectrum contribution to the tSZ -- CMB lensing cross-power spectrum covariance matrix is~\cite{Komatsu-Seljak2002,Hill-Pajer2013}
\be
\label{eq.Tllyphi}
T_{\ell\ell'}^{y\phi} = \int dz \frac{d^2V}{dz d\Omega} \int dM \frac{dn(M,z)}{dM} \left| \tilde{y}_{\ell}(M,z) \right|^2 | \tilde{\phi}_{\ell'}(M,z) |^2 \,.
\ee
We also compute the trispectrum contribution to the tSZ auto-power spectrum when including it in our constraints, using an analogous expression to Eq.~(\ref{eq.Tllyphi}) (note that the tSZ trispectrum is not mentioned in~\cite{Planck2013ymap}).  Due to computational constraints, we only compute the
diagonal elements of $T_{\ell\ell'}^{y\phi}$ and $T_{\ell\ell'}^{yy}$, as the trispectra must be re-calculated for each variation in the cosmological parameters
or ICM physics model.  This choice also allows us to work in the limit in which the multipole bins in our measurement are uncorrelated.  Based on the results
of~\cite{Hill-Pajer2013}, the neglect of the off-diagonal elements should only have an effect for very large angular scales ($\ell \lsim 200$), and is likely only
important for the tSZ auto-spectrum (not the cross-spectrum).  Future high-precision measurements of the tSZ auto-spectrum (and possibly the tSZ -- CMB lensing cross-spectrum) for which the sample variance is comparable to the statistical errors will require more careful treatment and more efficient computational implementation of the trispectrum.

\section{Thermal SZ Reconstruction}
\label{sec:tSZrecon}
\subsection{Data and Cuts}
\label{sec:data}
Our thermal SZ reconstruction is based on the nominal mission maps from the first 15.5 months of operation of the Planck satellite~\cite{Planck2013overview}.  In particular, our ILC pipeline uses the data from the 100, 143, 217, 353, and 545 GHz channels of the Planck High-Frequency Instrument (HFI)~\cite{Planck2013HFI}.  Note that these frequencies span the zero point of the tSZ spectral function at 217 GHz (the spectral function is given after Eq.~(\ref{eq.tSZdef})).  We use the HFI 857 GHz map as an external tracer of dust emission, both from the Galaxy and from the CIB.  The FWHM of the beams in the HFI channels ranges from $9.66$ arcmin at 100 GHz to $4.63$ arcmin at 857 GHz. We do not use the data from the Low-Frequency Instrument (LFI) channels due to their significantly larger beams~\cite{Planck2013LFI}.  We work with the Planck maps in {\tt HEALPix}\footnote{{\tt http://www.healpix.jpl.nasa.gov }} format at the provided resolution of $N_{\rm side} = 2048$, for which the pixels have a typical width of $1.7$ arcmin.  At all HFI frequencies, we use the Zodi-corrected maps provided by the Planck collaboration.  Also, when needed for the 545 and 857 GHz maps, we convert from MJy/sr to ${\rm K_{CMB}}$ using the values provided in the Planck explanatory supplement.\footnote{We use the ``545-avg'' and ``857-avg'' values from {\tt http://www.sciops.esa.int/wikiSI/planckpla/index.php?title=UC\char`_CC\char`_Tables}.}

Our analysis pipeline is designed to follow the approach of~\cite{Planck2013ymap} to a large degree.  As in~\cite{Planck2013ymap}, we approximate the Planck beams as circular Gaussians, with FWHM values given in Table 1 of~\cite{Planck2013ymap}.  The first step of our analysis pipeline is to convolve all of the HFI maps to a common resolution of $10$ arcmin, again following~\cite{Planck2013ymap}.  This value is dictated by the angular resolution of the 100 GHz map, and is necessary in order to apply the ILC algorithm described in the following section. 

A key difference between our analysis and~\cite{Planck2013ymap} is that our ILC algorithm relies on the use of cross-correlations between independent ``single-survey'' maps, which prevents any auto-correlations due to noise in the input channel maps leaking into the output Compton-$y$ map --- effectively, the ILC minimizes ``non-tSZ'' signal in the reconstructed $y$-map, rather than ``non-tSZ + instrumental noise''.  Thus, instead of using the full Planck HFI nominal mission maps, we work with the ``survey 1'' (S1) and ``survey 2'' (S2) maps; we also include the associated masks for the S1 and S2 maps in our analysis below, as neither survey completely covers the full sky.  The only exception to this choice is our use of the full nominal mission 857 GHz map, which, as mentioned earlier, is used only as an external dust tracer rather than as a component in the ILC pipeline. 

We use the 857 GHz map to construct a mask that excludes regions of the sky with the brightest dust emission.  This mask is primarily meant to remove Galactic dust emission, but also removes a small fraction of the CIB emission.  We construct the mask by simply thresholding the 857 GHz map until the desired sky fraction remains.  Our fiducial results use a mask in which $70$\% of the sky is removed, i.e., $\fsky = 0.30$.  We also consider $\fsky = 0.20$ and $0.40$ cases as tests for our primary results. 

We construct a point source mask using the same approach as in~\cite{Planck2013ymap}.  We take the union of the individual point source masks provided at each of the LFI and HFI frequencies~\cite{Planck2013sources}.  In order to mask sources as thoroughly as possible, we use the $5\sigma$ catalogs rather than the $10\sigma$ catalogs.  The authors of \cite{Planck2013ymap} verify that this procedure removes all resolved radio sources, as well as an unknown number of IR sources.  Unresolved radio and IR point sources will still be present in the derived Compton-$y$ map, but this masking prevents their emission from significantly biasing the ILC algorithm. 

In our tSZ reconstruction pipeline, we take the union of the 857 GHz Galactic dust mask, the point source mask, and all of the masks associated with the S1 and S2 channel maps.  We will refer to this mask as the ``$y$-map mask''.  For our baseline results, which use the $\fsky = 0.30$ Galactic dust mask, the total sky fraction left in the $y$-map mask is $\fsky = 0.25180$. 

When estimating the tSZ -- CMB lensing cross-spectrum, we also must account for the mask associated with the Planck CMB lensing potential map.  The construction of this mask is described in full detail in~\cite{Planck2013lensing}.  In addition to masking Galactic dust and point sources, this mask also removes regions contaminated by CO emission, as well as extended nearby objects, such as the Magellanic Clouds.  Note that although some tSZ clusters are masked in the 143 GHz lensing reconstruction, they are not masked in the 217 GHz reconstruction, and thus in the publicly released map --- a minimum-variance combination of 143 and 217 GHz --- these tSZ clusters are not masked.  (The noise levels at the location of these clusters will be somewhat higher than elsewhere in the lensing potential map, since the signal there is only reconstructed from one of the two channels, but we neglect this small effect in our analysis.)  The $y$-map mask is sufficiently thorough that the inclusion of the lensing mask only covers a small amount of additional sky --- the total sky fraction left in the combined mask for our fiducial results is $\fsky = 0.25177$. 

We apodize all masks used in the analysis by smoothing them with a Gaussian beam of FWHM $10$ arcmin.  The apodization prevents excessive ringing or other artifacts when we compute power spectra.  The effective sky fraction is reduced by $<0.01$\% by the apodization (compared to using an unapodized mask), but we nonetheless take this into account by computing $\fsky$ via
\be
\label{eq.fskyapod}
f_{\rm sky} = \frac{1}{4\pi} \int d^2 \hat{n} M(\hat{n}) \,,
\ee
where $M(\hat{n})$ is the apodized mask.  Note that when estimating power spectra (see below), we correct for the effects of the mask using the average value of the square of the apodizing mask, rather than its mean~\cite{Planck2013lensing}:
\be
\label{eq.fsky2apod}
f_{\rm sky,2} = \frac{1}{4\pi} \int d^2 \hat{n} M^2(\hat{n}) \,.
\ee

\subsection{ILC}
\label{sec:ILC}
In order to construct a tSZ map over a large fraction of the sky using the Planck HFI maps, we implement a slightly modified version of the standard ILC technique.  In the ILC approach, one assumes that the observed temperature $T$ at each pixel $p$, of which there are $N_{\rm pix}$, in all $N_{\rm obs}$ maps (indexed by $i$ in the following) can be written as a linear combination of the desired signal ($y$) and noise ($n$):
\be
\label{eq.ILCdef}
T_i(p) = a_i y(p) + n_i(p) \,, 
\ee
where $a_i$ is the product of $T_{\rm CMB} = 2.726$ K and the tSZ spectral function at the $i^{\rm th}$ frequency, which is computed by integrating over the relevant bandpass for each channel.  We use the values of $a_i$ given in Table 1 of~\cite{Planck2013ymap}; we obtain values consistent with these when integrating the tSZ spectral function over the publicly released Planck HFI bandpasses~\cite{Planck2013HFIbands}.  The effect of the bandpass integration compared to simply using the central frequency of each channel is fairly small except for the 217 GHz channel, because it spans the tSZ null frequency (as emphasized in~\cite{Planck2013CIBxlens}).  For the 217 GHz channel, a na\"{i}ve calculation gives a result with the wrong sign and an amplitude that is incorrect by roughly an order of magnitude; thus, accounting for the bandpass in quite important at this frequency.

The standard ILC approach computes an estimate of the desired signal in each pixel $\hat{y}(p )$ by constructing the minimum-variance linear combination of the observed maps that simultaneously satisfies the constraint of unit response to the signal of interest.  Defining the ILC weights $w_i$ via $\hat{y}(p ) = w_i T_i(p )$, we thus seek to minimize the variance $\sigma_y^2 = N_{\rm pix}^{-1} \sum_p \left( \hat{y}(p ) - \langle \hat{y} \rangle \right)^2$ while enforcing the constraint $w_i a_i = 1$ (summation over repeated indices is implied throughout).  A simple derivation using Lagrange multipliers yields the desired result~\cite{Eriksenetal2004}:
\be
\label{eq.ILCweightsstandard}
w_j = \frac{a_i (\hat{R}^{-1})_{ij}}{a_k (\hat{R}^{-1})_{kl} a_l} \,,
\ee
where $\hat{R}_{ij} = N_{\rm pix}^{-1} \sum_p  \left( T_i(p ) - \langle T_i \rangle \right) \left( T_j(p ) - \langle T_j \rangle \right)$ is the empirical covariance matrix of the (masked) observed maps.  Note that the weights $w_{i}$ have units of ${\rm K}_{\rm CMB}^{-1}$ in this formulation.  Also note that it is important to mask the Galaxy and point sources \emph{before} applying the ILC algorithm, as otherwise the strong emission from these sources can heavily bias the derived weights.  We apply the full $y$-map mask described in the previous section before running our ILC algorithm.

This standard ILC approach can be extended to explicitly prevent any signal from the primordial CMB leaking into the derived $y$-map.  This step is facilitated by the fact that the CMB spectrum is known to be a blackbody to extremely good precision.  Eq.~(\ref{eq.ILCdef}) is now modified to explicitly include the CMB as a second signal, in addition to the tSZ signal:
\be
\label{eq.CILCdef}
T_i(p) = a_i y(p) + b_i s(p ) + n_i(p ) \,, 
\ee
where $s(p )$ is the CMB signal in pixel $p$ and $b_i$ is the CMB spectrum evaluated at each map's frequency, which is simply unity for maps in ${\rm K_{CMB}}$ units (as ours are).  Imposing the condition that the ILC $y$-map has zero response to the CMB signal, i.e., $w_i b_i = 0$, and solving the system with Lagrange multipliers leads to the following modified version of Eq.~(\ref{eq.ILCweightsstandard})~\cite{Remazeillesetal2011}:
\be
\label{eq.CILCweights}
w_j = \frac{\left( b_k (\hat{R}^{-1})_{kl} b_l \right) a_i (\hat{R}^{-1})_{ij} - \left( a_k (\hat{R}^{-1})_{kl} b_l \right) b_i (\hat{R}^{-1})_{ij}}{\left( a_k (\hat{R}^{-1})_{kl} a_l \right) \left( b_m (\hat{R}^{-1})_{mn} b_n \right) - \left( a_k (\hat{R}^{-1})_{kl} b_l \right)^2} \,.
\ee
Following~\cite{Remazeillesetal2011}, we will refer to this approach as the ``constrained'' ILC (CILC).  This method ensures that no CMB signal leaks into the derived $y$-map.  We implement Eq.~(\ref{eq.CILCweights}) in our tSZ reconstruction pipeline, with additional modifications described in the following.

Our method modifies the standard ILC or CILC approach in two ways.  First, we note that the variance $\sigma_y^2$ can be written as a sum over the power spectrum $C_{\ell}^y$ as
\be
\sigma_y^2 = \sum_{\ell=0}^{\infty} \frac{2\ell +1}{4\pi} C_{\ell}^y \,.
\ee
Thus, the minimization of the variance in the ILC map is the minimization of this sum, taken over all $\ell$.  Our goal is to produce an ILC $y$-map to be used for cross-correlation with the Planck CMB lensing potential map (and possibly other extragalactic maps).  It is therefore most important to minimize the extragalactic contamination, especially from the CIB and IR sources, rather than contamination from Galactic dust.   Hence, we modify the ILC to minimize only a restricted sum over the power spectrum:
\be
\label{eq.restrictedCellsum}
\tilde{\sigma}_y^2 = \sum_{\ell=\ell_a}^{\ell_b} \frac{2\ell +1}{4\pi} C_{\ell}^y \,.
\ee
We choose $\ell_a = 300$ and $\ell_b = 1000$, as Galactic dust emission is subdominant to the CIB over this $\ell$ range (after heavily masking the Galaxy), but the tSZ signal is still significant.  In general, even after thorough masking, Galactic dust dominates over the tSZ at low-$\ell$, while at high-$\ell$ the CIB takes over~\cite{Planck2013ymap}.  We verify that our results are stable to modest variations in these values.

In addition to minimizing a restricted sum over the power spectrum, we implement a second modification to the ILC algorithm which prevents instrumental noise from contributing to the ILC weights.  Instead of minimizing $\tilde{\sigma}_y^2$ computed from the auto-spectra of the channel maps, we compute this quantity using cross-spectra of the S1 and S2 channel maps.  The minimized quantity is now a cross-statistic:
\be
\label{eq.restrictedCellsumcross}
\tilde{\sigma}_{y_{12}}^2 = \sum_{\ell=\ell_a}^{\ell_b} \frac{2\ell +1}{4\pi} C_{\ell}^{y_1 y_2} \,,
\ee
where the ``1'' and ``2'' subscripts refer to S1 and S2.  Note that this approach now implies that we construct both an S1 and S2 ILC $y$-map, but using the same weights $w_i$, i.e., $y_1(p ) = w_i T^1_i(p )$ and $y_2(p ) = w_i T^2_i(p )$, where $T^1_i$ and $T^2_i$ are the $i^{\rm th}$ S1 and S2 channel maps, respectively.

Our ILC approach thus consists of finding the linear combination of maps that minimizes $\tilde{\sigma}_{y_{12}}^2$ in Eq.~(\ref{eq.restrictedCellsumcross}) while simultaneously requiring unit response to a tSZ spectrum and zero response to a CMB spectrum.  Straightforward algebra using Lagrange multipliers yields the final expression for the weights --- it is identical to the CILC weights in Eq.~(\ref{eq.CILCweights}), except that the covariance matrix $\hat{R}_{ij}$ is now replaced by a slightly modified quantity:
\be
\label{eq.modifiedcovmatrix}
\hat{\tilde{R}}_{ij} = \sum_{\ell=\ell_a}^{\ell_b} \frac{2\ell + 1}{2\pi} C_{\ell, 12}^{ij} \,,
\ee
where $C_{\ell, 12}^{ij}$ is the cross-power spectrum of the $i^{\rm th}$ S1 channel map and the $j^{\rm th}$ S2 channel map.  Eqs.~(\ref{eq.CILCweights}) and~(\ref{eq.modifiedcovmatrix}) define our modified CILC approach to constructing a $y$-map.

As mentioned above, the weights determined by Eqs.~(\ref{eq.CILCweights}) and~(\ref{eq.modifiedcovmatrix}) actually yield two $y$-maps: one constructed by applying the weights to the S1 channel maps and one from the S2 channel maps.  In Section~\ref{sec:Clyy} we compute the tSZ auto-power spectrum by taking the cross-power spectrum of the S1 and S2 $y$-maps.  In order to compute the cross-power spectrum of the tSZ and CMB lensing signals, we co-add the S1 and S2 $y$-maps to obtain a final $y$-map which we then cross-correlate with the CMB lensing potential map.  The co-addition of the S1 and S2 $y$-maps is performed with an inverse variance weighting to obtain the minimum-variance combination of the two maps.

\begin{figure}
\centering
\includegraphics[width=\textwidth]{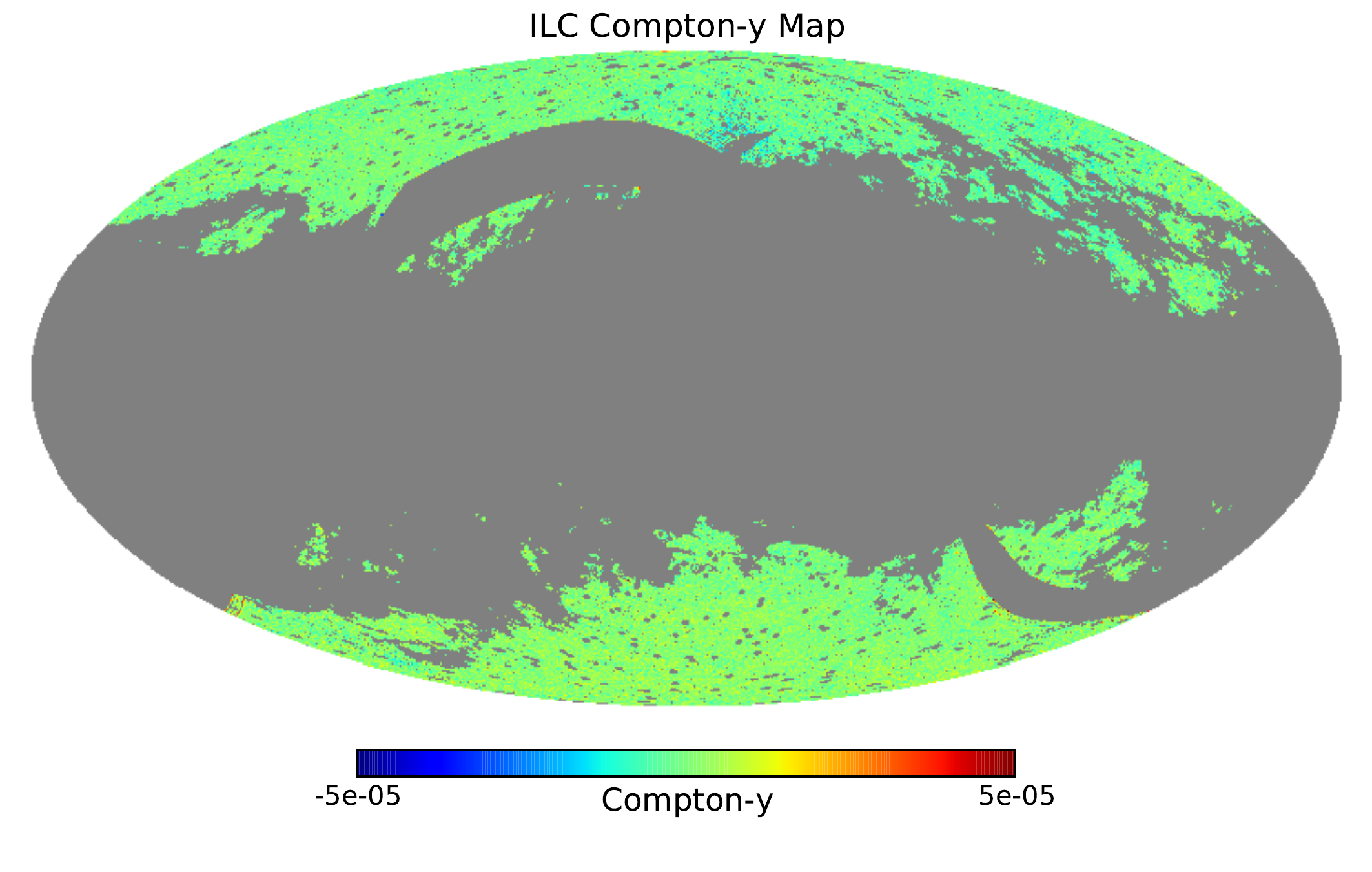}
\caption{The Compton-$y$ map reconstructed by our ILC pipeline for the fiducial $\fsky = 0.30$ case, plotted in Galactic coordinates.  Note that the mean of the map has been removed before plotting for visual clarity.
\label{fig.ymap}}
\end{figure}

\begin{figure}
\centering
\includegraphics [totalheight=0.4\textheight]{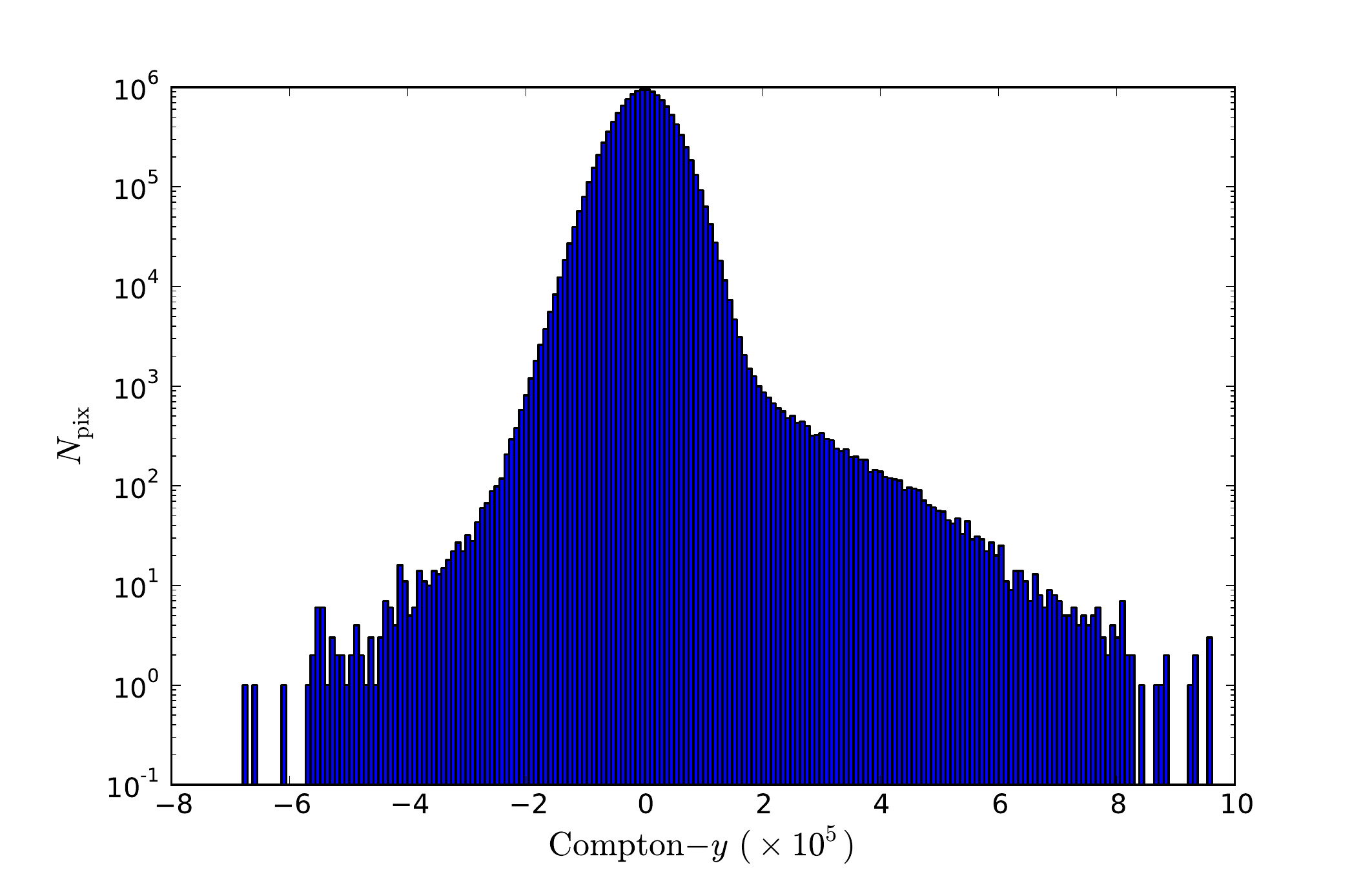}
\caption{Histogram of Compton-$y$ values in our fiducial ILC $y$-map.  The non-Gaussian tail extending to positive $y$-values provides some evidence that the map does indeed contain tSZ signal.  The negative tail is likely due to residual Galactic dust in the map.  See~\cite{Wilsonetal2012,Hill-Sherwin2013,Hilletal2013} for related work on interpreting the moments of Compton-$y$ histograms.
\label{fig.ymaphist}}
\end{figure}

\begin{figure}
\centering
\includegraphics[totalheight=0.5\textheight]{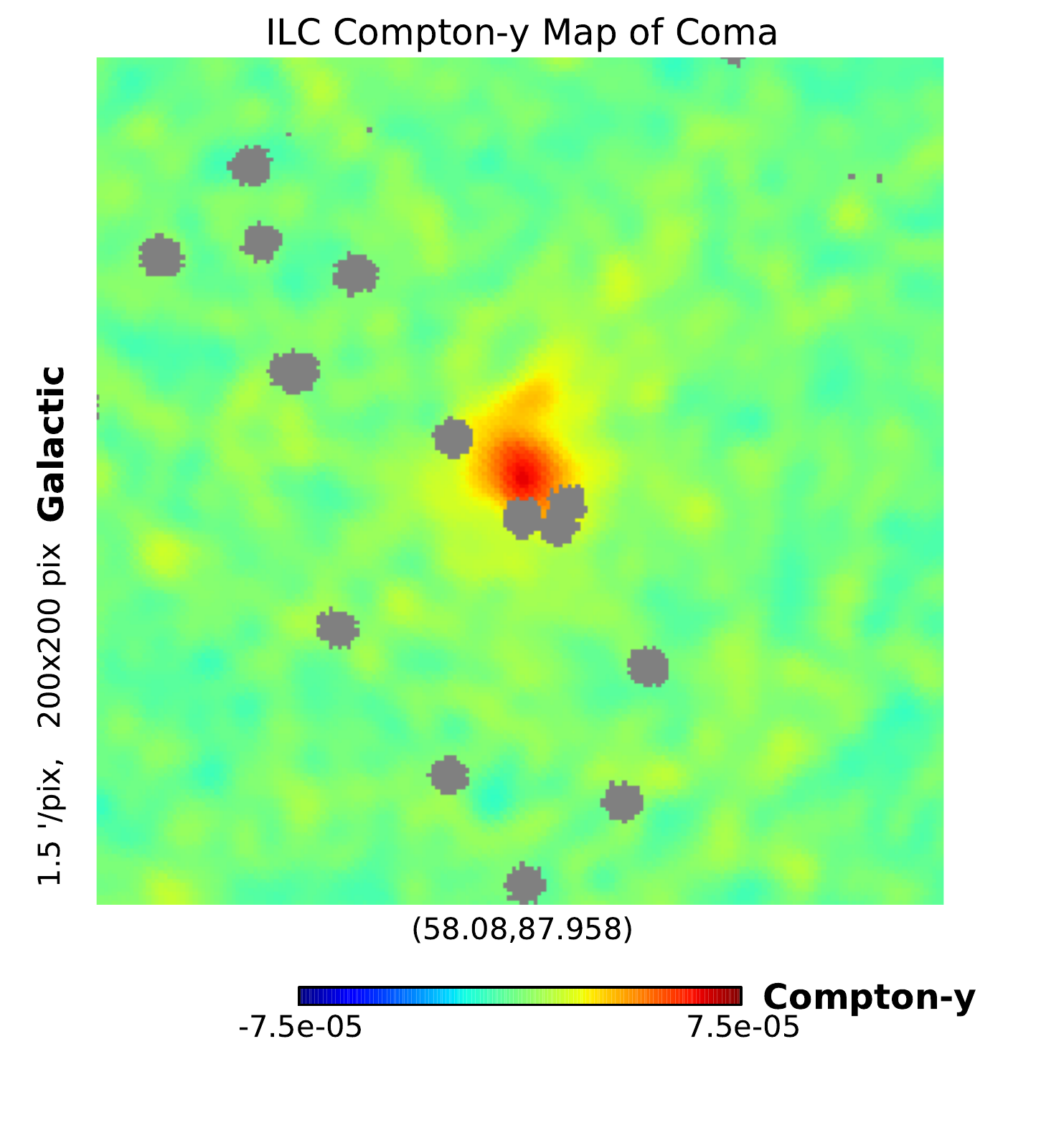}
\caption{Sub-map of a $5^{\circ}$-by-$5^{\circ}$ region centered on the Coma cluster in the ILC $y$-map (after removing the map mean).  The masked areas are the locations of point sources in this field.
\label{fig.ymapComa}}
\end{figure}

Fig.~\ref{fig.ymap} shows the final co-added $y$-map obtained from our modified CILC pipeline for the fiducial $\fsky = 0.30$ (note that the actual final $\fsky = 0.25180$ after accounting for the S1 and S2 masks and the point source mask, as mentioned in Section~\ref{sec:data}).  The mean has been removed from this map before plotting, in order to allow better visual clarity.  Fig.~\ref{fig.ymaphist} shows the histogram of Compton-$y$ values in this map.  Apart from the dominant Gaussian noise component, there is a clear non-Gaussian tail extending to positive Compton-$y$ values, which provides some evidence that there is indeed tSZ signal in the map.  A smaller non-Gaussian tail extending to negative values is likely caused by residual Galactic dust in the map.  Interpreting the moments of this histogram~\cite{Wilsonetal2012,Hill-Sherwin2013,Hilletal2013} is an interesting prospect for future work, but will require careful consideration of the non-tSZ components, which is beyond the scope of our analysis here.  As further evidence of the success of our reconstruction, we show a sub-map of a region centered on the Coma cluster in Fig.~\ref{fig.ymapComa}.  Coma is the most significant tSZ cluster in the Planck sky~\cite{Planck2012Coma}.  The central pixel's $y$-value in Coma in our (mean-subtracted) ILC map is $\approx 7 \times 10^{-5}$, which is fairly consistent with the determination in~\cite{Planck2012Coma} (see their Figs.~4 or 5).  Although we cannot easily estimate the error on this quantity without simulations, this result provides additional evidence for the overall success of our ILC reconstruction.

The primary disadvantage of the ILC technique is its assumption that the signal of interest ($y$ in Eq.~(\ref{eq.ILCdef})) is completely uncorrelated with the noise and foregrounds at each frequency ($n_i$ in Eq.~(\ref{eq.ILCdef})).  In the case of the tSZ signal, this assumption is likely violated by the correlation between the tSZ and IR sources (e.g., dusty star-forming galaxies)~\cite{Addisonetal2012}, for which some evidence has recently been found~\cite{Mesingeretal2012,Hincksetal2013}.  However, given the indirect nature of this evidence and difficulty in assessing the amplitude of the correlation, we neglect it for the present time (as in~\cite{Planck2013ymap}).  Future tSZ reconstructions at higher SNR --- for example, using the complete Planck data set --- may need to consider its implications.

Finally, we note that our ILC approach differs somewhat from those employed in the Planck $y$-map analysis~\cite{Planck2013ymap}.  In particular, we do not implement a method that combines reconstructions at varying angular scales, such as the Needlet ILC (NILC)~\cite{Remazeillesetal2013}.  Since the Planck HFI maps are smoothed to a common 10 arcmin resolution before the tSZ reconstruction is performed in both~\cite{Planck2013ymap} and this work, it seems difficult to take advantage of the additional power of a multi-scale approach.  In future work we plan to combine the novel elements of our ILC pipeline described above with a multi-scale reconstruction pipeline, allowing the simultaneous use of data from Planck (HFI and LFI), WMAP, and ground-based experiments such as ACT and SPT.

Although the Compton-$y$ map reconstructed from our pipeline is not free of residual contamination, as evident in the histogram and power spectrum (see the following section), we believe that it should nonetheless be useful for cross-correlation studies if treated carefully.  Thus, we make available to the community a {\tt HEALPix} version of our fiducial $\fsky = 0.30$ map\footnote{{\tt http://www.astro.princeton.edu/\textasciitilde jch/ymapv1/ }}.


\subsection{Thermal SZ Auto-Power Spectrum}
\label{sec:Clyy}
We compute the tSZ auto-power spectrum by cross-correlating the S1 and
S2 Compton-$y$-maps obtained with the pipeline described in Section~\ref{sec:ILC}.  Given a reconstructed Compton-$y$ map, $\hat{y}(\hat{n})$, we perform a spherical harmonic transform via $\hat{y}_{\ell m} = \int d^2 \hat{n} \, Y^{*}_{\ell m} (\hat{n}) \, \hat{y}(\hat{n})$ and compute the tSZ power spectrum using a simple pseudo-$C_{\ell}$ estimator:
\be
\label{eq.pseudoCl}
\hat{C}_{\ell}^{yy} = \frac{f_{\rm sky,2}^{-1}}{2\ell + 1} \sum_m \hat{y}^{1}_{\ell m} \hat{y}^{2*}_{\ell m} \,,
\ee
where $f_{\rm sky,2}$ is given by Eq.~(\ref{eq.fsky2apod}), computed
using only the $y$-map mask (clearly the lensing potential is not
involved in this analysis), and $y^{1}$ and $y^{2}$ refer to the S1
and S2 $y$-maps, respectively.  Our power spectrum estimate also accounts for the deconvolution of the 10 arcmin smoothing applied to all channel maps before the ILC reconstruction pipeline.  We ``whiten'' the estimated power spectrum by multiplying by $\ell(\ell+1)/(2\pi)$ and then bin it using bins identical to those chosen in~\cite{Planck2013ymap} in order to allow a direct comparison of the results.  We follow~\cite{Planck2013lensing,Planck2013CIBxlens,Planck2013ymap} in neglecting any correlations arising in the error bars due to mode coupling induced by the mask; we simply correct for the power lost through masking with the factor $f_{\rm sky,2}^{-1}$.

\begin{figure}
\centering
\includegraphics[totalheight=0.5\textheight]{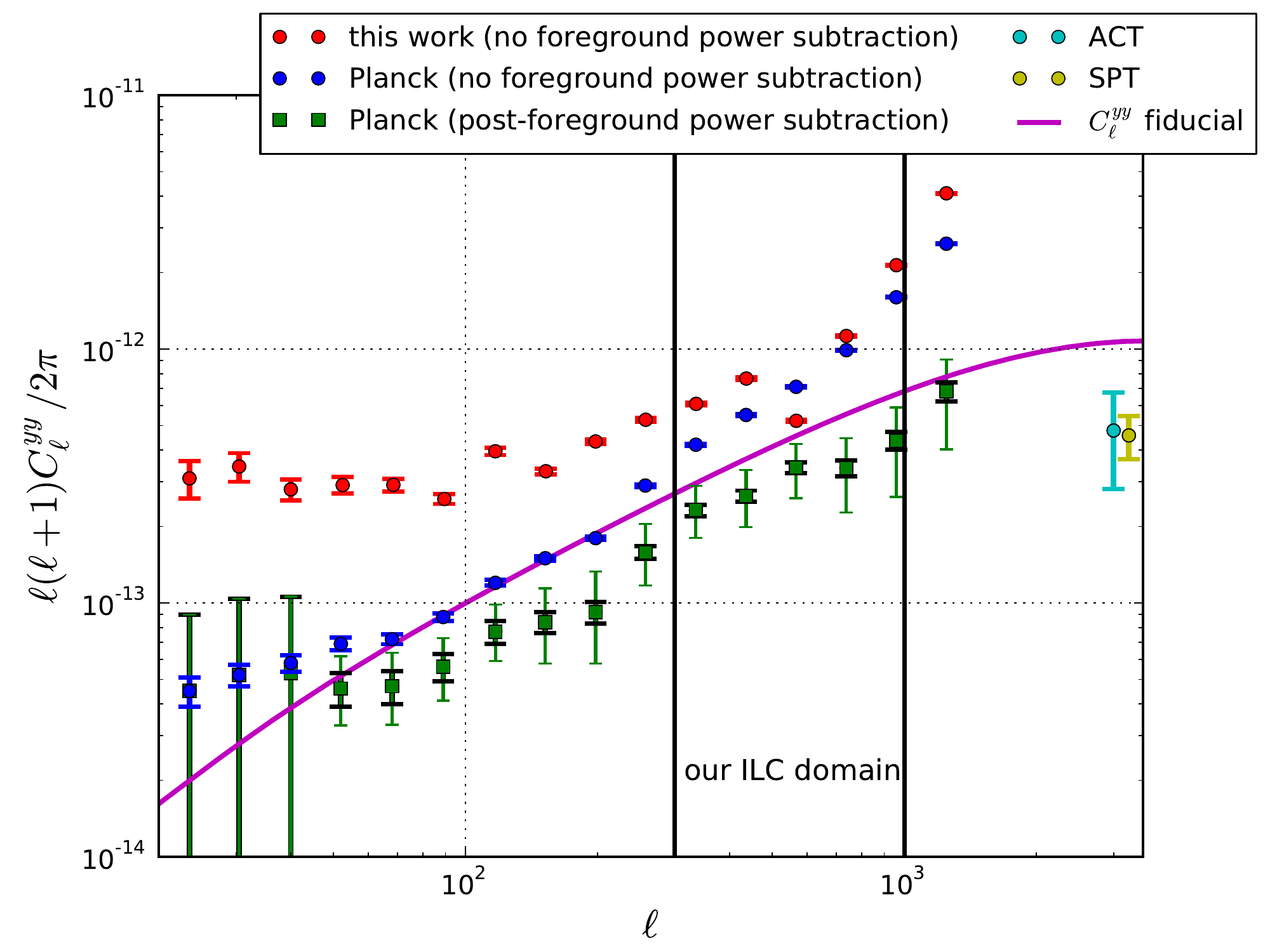}
\caption{The tSZ power spectrum estimated from our fiducial S1 and S2 Compton-$y$ maps (red circles), with statistical errors only.  The Planck results from~\cite{Planck2013ymap} are shown as blue circles and green squares --- the green squares have been corrected to subtract residual power from leakage of CIB and IR and radio point sources into the $y$-map constructed in~\cite{Planck2013ymap}, while the blue circles are the uncorrected results (with statistical errors only).  We have not attempted to subtract the residual power due to these foregrounds; the red circles are the raw cross-spectrum of our S1 and S2 ILC $y$-maps, which are in reasonable agreement with the uncorrected Planck points (blue circles) over the region in which our ILC algorithm minimizes residual contamination ($300 < \ell < 1000$, as delineated on the plot; see Section~\ref{sec:ILC}).  The thick black error bars are the statistical errors alone on the foreground-corrected Planck points, while the thin green error bars are the combined statistical and foreground-subtraction uncertainties (from Table 3 of~\cite{Planck2013ymap}).  The solid magenta curve shows the tSZ power spectrum predicted by our fiducial model.  The cyan and yellow points are the most recent ACT~\cite{Sieversetal2013} and SPT~\cite{Crawfordetal2013} constraints on the tSZ power spectrum amplitude at $\ell=3000$, respectively (the SPT point is slightly offset for visual clarity).  Note that the ACT and SPT constraints are derived in a completely different data analysis approach, in which the tSZ signal is constrained through its indirect contribution to the total CMB power measured at $\ell = 3000$ by these experiments; in other words, a $y$-map is not constructed.  We use the corrected Planck points with their full errors when deriving cosmological and astrophysical constraints in Section~\ref{sec:interp}.
\label{fig.Clyymeas}}
\end{figure}

Fig.~\ref{fig.Clyymeas} shows the tSZ power spectrum estimated from the cross-spectrum of our fiducial $f_{\rm sky} = 0.30$ S1 and S2 ILC Compton-$y$ maps, as well as the Compton-$y$ power spectrum estimated in~\cite{Planck2013ymap} before and after subtracting residual contributions to the power spectrum from clustered CIB, IR point sources, and radio point sources.  Note that we have not attempted to subtract these residual contributions from our tSZ power spectrum shown in Fig.~\ref{fig.Clyymeas}.  Our data points (red circles) should be compared directly to the uncorrected Planck points shown as blue circles; moreover, they should only be compared over the region in which our ILC is designed to remove contamination, i.e., $300 < \ell < 1000$ (as delineated in the figure).  Within this region, our results are in reasonable agreement with the uncorrected Planck results.  The green squares show the Planck points after power from residual foregrounds is subtracted.  In general, the uncorrected tSZ power spectra are contaminated by power from residual Galactic dust at low multipoles and from residual CIB, IR sources, and radio sources at high multipoles~\cite{Planck2013ymap}.  Simple tests varying the sky fraction used in our ILC Compton-$y$ reconstruction give power spectra broadly consistent with that shown in Fig.~\ref{fig.Clyymeas}; the $\fsky = 0.20$ results appear to have slightly more CIB leakage and less Galactic dust, while the opposite is true for the $\fsky = 0.40$ results.

It is also important to note that the error bars shown on our data points and the uncorrected Planck data points in Fig.~\ref{fig.Clyymeas} are statistical errors only:
\be
\label{eq.Clyymeaserr}
\left( \Delta \hat{C}_{\ell}^{yy} \right)^2 = \frac{1}{f_{\rm sky,2}} \frac{2}{(2\ell + 1) \Delta \ell} \left(\hat{C}_{\ell}^{yy}\right)^2 \,,
\ee
where $\Delta \ell$ is the width of a given multipole bin centered at $\ell$.  Note that $\hat{C}_{\ell}^{yy}$ is the measured tSZ auto-power spectrum, i.e., it includes the noise bias.  We show both the statistical-only and statistical+foreground errors on the corrected Planck points.  Neither set of error bars includes contributions from sample variance, although we will include this effect in our cosmological analysis in Section~\ref{sec:interp}.  The uncertainties arising from the subtraction of residual foreground contributions dominate the total errors on the corrected Planck data points~\cite{Planck2013ymap}.  We emphasize again that the visual discrepancy between the red and green points in Fig.~\ref{fig.Clyymeas} is not a sign that our $y$-map is significantly more contaminated than that in~\cite{Planck2013ymap}, but only a reflection of the fact that we have not subtracted residual foreground contributions to the auto-spectrum of the $y$-map.  Within $300 < \ell < 1000$, the raw power spectrum of our ILC $y$ reconstruction is quite similar to that of~\cite{Planck2013ymap}, as seen in the similarity between the red and blue points over this multipole range in Fig.~\ref{fig.Clyymeas}. Our aim is to measure the cross-correlation of the tSZ and CMB lensing signals, rather than the tSZ auto-spectrum; the latter requires a higher threshold for removing foregrounds and contamination.  We do not strive to directly repeat the foreground analysis of~\cite{Planck2013ymap} here, and instead will simply use their tSZ power spectrum in combination with our tSZ -- CMB lensing cross-power spectrum in the cosmological interpretation of our results in Section~\ref{sec:interp}.

\section{Thermal SZ -- CMB Lensing Cross-Power Spectrum}
\label{sec:tSZxlensing}
\subsection{CMB Lensing Potential Map}
\label{sec:lensmap}
To lowest order in the CMB lensing potential $\phi$, gravitational lensing introduces a correlation between the lensed temperature field and the gradient of the unlensed temperature field (see Eq.~(\ref{eq.Tlensed})).  This correlation leads to statistical anisotropy in the observed (lensed) CMB temperature field on the arcminute angular scales where lensing effects become detectable (the RMS deflection of a CMB photon due to lensing is $\sim 2$--$3$ arcmin).  Using this property of the lensed CMB, it is possible to reconstruct a map of the lensing potential itself, $\phi(\hat{n})$~\cite{Okamoto-Hu2003,Hirata-Seljak2003}.  The Planck team performed a detailed study of these effects in the nominal mission data, including reconstructions of the lensing potential at 100, 143, and 217 GHz and a $25\sigma$ detection of the power spectrum of the lensing potential~\cite{Planck2013lensing}.

In our measurement of the tSZ -- CMB lensing cross-power spectrum, we utilize the publicly released Planck CMB lensing potential map, which is a minimum-variance combination of the 143 and 217 GHz reconstructions.  As described in Section~\ref{sec:data}, we combine the mask associated with the CMB lensing potential map with the mask associated with our ILC $y$-map when computing the tSZ -- CMB lensing cross-power spectrum.  We refer the reader to~\cite{Planck2013lensing} for a complete description of the Planck CMB lensing reconstruction.  We have verified that our pipeline produces a lensing potential power spectrum estimate from the publicly released map which is broadly consistent with that presented in~\cite{Planck2013lensing}.

Note that the inclusion of the 217 GHz channel in the publicly released map is crucial, as it allows reconstruction to be performed in regions of the sky centered on tSZ clusters that must be masked in the 143 GHz reconstruction.  The 143 GHz masking of the tSZ-detected clusters does imply that some small effects could be present at these locations in the combined 143+217 GHz map, such as changes in the effective noise level and estimator normalization.  However, given that only the most massive clusters in the universe are masked in this procedure ($\sim 500$--$600$ objects\footnote{These are SNR $>5$ clusters detected with the MMF1 or MMF3 pipelines~\cite{Planck2013SZcatalog}.}), we do not expect it to significantly affect our measurement of the tSZ -- CMB lensing cross-spectrum, which is dominated by clusters well below the Planck detection threshold (see Section~\ref{sec:theoryPS}).  Fig.~\ref{fig.Clmasscontribs} indicates that the clusters masked in the 143 GHZ lensing reconstruction contribute $\approx 10$\% of the tSZ -- CMB lensing cross-power spectrum signal.  The change in the effective normalization of the lensing estimator in the combined 143+217 GHz map at these locations is a factor of $\approx 2$ due to the masking, and thus it is possible that our treatment has missed an error of roughly this factor on $\approx 10$\% of the signal.  This level of error is well below the statistical error on our detection (see the next two sections), and thus we do not consider these effects further.  Future high-signal-to-noise detections of this cross-spectrum will need to carefully consider the effects of tSZ masking in the lensing reconstruction --- performing the lensing reconstruction on either a 217 GHz map or an ILC CMB map constructed to have zero tSZ signal would likely solve this problem, but simulations should be used to completely characterize the effects.

\subsection{Measurement}
\label{sec:yphimeas}

\begin{figure}
\centering
\includegraphics[width=\textwidth]{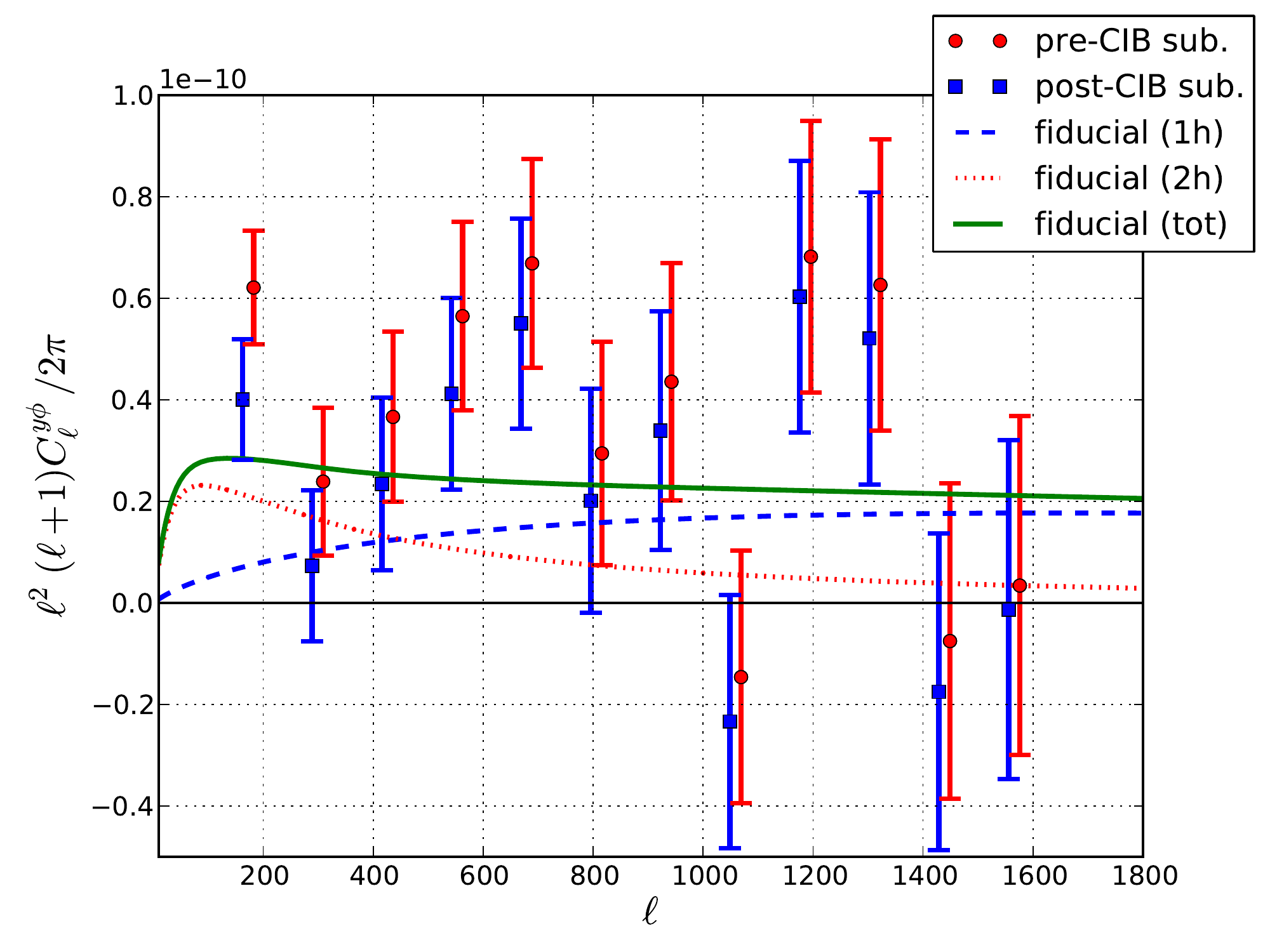}
\caption{The tSZ -- CMB lensing cross-power spectrum estimated from our fiducial co-added Compton-$y$ map and the publicly released Planck lensing potential map is shown in the red circles, with statistical errors computed via Eq.~(\ref{eq.Clyphimeaserr}).  The blue squares show the cross-power spectrum after a correction for CIB leakage into the $y$-map has been applied (see Section~\ref{sec:CIBsub} for a description of the subtraction procedure).  The errors on the blue squares include statistical uncertainty and additional systematic uncertainty arising from the CIB subtraction.  The final significance of the CIB-corrected measurement of $C_{\ell}^{y\phi}$ is $6.2\sigma$.  The solid green curve shows the theoretical prediction of our fiducial model for the tSZ -- CMB lensing cross-spectrum (see Section~\ref{sec:theory}) --- we emphasize that this curve is not a fit to the data, but rather an independent prediction.  It agrees well with our measurement: $\chi^2 = 14.6$ for $12$ degrees of freedom.  The dashed blue and dotted red curves show the one- and two-halo contributions to the fiducial theoretical cross-spectrum, respectively.  Our measurement shows clear evidence of contributions from both terms.
\label{fig.Clyphimeas}}
\end{figure}

We use the co-added $y$-map described in Section~\ref{sec:ILC} and the CMB lensing potential map described in Section~\ref{sec:lensmap} to measure the tSZ -- CMB lensing cross-power spectrum for our fiducial $f_{\rm sky} = 0.3$ mask.  We compute the cross-power spectrum using a simple pseudo-$C_{\ell}$ estimator analogous to that in Eq.~(\ref{eq.pseudoCl}):
\be
\label{eq.pseudoClyphi}
\hat{C}_{\ell}^{y\phi} = \frac{f_{\rm sky,2}^{-1}}{2\ell + 1} \sum_m \hat{y}_{\ell m} \hat{\phi}^{*}_{\ell m} \,,
\ee
where $f_{\rm sky,2} = 0.24281$ is given by Eq.~(\ref{eq.fsky2apod}), computed
using the union of the $y$-map mask and the lensing map mask.  We multiply the cross-power spectrum by $\ell^2(\ell+1)/(2\pi)$ and bin it using twelve linearly spaced bins over the range $100 < \ell < 1600$.  These bins are identical to those chosen in~\cite{Planck2013CIBxlens}, with the exception that our upper multipole limit is slightly lower than theirs ($1600$ instead of $2000$) --- we conservatively choose a lower multipole cutoff due to our simplified treatment of the Planck beams.  As in Section~\ref{sec:Clyy}, we follow~\cite{Planck2013lensing,Planck2013CIBxlens,Planck2013ymap} in neglecting any correlations arising in the error bars due to mode coupling induced by the mask, and simply correct for the power lost through masking with the factor $f_{\rm sky,2}^{-1}$.

We compute statistical error bars on the cross-power spectrum using the standard analytic prescription (e.g.,~\cite{Planck2013CIBxlens}):
\be
\label{eq.Clyphimeaserr}
\left( \Delta \hat{C}_{\ell}^{y\phi} \right)^2 = \frac{1}{f_{\rm sky,2}} \frac{1}{(2\ell + 1) \Delta \ell} \left(\hat{C}_{\ell}^{yy} \hat{C}_{\ell}^{\phi\phi} + \left( C_{\ell}^{y\phi} \right)^2 \right) \,,
\ee
where $\hat{C}_{\ell}^{yy}$ and $\hat{C}_{\ell}^{\phi\phi}$ are the measured auto-spectra of the co-added $y$-map- and the $\phi$-map, respectively (i.e., these power spectra include the noise bias), and $C_{\ell}^{y\phi}$ is the fiducial theoretical cross-spectrum.  The contribution from the second term is typically $3$--$5$ orders of magnitude smaller than the contribution from the first --- thus any uncertainty in the theoretical modeling of $C_{\ell}^{y\phi}$ is completely negligible for the purposes of Eq.~(\ref{eq.Clyphimeaserr}).  This fact arises because the individual $y$- and $\phi$-maps are quite noisy, and thus the approximation that they are nearly uncorrelated (which is implicit in Eq.~(\ref{eq.Clyphimeaserr})) is valid.  Finally, we note that five additional error calculation methods were tested in~\cite{Planck2013CIBxlens} (see their Appendix A) using both simulations and data, and all were found to give results consistent with Eq.~(\ref{eq.Clyphimeaserr}).  Given the similarity between our analysis and theirs, we expect that Eq.~(\ref{eq.Clyphimeaserr}) should be quite accurate for our measurement as well.  In future work we will assess this issue more carefully using forthcoming public lensing simulations from the Planck team.

Fig.~\ref{fig.Clyphimeas} shows our initial measurement of the tSZ -- CMB lensing power spectrum (red circles).  A strong signal is clearly detected.  We quantify the detection significance according to:
\be
\label{eq.SNRdef}
{\rm SNR} = \sqrt{\sum_{i=1}^{12} \left( \frac{\hat{C}_{i}^{y\phi}}{\Delta C_i^{y\phi}} \right)^2} \,,
\ee
where $i$ indexes each of the twelve multipole bins in our measurement.  The SNR of our initial detection is $8.7\sigma$.  However, as described in detail in the following section, the measurement is subject to contamination from CIB leakage into the $y$-map.  The CIB-subtracted results are shown as blue squares in Fig.~\ref{fig.Clyphimeas}.

\subsection{CIB Contamination Correction}
\label{sec:CIBsub}

The primary systematic affecting our measurement of the tSZ -- CMB lensing cross-power spectrum is leakage of CIB emission into the ILC $y$-map.  Not only is the CIB emission difficult to completely remove from the $y$-map~\cite{Planck2013ymap}, it also correlates very strongly with the CMB lensing signal due to the similar mass and redshift kernels sourcing the two fields~\cite{Songetal2003,Planck2013CIBxlens}.  Thus, it is clear that we need to investigate the possible contamination of our tSZ -- CMB lensing cross-spectrum measurement due to residual CIB emission correlating with the CMB lensing potential.

We adopt a primarily empirical approach based on the Planck 857 GHz map to address this contamination, although we also test an alternative method in Section~\ref{sec:errors} and find that it gives consistent results.  We use the 857 GHz map as a tracer of dust emission from both the Galaxy and the CIB.  We cross-correlate the $y$-map constructed in Section~\ref{sec:ILC} with the 857 GHz map and use the resulting cross-power spectrum to assess the level of CIB leakage into the $y$-map.  Specifically, we assume that the 857 GHz signal $T_{857}(p )$ in pixel $p$ is given by
\be
\label{eq.857model}
T_{857}(p ) = T_{\rm CIB}(p ) + T_{\rm Gal}(p ) \,,
\ee
where $T_{\rm CIB}(p )$ and $T_{\rm Gal}(p )$ are the CIB and Galactic dust emission in pixel $p$.  (The CMB, tSZ, and other sky components are completely overwhelmed by dust emission at 857 GHz~\cite{Planck2013CIBxlens,Planck2013CIB}.)  Neglecting any other contaminants in the ILC $y$-map, we assume that the observed signal $\hat{y}(p )$ in pixel $p$ is given by
\be
\label{eq.y857model}
\hat{y}(p ) = y(p ) + \alpha_{\rm CIB} T_{\rm CIB}(p ) + \alpha_{\rm Gal} T_{\rm Gal}(p ) \,,
\ee
where $y(p )$ is the ``true'' uncontaminated Compton-$y$ signal and $\alpha_{\rm CIB}$ and $\alpha_{\rm Gal}$ are free parameters that describe the mean leakage of CIB and Galactic dust emission into the reconstructed $y$-map, respectively.  Note that the definitions in Eqs.~(\ref{eq.857model}) and~(\ref{eq.y857model}) imply that we have normalized the frequency dependence of the CIB and Galactic dust emission by their values at 857 GHz.  An important assumption implicit in the following is that the CIB emission at the lower HFI frequencies is perfectly correlated with that at 857 GHz; while this correlation is not perfect in reality, it is in fact very strong ($\approx 80$--$90$\% --- see Fig.~13 in~\cite{Planck2013CIBxlens}), which suffices for our purposes here.\footnote{We thank U. Seljak for emphasizing this point.}

Using Eq.~(\ref{eq.857model}) and assuming that the CIB and Galactic dust emission is uncorrelated (which physically must be true given their disparate origins), the auto-power spectrum of the 857 GHz map is
\be
\label{eq.857autospec}
C_{\ell}^{857} = C_{\ell}^{\rm CIB} + C_{\ell}^{\rm Gal} \,,
\ee
where $C_{\ell}^{\rm CIB}$ and $C_{\ell}^{\rm Gal}$ are the power spectra of the CIB and Galactic dust emission, respectively.  Similarly, the cross-power spectrum between our ILC $y$-map and the Planck 857 GHz map is
\be
\label{eq.y857crossspec}
C_{\ell}^{y \times 857} = \alpha_{\rm CIB} C_{\ell}^{\rm CIB} + \alpha_{\rm Gal} C_{\ell}^{\rm Gal} \,,
\ee
where we have neglected any possible correlation between the CIB emission and the tSZ signal, as in Section~\ref{sec:ILC} (following~\cite{Planck2013ymap}).  Such a correlation likely does exist~\cite{Addisonetal2012}, but in this analysis it would be very difficult to separate from the correlation between CIB leakage in the $y$-map and the CIB emission in the 857 GHz map, given the similar shape of the signals in multipole space.  The $\alpha_{\rm CIB}$ parameter essentially captures a combination of the physical tSZ -- CIB correlation and the spurious correlation due to leakage of CIB into the $y$-map.  For simplicity, we assume that the spurious correlation dominates.  Given a determination of $\alpha_{\rm CIB}$, it is straightforward to correct our measurement of the tSZ -- CMB lensing cross-spectrum:
\be
\label{eq.Clyphicorr}
\hat{C}_{\ell}^{y\phi, \mathrm{corr}}  = \hat{C}_{\ell}^{y\phi} - \alpha_{\rm CIB} \hat{C}_{\ell}^{\phi \times 857} \,,
\ee
where $\hat{C}_{\ell}^{\phi \times 857}$ is the cross-power spectrum of the CMB lensing potential and the CIB emission at 857 GHz, which was measured at $16\sigma$ in~\cite{Planck2013CIBxlens} (including all statistical and systematic errors).  If we considered both the physical tSZ -- CIB correlation and the spurious correlation due to leakage of CIB into the $y$-map in this analysis, then the correction computed in Eq.~(\ref{eq.Clyphicorr}) would be smaller than what we find below.  The results of Section~\ref{sec:interp} suggest that it is unlikely that we have overestimated the CIB contamination correction, and thus our current neglect of the physical tSZ -- CIB correlation seems reasonable.  Nevertheless, this is clearly an area in need of improvement in future tSZ analyses.

\begin{figure}
\begin{minipage}[b]{0.495\linewidth}
\centering
\includegraphics[width=\textwidth]{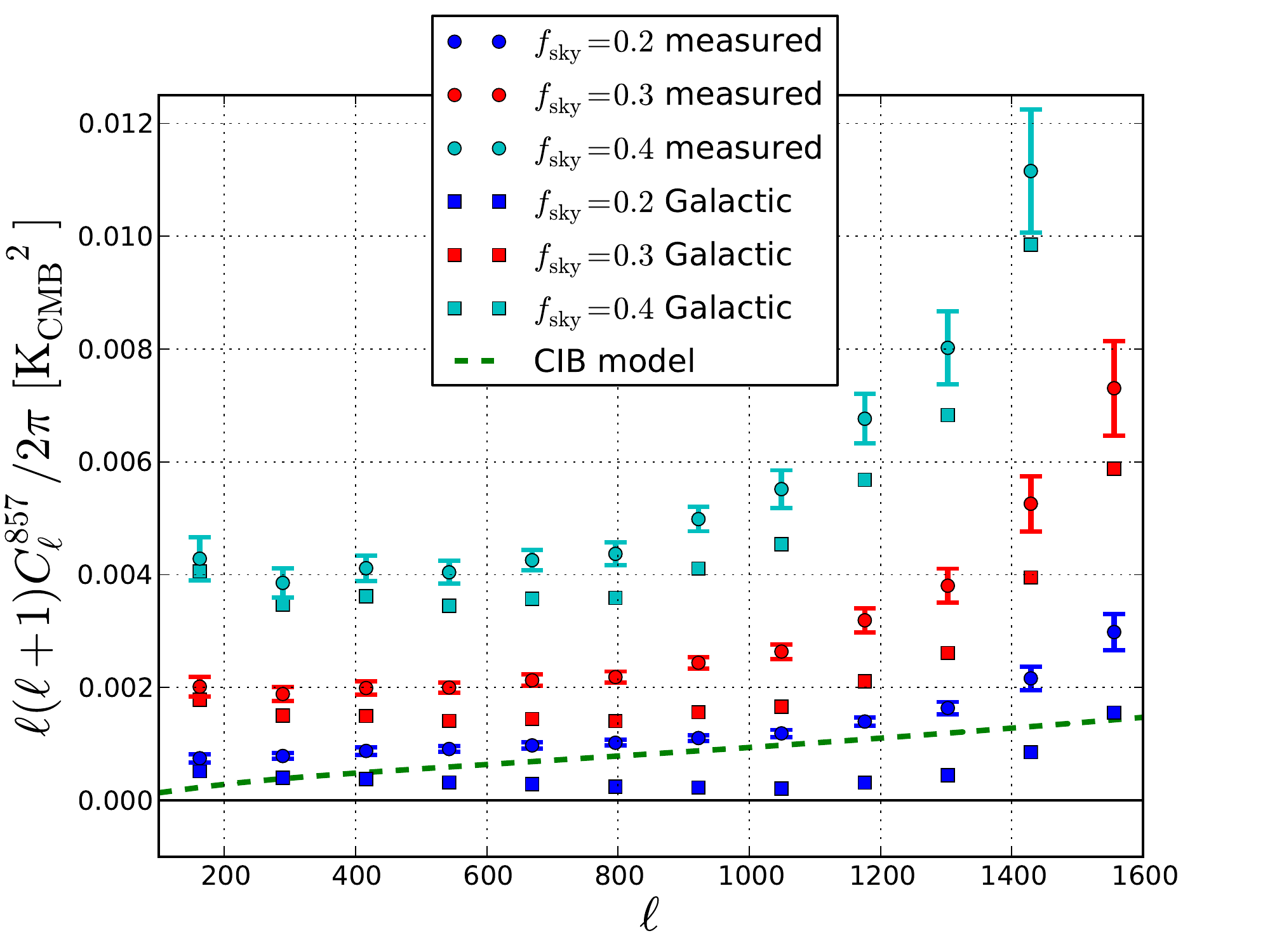}
\end{minipage}
\begin{minipage}[b]{0.495\linewidth}
\centering
\includegraphics[width=\textwidth]{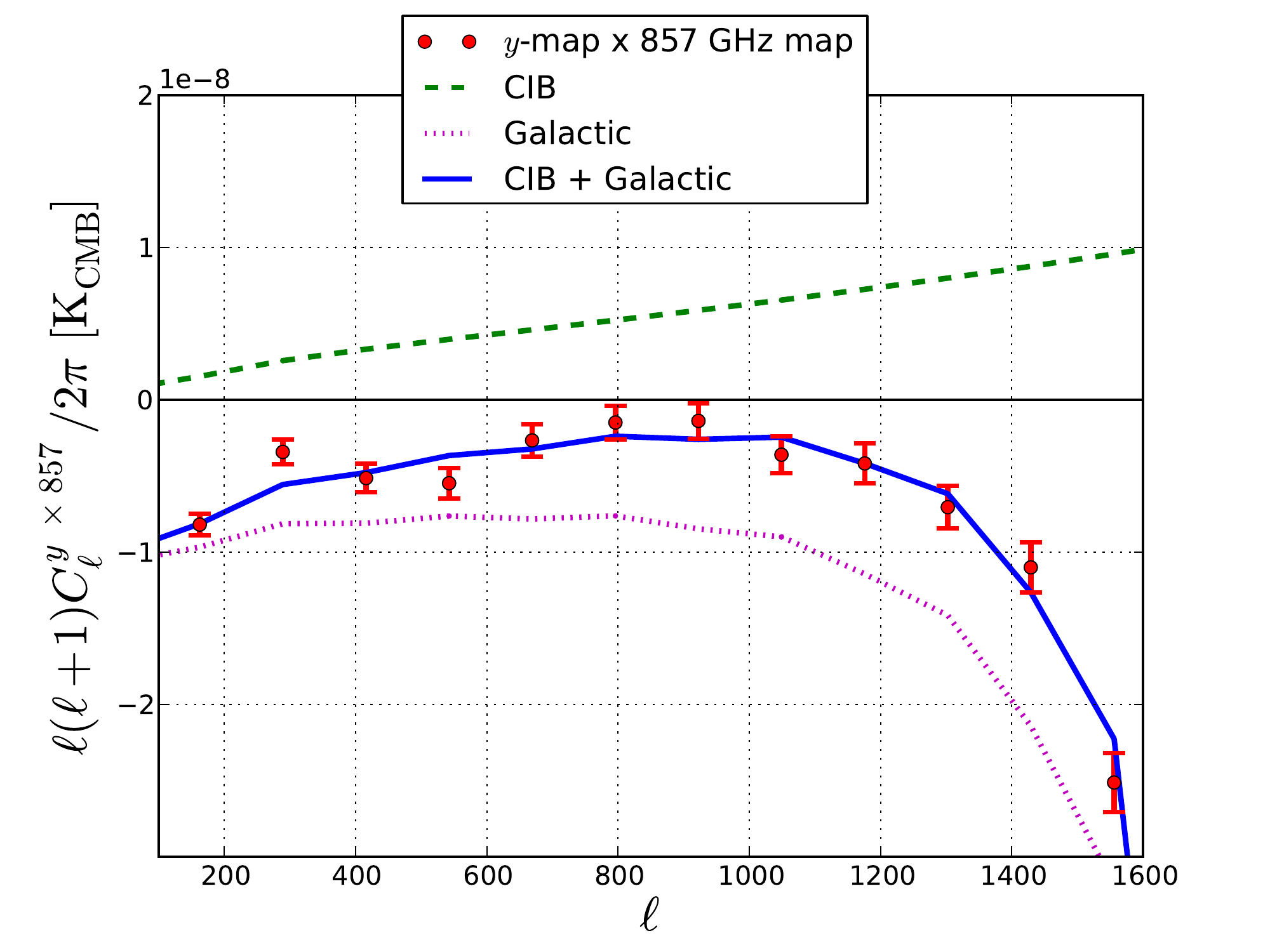}
\end{minipage}
\caption{The left panel shows the auto-power spectrum of the Planck 857 GHz map (circles with error bars), computed for three sky cuts --- note that the fiducial case is $f_{\rm sky} = 0.3$ (red circles).  The right panel shows the cross-power spectrum between our fiducial co-added $y$-map and the Planck 857 GHz map (for $f_{\rm sky} = 0.3$ only).  These measurements provide a method with which to assess the level of residual contamination in the $y$-map due to dust emission from the CIB and from the Galaxy.  We model the 857 GHz map as a combination of CIB emission (computed from the best-fit model in~\cite{Addisonetal2013}, shown as a dashed green curve in the left panel) and Galactic dust emission (computed by subtracting the CIB model from the measured 857 GHz power spectrum, shown as squares in the left panel).  We then model the $y$-map -- 857 GHz cross-power spectrum as a linear combination of these emission sources using two free amplitude parameters (see the text for further information).  We fit these parameters using the measured cross-power spectrum; the results of the fit are shown as dotted magenta (Galactic) and dashed green (CIB curves), with the sum of the two in solid blue.  The model provides a reasonable fit to the cross-power spectrum ($\chi^2 = 16.9$ for $12-2=10$ degrees of freedom) and allows us to effectively constrain the level of CIB leakage into the $y$-map.
\label{fig.cly857}}
\end{figure}

The first step in the CIB correction pipeline is measuring the auto-power spectrum of the 857 GHz map and separating the CIB and Galactic components using their different shapes in multipole space.  We adopt a model for the CIB emission at 857 GHz from~\cite{Addisonetal2013}, which is based on a simultaneous fit to power spectrum measurements from Planck, Herschel, ACT, and SPT, as well as number count measurements from Spitzer and Herschel.  We refer the reader to~\cite{Addisonetal2013} for a complete description of the model --- the values used in this work are shown in the ``Planck 857 GHz'' panel of Fig.~1 in~\cite{Addisonetal2013}\footnote{We are grateful to G.\ Addison for providing the CIB model fitting results from~\cite{Addisonetal2013}.}.  We subtract the CIB model power spectrum from the measured 857 GHz power spectrum to obtain an estimate of the power spectrum of the Galactic dust emission.  The results of this procedure are shown in the left panel of Fig.~\ref{fig.cly857} for three different sky cuts, including our fiducial $f_{\rm sky} = 0.3$ case (the other cases will be considered when we perform null tests in Section~\ref{sec:errors}).  It is clear in the figure that the total power at 857 GHz increases as the sky fraction increases, simply because more Galactic dust emission is present.  (Note that these power spectra have been corrected with the $f_{\rm sky,2}$ factor described in Eq.~(\ref{eq.fsky2apod}).)  The only case in which the CIB power partially dominates the total measured power is for $f_{\rm sky} = 0.2$, primarily around $\ell=1000$; however, the CIB contribution is clearly non-negligible in all cases.  The key result from this analysis is an estimate for both $C_{\ell}^{\rm CIB}$ and $C_{\ell}^{\rm Gal}$ in Eq.~(\ref{eq.857autospec}).

The second step in the CIB correction pipeline is applying these results to the cross-spectrum of the ILC $y$-map and the 857 GHz map in order to measure $\alpha_{\rm CIB}$ and $\alpha_{\rm Gal}$, as described in Eq.~(\ref{eq.y857crossspec}).  The results of this procedure are shown in the right panel of Fig.~\ref{fig.cly857}, for the fiducial $f_{\rm sky} = 0.3$ case only.  The red circles in the plot are the measured cross-power spectrum between our ILC $y$-map and the Planck 857 GHz map, while the curves show the best-fit result when applying Eq.~(\ref{eq.y857crossspec}) to these measurements.  For our fiducial $f_{\rm sky} = 0.3$ analysis, we measure
\ba
\alpha_{\rm CIB} & = & (6.7 \pm 1.1) \times 10^{-6} \,\, \mathrm{K^{-1}_{CMB}} \nonumber \\
\alpha_{\rm Gal} & = & (-5.4 \pm 0.4) \times 10^{-6} \,\, \mathrm{K^{-1}_{CMB}} \,,
\label{eq.y857fitresults}
\ea
with marginalized $1\sigma$ uncertainties given for both parameters.  The model provides a reasonable fit to the data, with $\chi^2 = 16.9$ for $12-2=10$ degrees of freedom.  Note that the results of the fit also imply that our ILC weights yield a positive response when applied to a CIB-like spectrum, and a negative response when applied to a Galactic-dust-like spectrum, a result which was also found in~\cite{Planck2013ymap,Planck2013SZcatalog}.  Finally, we note that while this approach is sufficient for correcting CIB leakage into the tSZ -- CMB lensing cross-power spectrum, it is unlikely to suffice for correcting CIB leakage into quadratic or higher-order tSZ statistics computed from the $y$-map, such as the tSZ auto-power spectrum.  The crucial point is that the tSZ -- CMB lensing cross-spectrum is linear in $y$, and hence only an estimate of the mean CIB leakage into the $y$-map is needed to compute the correction, which our cross-correlation method provides.  Fluctuations around this mean leakage will clearly contribute residual power to the tSZ auto-spectrum at a level which is not constrained in this approach.  Physically, this arises because the spectrum of the CIB emission varies from source to source; thus, although we can remove the mean leakage into the $y$-map with our approach, fluctuations around this mean will still exist and contribute to the auto-power spectrum and higher-order statistics.  The Planck team used simulations to assess the residual amount of non-tSZ power in their $y$ auto-spectrum~\cite{Planck2013ymap}; it seems likely that a combination of simulations and more refined empirical methods will be useful in future studies.  For now we do not attempt to provide a cleaned tSZ auto-power spectrum from our $y$-maps.

We use the measured value of $\alpha_{\rm CIB}$ to correct the tSZ -- CMB lensing cross-power spectrum following Eq.~(\ref{eq.Clyphicorr}).  We propagate the uncertainty resulting from the fit, as well as the uncertainties on $\hat{C}_{\ell}^{\phi \times 857}$ given in~\cite{Planck2013CIBxlens}, into the final error bars on $\hat{C}_{\ell}^{y\phi, \mathrm{corr}}$.  Our final, CIB-corrected measurement of the tSZ -- CMB lensing cross-power spectrum is shown in Fig.~\ref{fig.Clyphimeas} as the blue squares.  The final detection significance is $6.2\sigma$ after the CIB correction.  The solid green curve in Fig.~\ref{fig.Clyphimeas} shows the theoretical prediction of our fiducial model described in Section~\ref{sec:theory}.  This curve is not a fit to the measurements, but already agrees well with the data: $\chi^2 = 14.6$ for $12$ degrees of freedom.  The figure also shows the one- and two-halo terms for the fiducial model; it is clear that our measurement shows evidence of contributions from both terms.


The detection significance is somewhat lower for the tSZ -- CMB lensing cross-power spectrum than for either of the auto-spectra, which are detected at $12.3\sigma$~\cite{Planck2013ymap} and $25\sigma$~\cite{Planck2013lensing}, respectively.  If the $y$ and $\phi$ fields were perfectly correlated, i.e., $r_{\ell}^{y\phi} = 1$, and if the contributions to their individual SNR were over identical multipole ranges, then we could expect a $\sqrt{12 \times 25} \approx 17\sigma$ significance for the detection of $C_{\ell}^{y\phi}$.  However, since $\langle r_{\ell}^{y\phi} \rangle \approx 0.33$ over the relevant multipole range for our measurement, we obtain a detection significance of $17 \times 0.33 \approx 6\sigma$.  The fact that this estimate is in rough agreement with our actual measurement indicates that the multipole ranges over which the tSZ and CMB lensing auto-spectra have significant SNR are fairly similar.

Table~\ref{tab.bandpowers} gives our final CIB-corrected bandpowers and the associated uncertainties.

\begin{table}[htbp]
\centering
\begin{tabular}{||c|rrrrrrrrrrrr||}
\hline
\hline
$\ell_{\rm mean}$ & $163$ & $290$ & $417$ & $543$ & $670$ & $797$ & $923$ & $1050$ & $1177$ & $1303$ & $1430$ &$1557$ \\
\hline
$\ell^2(\ell+1)C_{\ell}^{y\phi}/(2\pi) \times 10^{11}$ & $4.01$ & $0.73$ & $2.34$ & $4.12$ & $5.50$ & $2.01$ & $3.39$ & $-2.34$ & $6.03$ & $5.21$ & $-1.75$ & $-0.13$ \\
$\Delta(\ell^2(\ell+1)C_{\ell}^{y\phi}/(2\pi)) \times 10^{11}$ & $1.19$ & $1.49$ & $1.70$ & $1.89$ & $2.07$ & $2.21$ & $2.35$ & $2.49$ & $2.68$ & $2.88$ & $3.12$ & $3.34$ \\
\hline
\hline
\end{tabular}
\caption{\label{tab.bandpowers}Measured tSZ -- CMB lensing cross-spectrum bandpowers.  All values are dimensionless.  These bandpowers have been corrected for contamination from CIB emission following the procedure described in Section~\ref{sec:CIBsub}; they correspond to the blue squares shown in Fig.~\ref{fig.Clyphimeas}.  Uncertainties due to the CIB subtraction have been propagated into the final errors provided here.  Note that the bins have been chosen to match those in~\cite{Planck2013CIBxlens}.}
\end{table}

\subsection{Null Tests and Systematic Errors}
\label{sec:errors}

We perform a series of null tests designed to search for any residual systematic errors that might affect our measurement of the tSZ -- CMB lensing cross-power spectrum.  Our calculation of the statistical errors on both the fiducial measurement and the null tests follows Eq.~(\ref{eq.Clyphimeaserr}) throughout.  We refer the reader to~\cite{Planck2013lensing,Planck2013CIBxlens} for a complete discussion of possible systematics that can affect the CMB lensing reconstruction.  Essentially all of these effects are far smaller than our error bars, the smallest of which is $\approx 30$\% (this is the fractional error on our lowest $\ell$-bin).  A particular cause for concern, however, is spurious lensing signal induced by other sources of statistical anisotropy in the temperature maps, such as the Galactic and point source masks.  We mitigate this ``mean-field'' bias by discarding all data below $\ell = 100$ in our analysis, following~\cite{Planck2013CIBxlens}.

\begin{figure}
\centering
\includegraphics[totalheight=0.4\textheight]{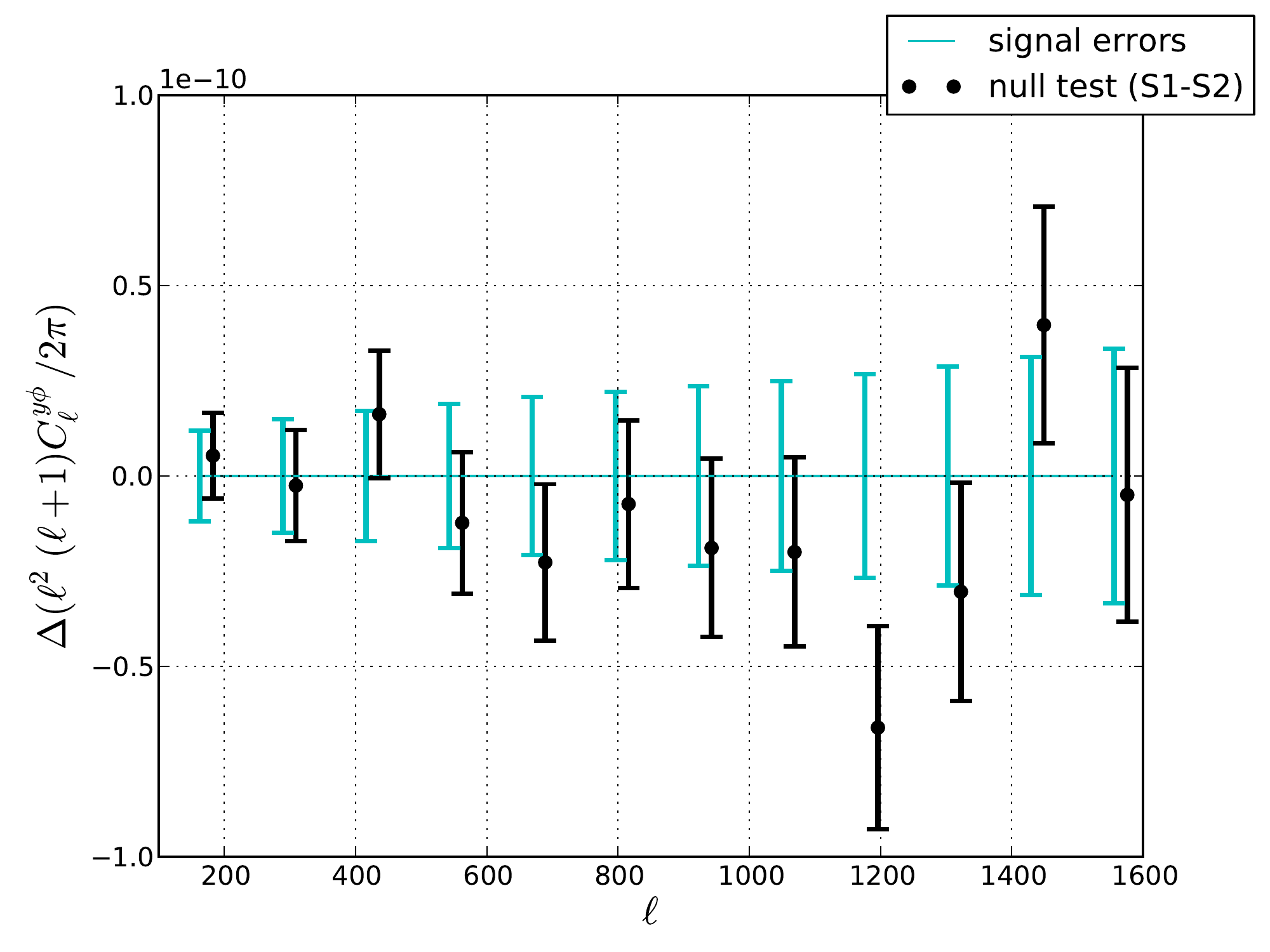}
\caption{Cross-power spectrum between the null Compton $y$-map and the CMB lensing potential map (black points).  The null $y$-map is constructed by subtracting (rather than co-adding) the S1 and S2 season $y$-maps output by our ILC pipeline in Section~\ref{sec:ILC}.  The error bars, shown in black, are computed by applying Eq.~(\ref{eq.Clyphimeaserr}) to the null $y$-map and CMB lensing potential map.  The result is consistent with a null signal: $\chi^2 = 13.2$ for $12$ degrees of freedom.  The cyan error bars show the $1\sigma$ errors on our fiducial tSZ -- CMB lensing cross-spectrum (i.e., the error bars on the blue squares in Fig.~\ref{fig.Clyphimeas}), which allow a quick assessment of the possible importance of any systematic effect in our measurement.
\label{fig.nulltestS1S2diff}}
\end{figure}

The first null test we perform is to cross-correlate a null Compton-$y$ map with the CMB lensing potential map.  The null $y$-map is constructed by subtracting, rather than co-adding, the S1 and S2 $y$-maps that are produced by our ILC pipeline, as described in Section~\ref{sec:ILC}.  The subtraction is performed using the same inverse-variance combination that is applied in the co-addition, guaranteeing that the noise level in the subtracted map is identical to that in the co-added map.  The null map should be free of any astrophysical emission, including tSZ, CIB, or Galactic dust signals.  The cross-power spectrum of the null $y$-map with the CMB lensing potential map is shown in Fig.~\ref{fig.nulltestS1S2diff} (black points with solid black error bars).  The test is consistent with a null signal: $\chi^2 = 13.2$ for $12$ degrees of freedom.  For reference, Fig.~\ref{fig.nulltestS1S2diff} also shows the $1\sigma$ error bars on our fiducial signal in cyan.  The consistency between the errors, as well as the reasonable value of $\chi^2$, provides additional support for the robustness of our statistical error calculation.

\begin{figure}
\begin{minipage}[b]{0.495\linewidth}
\centering
\includegraphics[width=\textwidth]{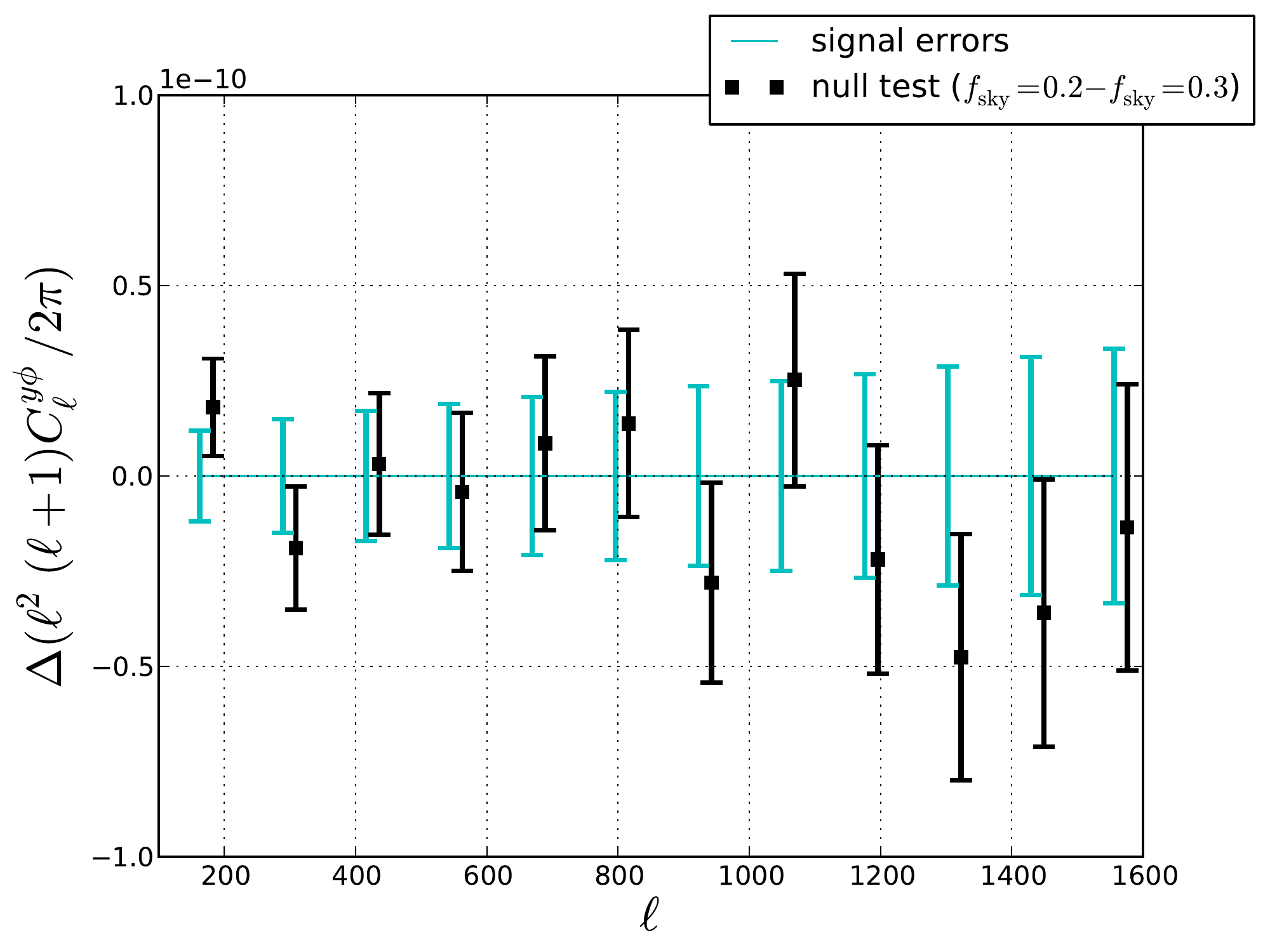}
\end{minipage}
\begin{minipage}[b]{0.495\linewidth}
\centering
\includegraphics[width=\textwidth]{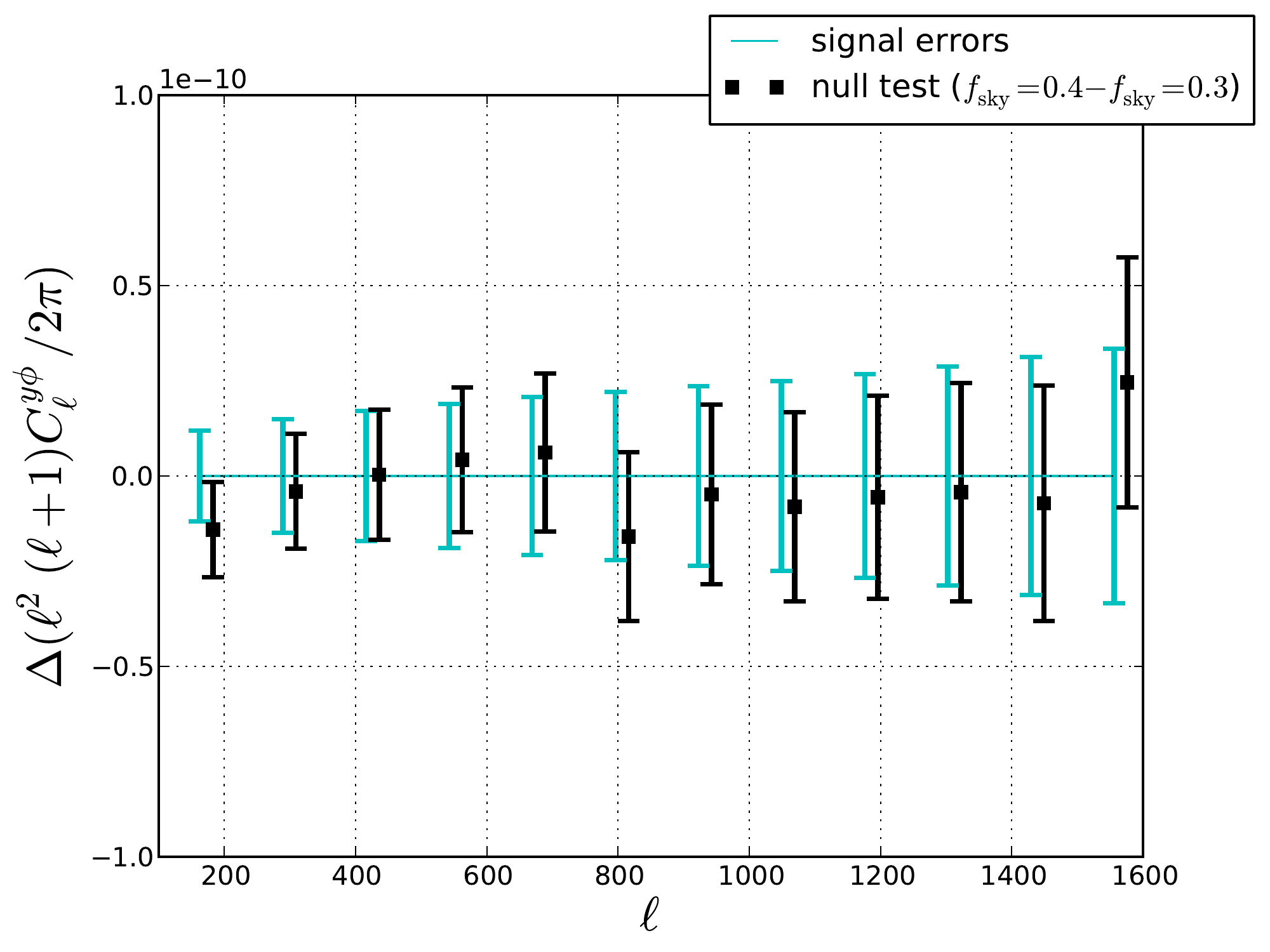}
\end{minipage}
\caption{Difference between the tSZ -- CMB lensing cross-power spectrum measured using our fiducial $f_{\rm sky} = 0.3$ Galactic mask and using an $f_{\rm sky} = 0.2$ mask (black squares in left panel) or $f_{\rm sky} = 0.4$ mask (black squares in right panel).  The error bars on the black points are computed by applying Eq.~(\ref{eq.Clyphimeaserr}) to the $f_{\rm sky} = 0.2$ or $0.4$ case, i.e., using the same procedure as for the fiducial $f_{\rm sky} = 0.3$ analysis.  In both cases, the difference between the fiducial cross-power spectrum and the $f_{\rm sky}$ variation cross-power spectrum is consistent with null: $\chi^2 = 9.7$ ($12$ degrees of freedom) for the $f_{\rm sky} = 0.2$ case, and $\chi^2 = 2.8$ ($12$ degrees of freedom) for the $f_{\rm sky} = 0.4$ case.  The errors on the fiducial signal are shown in cyan, as in Fig.~\ref{fig.nulltestS1S2diff}.  Note that the entire pipeline described in Sections~\ref{sec:tSZrecon} and~\ref{sec:tSZxlensing}, including the CIB correction, is re-run to obtain the $f_{\rm sky} = 0.2$ and $0.4$ results.  The ILC weights for each sky cut are given in Table~\ref{tab.weights}.
\label{fig.nulltestfsky}}
\end{figure}

The second null test we perform is designed to search for residual contamination from the Galaxy.  If Galactic emission (e.g., synchrotron, free-free, or dust) leaks into both the $y$-map and the CMB lensing potential map, it could produce a spurious correlation signal in our analysis.  We investigate this possibility by running our entire analysis pipeline described thus far --- including the combination with the point source mask, the ILC $y$ reconstruction, the cross-power spectrum estimation, and the CIB correction --- on two additional sky masks, one more conservative than our fiducial choice, and one more aggressive.  These masks are constructed by thresholding the 857 GHz map in the same manner as our fiducial $f_{\rm sky} = 0.3$ mask (see Section~\ref{sec:data}); the conservative mask has $f_{\rm sky} = 0.2$, while the aggressive mask has $f_{\rm sky} = 0.4$.  As seen in the left panel of Fig.~\ref{fig.cly857}, the Galactic emission varies strongly as a function of masking, and thus if our results are significantly contaminated by the Galaxy, a strong variation in the signal should be seen.  The results of these tests are shown in Fig.~\ref{fig.nulltestfsky} --- the left panel shows the difference between $\hat{C}_{\ell}^{y\phi}$ computed for $f_{\rm sky} = 0.2$ and for the fiducial $f_{\rm sky} = 0.3$, while the right panel shows the difference between the $f_{\rm sky} = 0.4$ result and $f_{\rm sky} = 0.3$ result.  In each case, the error bars shown in black are the errors on the $f_{\rm sky} = 0.2$ or $0.4$ analyses, while the cyan errors are those for the fiducial $f_{\rm sky} = 0.3$ case.  No significant variation is seen for either $f_{\rm sky} = 0.2$ or $f_{\rm sky} = 0.4$.  For the former, we find $\chi^2 = 9.7$ ($12$ degrees of freedom) when comparing the difference spectrum to a null signal; for the latter, we find $\chi^2 = 2.8$ ($12$ degrees of freedom).  The $\chi^2$ value for the $f_{\rm sky} = 0.4$ test is somewhat low because the error bars on $\hat{C}_{\ell}^{y\phi}$ do not decrease significantly as $f_{\rm sky}$ increases, likely because the foreground-dominated noise in the $y$-map does not decrease, which more than compensates for the beneficial effect of having more modes in the map.  Regardless, the null test is passed in both the $f_{\rm sky} = 0.2$ and $0.4$ cases.  For reference, we also provide the ILC channel map weights derived in each case in Table~\ref{tab.weights}.  The weights change slightly as $f_{\rm sky}$ is increased in order to compensate for the increased Galactic dust emission, but are generally fairly stable around the fiducial $f_{\rm sky} = 0.3$ case.

\begin{table}[h]
\centering
\begin{tabular}{||c|rrrrr||}
\hline
\hline
$f_{\rm sky}$ & $100$ GHz & $143$ GHz & $217$ GHz & $353$ GHz & $545$ GHz \\
\hline
$0.2$ & $0.41782$ & $-0.93088$ & $0.50931$ & $0.00701$ & $-0.00325$ \\
$0.3$ (fiducial) & $0.45921$ & $-1.00075$ & $0.54397$ & $0.00032$ & $-0.00275$ \\
$0.4$ & $0.51380$ & $-1.11722$ & $0.62638$ & $-0.02113$ & $-0.00182$ \\
\hline
\hline
\end{tabular}
\caption{\label{tab.weights}Channel map weights computed by our ILC pipeline for three different sky cuts (see Eqs.~(\ref{eq.CILCweights}) and~(\ref{eq.modifiedcovmatrix})).  All values are in units of $\mathrm{K^{-1}_{CMB}}$.  The fiducial case used throughout this paper is $f_{\rm sky} = 0.3$.  Note that these $f_{\rm sky}$ values are computed by simply thresholding the 857 GHz map (see Section~\ref{sec:data}); the final sky fractions used in each case are smaller than these values due to the inclusion of the point source mask and S1 and S2 season map masks.  The weights change slightly as $f_{\rm sky}$ is increased as the ILC adjusts to remove the increased Galactic dust emission.}
\end{table}

\begin{figure}
\centering
\includegraphics[totalheight=0.4\textheight]{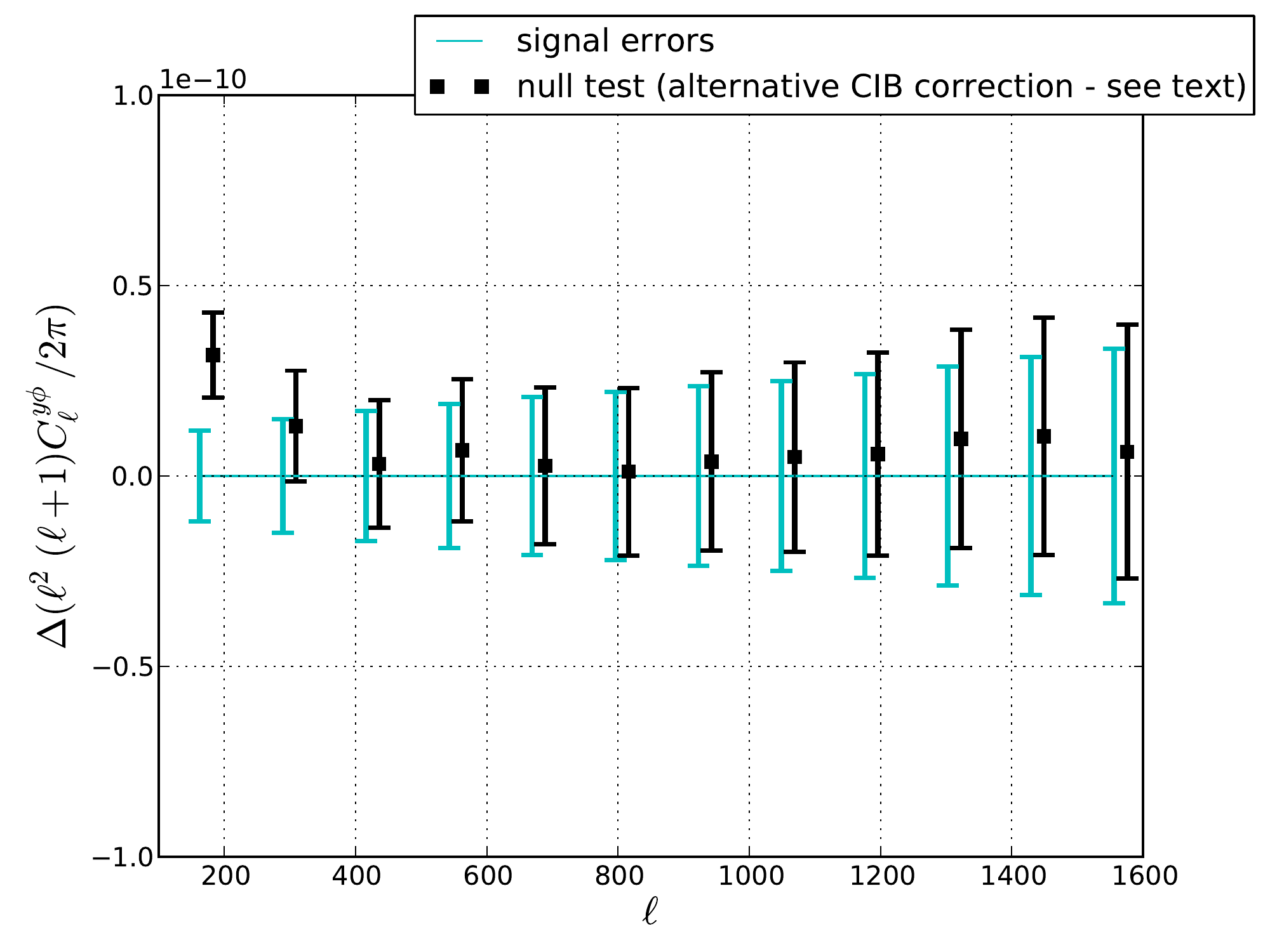}
\caption{Difference between the tSZ -- CMB lensing cross-power spectrum corrected for CIB contamination using our fiducial approach and corrected using an alternative approach based on the model fitting results presented in~\cite{Planck2013CIBxlens} (black squares).  In the alternative approach, we simply weight the $C_{\ell}^{\mathrm{CIB} \times \phi}$ results at each HFI frequency by the relevant ILC weight computed by our Compton-$y$ reconstruction pipeline (given in Table~\ref{tab.weights}), and then sum the results to get the total leakage into the $C_{\ell}^{y\phi}$ measurement.  The difference between our fiducial CIB-corrected cross-spectrum and the cross-spectrum corrected using this method is consistent with null: $\chi^2 = 9.4$ for $12$ degrees of freedom.  The slight disagreement at low-$\ell$ is likely due to discrepancies between the model fit from~\cite{Planck2013CIBxlens} and the measured CIB -- CMB lensing cross-spectrum at 100, 143, and 217 GHz --- see the text for discussion.  The errors on our fiducial tSZ -- CMB lensing cross-spectrum measurement are shown in cyan, as in Figs.~\ref{fig.nulltestS1S2diff} and~\ref{fig.nulltestfsky}.
\label{fig.nulltestaltCIBsub}}
\end{figure}

The third null test we perform uses an alternative method for calculating the CIB correction to the tSZ -- CMB lensing measurement.  Our standard approach is based on cross-correlating the ILC $y$-map with the 857 GHz map, as described in Section~\ref{sec:CIBsub}.  As an alternative, we consider a method based on the best-fit model to the CIB -- CMB lensing cross-power spectrum measurements obtained in~\cite{Planck2013CIBxlens}.  We simply weight their $C_{\ell}^{\mathrm{CIB} \times \phi}$ result at each HFI frequency between $100$ and $545$ GHz by the ILC relevant weight computed with our fiducial $f_{\rm sky} = 0.3$ pipeline (as given in Table~\ref{tab.weights}), and then sum the results to obtain the total contamination to our $C_{\ell}^{y\phi}$ measurement.  We work with the model fitting results (in particular, the ``$j$ reconstruction'' results shown in Fig.~15 of~\cite{Planck2013CIBxlens}\footnote{We are grateful to O.\ Dor\'{e} for providing the CIB -- CMB lensing model fits from~\cite{Planck2013CIBxlens}.}) rather than the direct measurements of $C_{\ell}^{\mathrm{CIB} \times \phi}$ at each frequency because the noise in the $100$ and $143$ GHz measurements renders them too imprecise for our purposes.  We compute and subtract the CIB contamination from our initial tSZ -- CMB lensing cross-spectrum measurement (the red circles in Fig.~\ref{fig.Clyphimeas}) to obtain a CIB-corrected measurement independent of our standard CIB-correction pipeline.  Fig.~\ref{fig.nulltestaltCIBsub} shows the difference between the CIB-corrected cross-spectrum obtained using this alternative procedure and the CIB-corrected cross-spectrum obtained using our standard procedure (described in Section~\ref{sec:CIBsub}).  The errors on the alternative-correction points do not include uncertainty from the subtraction procedure, as we do not have the full errors on the CIB model fits from~\cite{Planck2013CIBxlens}, but this is likely a very small correction.  The difference between the cross-spectra estimated from the two subtraction procedures is consistent with null: $\chi^2 = 9.4$ for $12$ degrees of freedom.  A slightly significant difference is seen between the cross-spectra in the lowest two $\ell$-bins; however, Fig.~15 in~\cite{Planck2013CIBxlens} indicates that the $C_{\ell}^{\mathrm{CIB} \times \phi}$ model fits severely underpredict the measured results at 100, 143, and 217 GHz for these low $\ell$ values.  Modest variations in the low-$\ell$ $C_{\ell}^{\mathrm{CIB} \times \phi}$ values at these frequencies can easily explain the discrepancy seen in Fig.~\ref{fig.nulltestaltCIBsub}, and thus we do not consider it a cause for concern.  The general consistency between this CIB subtraction procedure and our standard approach based on the $y$ -- 857 GHz cross-correlation gives us confidence that the signal in Fig.~\ref{fig.Clyphimeas} is indeed the tSZ -- CMB lensing cross-power spectrum.

These three null tests address the primary possible systematic errors in our measurement.  However, in principle, any component of the CMB temperature field that produces a non-zero $\langle T(\ell) T(\ell - \ell') T(-\ell') \rangle$ could contaminate our measurement.  Fortunately, all such bispectra have been examined in detail in~\cite{Planck2013CIBxlens} for each of the HFI channels.  One possible worry is contamination from extragalactic point sources, either due to radio emission at the lower HFI frequencies or infrared dust emission at the higher HFI frequencies.  The contamination of this shot-noise bispectrum to the measurement of $C_{\ell}^{\mathrm{CIB} \times \phi}$ is shown in Fig.~11 of~\cite{Planck2013CIBxlens}, where it can be seen that the induced bias is completely negligible at $\ell \lsim 1500$ for every HFI frequency.  Thus, we neglect it for the purposes of our analysis, which essentially lies completely below this multipole.  Similarly, we neglect any contamination from the clustered CIB bispectrum (recently measured in~\cite{Crawfordetal2013,Planck2013CIB}), as this was also found to be negligible in~\cite{Planck2013CIBxlens}, even after a dedicated search for its effects.  For such a bispectrum to affect our results, the CIB emission would have to leak into both the $y$-map and CMB lensing potential map; the ILC used to construct the $y$-map should suppress such leakage below the raw level estimated in~\cite{Planck2013CIBxlens} for the $C_{\ell}^{\mathrm{CIB} \times \phi}$ measurements.

The integrated Sachs-Wolfe (ISW) effect in the CMB presents another possible source of contamination, as it traces the same large-scale matter density fluctuations as the CMB lensing potential~\cite{Sachs-Wolfe1967}.  Crucially, however, the ISW-induced fluctuations in the CMB have the same frequency dependence as the primordial CMB fluctuations, i.e., a blackbody spectrum.  Since our ILC $y$-map construction explicitly removes all signal with a CMB blackbody spectrum (see Section~\ref{sec:ILC}), no ISW -- CMB lensing correlation should contaminate our measurement of $C_{\ell}^{y\phi}$.

A final worry is the direct leakage of tSZ signal into the CMB lensing potential reconstruction, which would of course correlate with the tSZ signal in the ILC $y$-map.  This leakage was recently considered in~\cite{Osborneetal2013,vanEngelenetal2013}, and we utilize their results here.  In particular, it is shown in~\cite{vanEngelenetal2013} that the bias induced in the CMB lensing power spectrum by tSZ leakage is $\lsim 10$\% for a completely unmasked reconstruction over the multipole range we consider (the bias is more problematic at high multipoles, e.g., $\ell > 2000$).  However, this is the bias in the CMB lensing auto-power spectrum; the bias in the lensing reconstruction itself is much smaller ($\lsim \sqrt{10}$\%).  Furthermore, given that tSZ-detected clusters have been masked in the 143 GHz Planck CMB lensing reconstruction, and given that we work with the publicly released minimum-variance combination of the 143 and 217 GHz lensing reconstructions, the tSZ-induced bias in the CMB lensing potential map should be completely negligible for the purposes of our analysis.  Follow-up analyses with detailed simulations (such as those used in~\cite{vanEngelenetal2013}) will easily test this assumption.

Having passed a number of null tests and demonstrated that essentially all non-CIB sources of systematic error are negligible, we are confident in the robustness of our detection of the tSZ -- CMB lensing cross-power spectrum.  The blue squares in Fig.~\ref{fig.Clyphimeas} show our final CIB-corrected results, which are also presented in Table~\ref{tab.bandpowers} with the associated uncertainties.  The final detection significance is $6.2\sigma$.

\section{Interpretation}
\label{sec:interp}

As demonstrated in Section~\ref{sec:theory}, the tSZ -- CMB lensing cross-correlation signal is a probe of both the background cosmology (primarily via $\sigma_8$ and $\Omega_m$) and the ICM physics in the groups and clusters responsible for the signal.  In this section, we first fix the cosmological parameters and interpret our measurement in terms of the ICM physics.  We then fix the ICM physics to our fiducial model and constrain the cosmological parameters.  Finally, we demonstrate a method to simultaneously constrain the ICM and cosmology by combining our measurement with the Planck measurement of the tSZ auto-power spectrum.

\subsection{ICM Constraints}
\label{sec:interpICM}

\begin{figure}
\begin{minipage}[b]{0.495\linewidth}
\centering
\includegraphics[width=\textwidth]{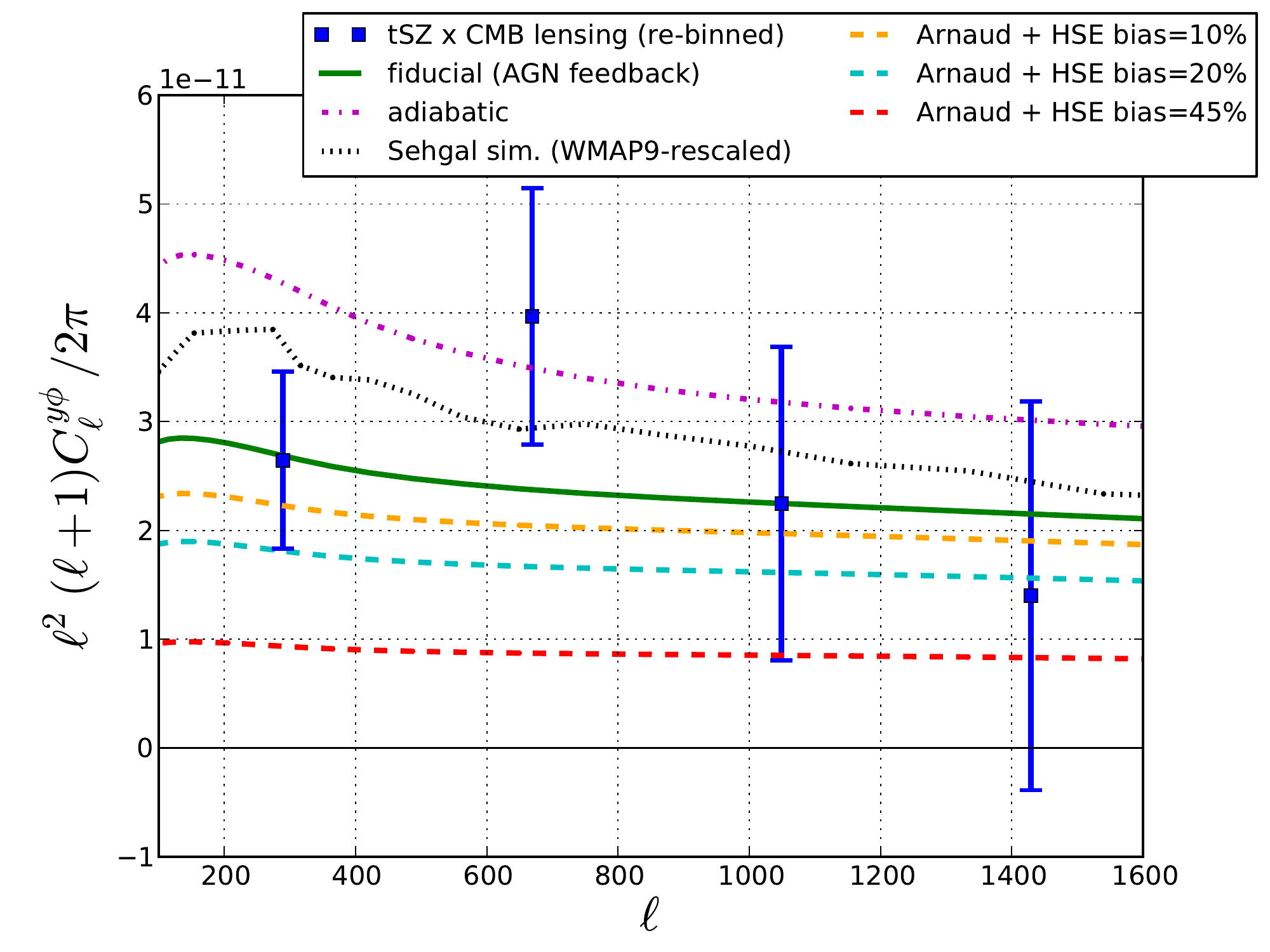}
\end{minipage}
\begin{minipage}[b]{0.495\linewidth}
\centering
\includegraphics[width=\textwidth]{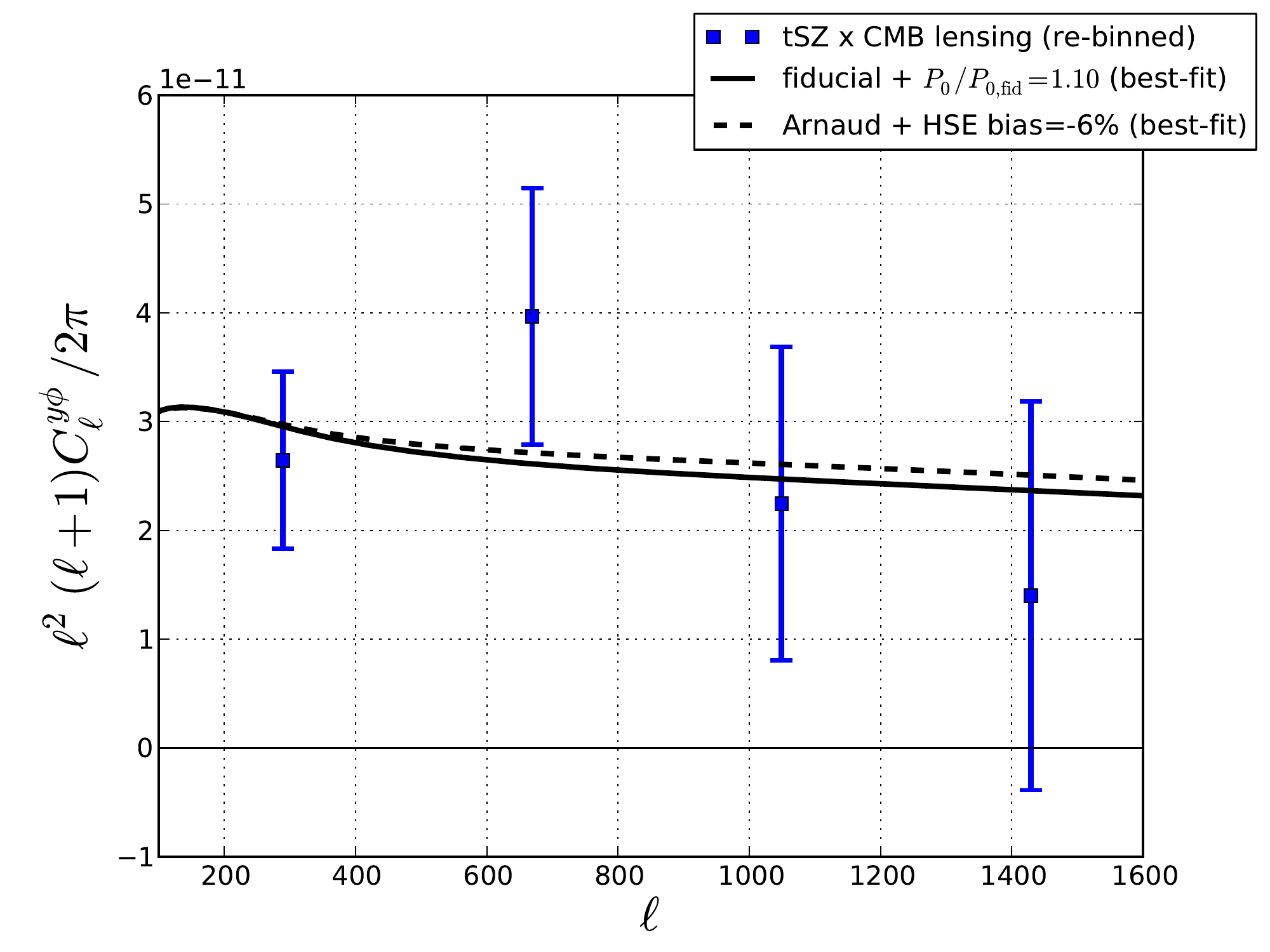}
\end{minipage}
\caption{ICM model comparison results.  In both panels, the blue squares show a re-binned version of the tSZ -- CMB lensing cross-power spectrum results presented in Fig.~\ref{fig.Clyphimeas}.  The re-binning was performed using inverse-variance weighting, and is only done for visual purposes; all model fitting results are computed using the twelve bins shown in Fig.~\ref{fig.Clyphimeas}.  The left panel shows a comparison between the predictions of several ICM physics models, all computed using a WMAP9 background cosmology.  The fiducial ``AGN feedback'' model~\cite{Battagliaetal2012} is the best fit, with $\chi^2 = 14.6$ for $12$ degrees of freedom (see the text for comparisons to the other models shown here).  The right panel shows the result of a fit for the amplitude of the pressure--mass relation $P_0$ with respect to its standard value in our fiducial model (solid black curve), as well as the result of a fit for the HSE mass bias in the UPP of Arnaud et al.~\cite{Arnaudetal2010} (dashed black curve).  Our measurements are consistent with the fiducial value of $P_0$ ($P_0/P_{0,{\rm fid}} = 1.10 \pm 0.22$) in our standard  ``AGN feedback'' model, and similarly with standard values of the HSE mass bias for the UPP, $(1-b) = 1.06^{+0.11}_{-0.14}$ (all results at $68$\% C.L.).
\label{fig.Clyphimodelcomp}}
\end{figure}

Fixing the background cosmological parameters to their WMAP9 values (in particular, $\sigma_8 = 0.817$ and $\Omega_m = 0.282$, i.e., our fiducial model), we derive constraints on ICM physics models from our measurement of the tSZ -- CMB lensing cross-power spectrum.  The ICM models are described in Section~\ref{sec:theory}, and include our fiducial ``AGN feedback'' model, the ``adiabatic'' model from~\cite{Battagliaetal2012}, the UPP of Arnaud et al.~\cite{Arnaudetal2010}, for which we consider various values of the HSE mass bias $(1-b)$ (see Section~\ref{sec:theory}), and the simulation results of~\cite{Sehgaletal2010}.  The results are shown in Fig.~\ref{fig.Clyphimodelcomp}, in which we have re-binned the points from Fig.~\ref{fig.Clyphimeas} for visual clarity (note that all fitting results and $\chi^2$ values are computed using the $12$ bins in Fig.~\ref{fig.Clyphimeas}).  We find that the fiducial model is the best fit to the data of the five models shown in Fig.~\ref{fig.Clyphimodelcomp}, with $\chi^2 = 14.6$ for $12$ degrees of freedom.  The adiabatic model is in some tension with our measurement, with $\chi^2 = 17.9$.  Fig.~\ref{fig.Clyphimodelcomp} also shows three versions of the UPP with differing amounts of HSE mass bias: $(1-b) = 0.9$ is a reasonable fit to the data ($\chi^2 = 15.7$), while $(1-b) = 0.8$ is in some tension ($\chi^2 = 17.8$).  Perhaps most interestingly, $(1-b) = 0.55$ is in serious tension with our measurement, with $\chi^2 = 25.1$, i.e., $\Delta \chi^2 = 10.5$ with respect to our fiducial model.  This value of $(1-b)$ was found to be necessary in~\cite{Planck2013counts} in order to reconcile the Planck SZ cluster count-derived cosmological parameters with those derived in the Planck primordial CMB analysis.  In the context of a fixed WMAP9 background cosmology, our results highly disfavor such an extreme value of the HSE mass bias for the UPP, although we are probing a lower-mass and higher-redshift subset of the cluster population than that used in the counts analysis.  Finally, the cross-spectrum derived from the Sehgal et al.~simulation~\cite{Sehgaletal2010} lies somewhat high but is a reasonable fit to the data, with $\chi^2 = 15.4$.  Note that this simulation result has been rescaled to our fiducial WMAP9 cosmological parameters, as described in Section~\ref{sec:theory}.

If we instead fix the background cosmology to the Planck + WMAP polarization parameters from~\cite{Planck2013params} ($\sigma_8 = 0.829$ and $\Omega_m = 0.315$), we obtain similar results.  In this case, the UPP with $(1-b) = 0.9$ is slightly preferred ($\chi^2 = 14.4$) over the fiducial ``AGN feedback'' model ($\chi^2 = 15.0$).  The UPP with $(1-b) = 0.8$ is also a reasonable fit, with $\chi^2 = 15.3$.  The ``adiabatic'' model is now highly disfavored, with $\chi^2 = 28.7$, while the Sehgal et al.~simulation is also in tension, with $\chi^2 = 20.7$.  Most strikingly, the UPP with $(1-b) = 0.55$ is still in serious tension with the data, with $\chi^2 = 22.4$.  Thus, regardless of the choice of background cosmology (WMAP9 or Planck), we do not find evidence for a value of the HSE mass bias in excess of the standard values predicted in numerical simulations, i.e., $(1-b) \approx 0.8$--$0.9$.

In addition to these general ICM model comparison results, we can use our tSZ -- CMB lensing measurements to fit an overall amplitude parameter for a given ICM model, holding all of the other parameters in the model (and the background cosmology) fixed.  For the fiducial ``AGN feedback'' model from~\cite{Battagliaetal2012}, we fit the overall amplitude of the normalization of the pressure--mass relation (keeping its mass and redshift dependence fixed), $P_0/P_{0,{\rm fid}}$, where $P_{0,{\rm fid}} = 18.1$ is the fiducial amplitude (see~\cite{Battagliaetal2012} for full details).  In the context of this model and a WMAP9 cosmology, we find
\be
P_0/P_{0,{\rm fid}} = 1.10 \pm 0.22 \, \, \, (68\% \mathrm{C.L.})
\label{eq.P0constraint}
\ee
using our tSZ -- CMB lensing cross-power spectrum results, with $\chi^2 = 14.4$ for $12$ degrees of freedom.  Thus, the fiducial model ($P_0/P_{0,{\rm fid}} = 1$) is consistent with our results.  Similarly, we constrain the HSE mass bias $(1-b)$ in the UPP~\cite{Arnaudetal2010}.  Holding all other UPP parameters fixed and using a WMAP9 cosmology, we find 
\be
(1-b) = 1.06^{+0.11}_{-0.14} \, \, \, (68\% \mathrm{C.L.})
\label{eq.HSEbiasconstraint}
\ee
with $\chi^2 = 14.4$ for $12$ degrees of freedom.  This constraint rules out extreme values of the HSE mass bias, if interpreted in the context of a fixed WMAP9 background cosmology.  In particular, we find $0.60 < (1-b) < 1.38$ at the $99.7$\% confidence level.

For comparison, we also perform the same analysis using the Planck collaboration measurement of the tSZ auto-power spectrum~\cite{Planck2013ymap} (see their Table 3 for the relevant data).  For the fiducial ``AGN feedback'' model and a WMAP9 cosmology, we find
\be
P_0/P_{0,{\rm fid}} = 0.80 \pm 0.05 \, \, \, (68\% \mathrm{C.L.})
\label{eq.P0constraintClyy}
\ee
using the Planck tSZ auto-power spectrum results, with $\chi^2 = 2.5$ for $16$ degrees of freedom (the low value of $\chi^2$ could be an indication that the additional error due to uncertainties in residual foreground power subtraction in~\cite{Planck2013ymap} could be too large) .  The fiducial model ($P_0/P_{0,{\rm fid}} = 1$) is in some tension with this constraint.  However, the fiducial ICM model can be brought within the $2\sigma$ allowed region (i.e., $P_0/P_{0,{\rm fid}} = 0.90 \pm 0.05$) if one fixes $\sigma_8 = 0.793$ rather than $\sigma_8 = 0.817$, which is within the $2\sigma$ allowed region of the WMAP9+eCMB+BAO+$H_0$ result~\cite{Hinshawetal2013}.  This is a clear manifestation of the usual ICM--cosmology degeneracy in tSZ constraints (see Section~\ref{sec:interpjoint} for a method to break this degeneracy).  For the HSE mass bias $(1-b)$ in the UPP, we find (again, for a fixed WMAP9 cosmology)
\be
(1-b) = 0.78^{+0.03}_{-0.04} \, \, \, (68\% \mathrm{C.L.})
\label{eq.HSEbiasconstraintClyy}
\ee
with $\chi^2 = 2.2$ for $16$ degrees of freedom (again, the low $\chi^2$ could be an indication that the foreground subtraction errors are over-estimated).  This result, as in Eq.~(\ref{eq.HSEbiasconstraint}), rules out extreme values of $(1-b$), although it is more dependent on the assumption of the fixed background WMAP9 cosmology, as the tSZ auto-power spectrum depends more sensitively on $\sigma_8$ and $\Omega_m$ than the tSZ -- CMB lensing cross-power spectrum. Overall, the constraints in Eqs.~(\ref{eq.P0constraint}) and~(\ref{eq.HSEbiasconstraint}) from our measurement of the tSZ -- CMB lensing cross-power spectrum are consistent with those obtained in Eqs.~(\ref{eq.P0constraintClyy}) and~(\ref{eq.HSEbiasconstraintClyy}) from the Planck collaboration measurement of the tSZ auto-power spectrum.  The lower values favored by the latter are a reflection of the low amplitude of the measured tSZ auto-power spectrum compared to fiducial theoretical expectations.  The low amplitude could be due to ICM physics, cosmology, or observational systematics --- this is discussed in more detail in the next section.


The slightly high amplitude of the cross-spectrum results in Eqs.~(\ref{eq.P0constraint}) and~(\ref{eq.HSEbiasconstraint}) could be explained in many different ways.  For example, a small amount of residual CIB contamination could be present in the tSZ -- CMB lensing cross-correlation measurement, though clearly neither constraint provides statistically significant evidence for this claim.  Alternatively, the results could suggest that perhaps our model has not fully accounted for all of the hot, ionized gas present in the universe --- in particular, highly diffuse, unbound gas is not included in the halo model computations described in Section~\ref{sec:theory}.  However, it is worth noting that in the simulations from which our fiducial ICM model is extracted, the baryon fraction within $r_{\rm vir}$ is found to be $\approx 80$\% (at the group scale, i.e, $M_{200c} \approx 10^{14} \, M_{\odot}/h$) to $95$\% (at the massive cluster scale, i.e., $M_{200c} \approx 10^{15} \, M_{\odot}/h$) of the cosmic mean, with the remainder of the gas located within $2$--$3 r_{\rm vir}$ (see Fig.~10 in~\cite{Battagliaetal2013}).  Thus, the ``missing baryons'' in this model constitute only a small fraction of the cosmic baryon budget.  The fact that the measured cross-power spectrum at low-$\ell$ lies slightly above the halo model predictions computed using our fiducial model might suggest a signal from these ``missing baryons'', but our results provide no statistically significant support for this claim.  We do clearly measure both the one- and two-halo terms in the tSZ -- CMB lensing cross-power spectrum (see Fig.~\ref{fig.Clyphimeas}), but by no means does this imply that the measurement shows signatures of diffuse, unbound gas.  Comparisons to full cosmological hydrodynamics simulations (such as those used to extract the pressure profile used in our fiducial model~\cite{Battagliaetal2012}) will be needed to assess the relative importance of halo-bound gas and diffuse, unbound gas in the tSZ -- CMB lensing cross-power spectrum, which likely has a strong angular scale dependence.  On the largest angular scales at which we measure the tSZ -- CMB lensing cross-power spectrum ($\ell \sim 100$, or, $\theta \sim 1^{\circ}$), the results are sensitive to the correlation between ionized gas and dark matter on comoving distance scales as large as $\sim 50 \, {\rm Mpc}/h$ (see the kernel in the left panel of Fig.~\ref{fig.dCldzell100}, which receives significant contributions at redshifts as large as $z \sim 2$).  In the halo model framework in which we interpret the measurement, these large-scale correlations physically arise from supercluster-sized dark matter structures that host many group- and cluster-sized collapsed halos.  The large, linear-regime dark matter structures imprint the CMB lensing distortions, while the hot electrons sourcing the tSZ signal are bound in the collapsed halos.  Whether the measured correlation also implies the existence of diffuse, unbound gas located outside of these collapsed halos is a question that will require further interpretation with simulations.

\subsection{Cosmological Constraints}
\label{sec:interpcosm}

Fixing the ICM physics prescription to our fiducial ``AGN feedback'' model~\cite{Battagliaetal2012}, we derive constraints on the cosmological parameters $\sigma_8$ and $\Omega_m$ using our measurement of the tSZ -- CMB lensing cross-power spectrum.  All other cosmological parameters are held fixed to their WMAP9+eCMB+BAO+$H_0$ maximum-likelihoood values~\cite{Hinshawetal2013}.  Note that we include the tSZ -- CMB lensing angular trispectrum (Eq.~(\ref{eq.Tllyphi})) in the covariance matrix when deriving these constraints, as discussed in Section~\ref{sec:theoryPS}.  The error bars given in Eq.~(\ref{eq.Clyphimeaserr}) thus include an additional term (e.g.,~\cite{Komatsu-Seljak2002,Hill-Pajer2013}):
\be
\label{eq.Clyphimeaserrwtrispec}
\left( \Delta \hat{C}_{\ell}^{y\phi} \right)^2 = \frac{1}{f_{\rm sky,2}} \frac{1}{(2\ell + 1) \Delta \ell} \left(\hat{C}_{\ell}^{yy} \hat{C}_{\ell}^{\phi\phi} + \left( C_{\ell}^{y\phi} \right)^2 \right) + \frac{1}{4\pi f_{\rm sky,2}} T_{\ell\ell}^{y\phi} \,.
\ee
The trispectrum term is highly subdominant to the noise arising from the $y$ and $\phi$ auto-spectra: for our fiducial ICM model and WMAP9 cosmology, the second term in Eq.~(\ref{eq.Clyphimeaserrwtrispec}) is $\lsim 0.2$\% of the first term.  However, we include it here for completeness.  Since the covariance matrix is now a function of cosmological parameters (i.e., $T_{\ell\ell}^{y\phi}$ depends on $\sigma_8$ and $\Omega_m$), simple $\chi^2$ minimization is no longer the correct procedure to use when deriving constraints (e.g.,~\cite{Eifleretal2009}).  The likelihood function is:
\be
\label{eq.yphiLike}
\mathcal{L}(\sigma_8,\Omega_m) = \frac{1}{\sqrt{(2\pi)^{N_b} \mathrm{det}(\mathrm{Cov})}} \mathrm{exp}\left(-\frac{1}{2} ( C_{\ell}^{y\phi} - \hat{C}_{\ell}^{y\phi} ) (\mathrm{Cov}^{-1})_{\ell\ell'} ( C_{\ell'}^{y\phi} - \hat{C}_{\ell'}^{y\phi} ) \right) \,,
\ee
where $N_b = 12$ is the number of bins in the measurement and $\mathrm{Cov}_{\ell\ell'} = \left( \Delta \hat{C}_{\ell}^{y\phi} \right)^2 \delta_{\ell\ell'}$ is the covariance matrix, with $\left( \Delta \hat{C}_{\ell}^{y\phi} \right)^2$ given in Eq.~(\ref{eq.Clyphimeaserrwtrispec}).  Both $C_{\ell}^{y\phi}$ and $\mathrm{Cov}_{\ell\ell'}$ are computed as a function of $\sigma_8$ and $\Omega_m$.

Using the likelihood in Eq.~(\ref{eq.yphiLike}), we compute constraints on $\sigma_8$ and $\Omega_m$.  We most tightly constrain the combination $\sigma_8 (\Omega_m / 0.282)^{0.26}$:
\be
\sigma_8 \left(\frac{\Omega_m}{0.282} \right)^{0.26} = 0.824 \pm 0.029 \, \, \, (68\% \mathrm{C.L.}).
\label{eq.Ommsig8constraint}
\ee
Note that here and elsewhere throughout the paper we report the best-fit value as the mean of the marginalized likelihood, while the lower and upper error bounds correspond to the $16\%$ and $84\%$ points in the marginalized cumulative distribution, respectively.  Marginalizing the likelihood in Eq.~(\ref{eq.yphiLike}) over $\Omega_m$, we obtain $\sigma_8 = 0.826^{+0.037}_{-0.036}$ at $68$\% C.L.  If we instead marginalize over $\sigma_8$, we find $\Omega_m = 0.281 \pm 0.033$ at $68$\% C.L.

We also re-analyze the Planck collaboration measurement of the tSZ auto-power spectrum presented in~\cite{Planck2013ymap}.  We use the bandpowers and (statistical + foreground) errors given in Table 3 of~\cite{Planck2013ymap}.  However, we also include the tSZ angular trispectrum in the covariance matrix, calculated via an expression analogous to Eq.~(\ref{eq.Tllyphi}).   The trispectrum contribution was neglected (or at least not discussed) in~\cite{Planck2013ymap}, but it is non-negligible on large angular scales, as shown in~\cite{Hill-Pajer2013} (see their Fig.~19).  For our fiducial ICM model and WMAP9 cosmology, we find that the tSZ trispectrum term is as much as $\sim 50$ times larger than the statistical + foreground error term in the covariance matrix (in terms of the error bar itself, this translates to a factor of $\sqrt{50} \approx 7$); this maximum occurs for the fourth lowest multipole bin shown in Fig.~\ref{fig.Clyymeas}, centered at $\ell \approx 52$, where the measurement errors are fairly small but the trispectrum contribution is enormous.  This effect is highly $\ell$-dependent: for the highest multipole bin shown in Fig.~\ref{fig.Clyymeas}, the trispectrum term is only $\approx 1$\% as large as the statistical + foreground error term in the covariance matrix, and is essentially negligible.  Note that the tSZ trispectrum contribution to the covariance matrix can be heavily suppressed by masking nearby, massive clusters when measuring the tSZ power spectrum~\cite{Shawetal2009,Hill-Pajer2013}; however, since this procedure was not used in~\cite{Planck2013ymap}, we do not apply it either, in order to obtain results that can be directly compared with those from~\cite{Planck2013ymap}.  It would likely be beneficial to consider such a masking procedure in future large-angular scale measurements of the tSZ power spectrum.

In addition to our inclusion of the tSZ trispectrum term in the covariance matrix, our re-analysis of the Planck collaboration measurement of the tSZ auto-power spectrum also differs in our choice of fiducial ICM model --- as mentioned throughout this paper, we use the ``AGN feedback'' model from~\cite{Battagliaetal2012}, whereas~\cite{Planck2013ymap} used the UPP of~\cite{Arnaudetal2010} with a fixed HSE mass bias $(1-b) = 0.80$.  

Implementing a likelihood for the tSZ power spectrum analogous to that in Eq.~(\ref{eq.yphiLike}), we compute constraints on $\sigma_8$ and $\Omega_m$.  Again, we find that the combination $\sigma_8 (\Omega_m / 0.282)^{0.26}$ is most tightly constrained:
\be
\sigma_8 \left(\frac{\Omega_m}{0.282} \right)^{0.26} = 0.771^{+0.012}_{-0.011} \, \, \, (68\% \mathrm{C.L.}).
\label{eq.Ommsig8constraintClyy}
\ee
This result is consistent with that presented in~\cite{Planck2013ymap}, although the Planck team found that a slightly different parameter combination was best constrained, obtaining $\sigma_8 \left(\Omega_m / 0.28 \right)^{0.395} = 0.784 \pm 0.016$.  We obtain a slightly lower value due to our choice of ICM model with which to interpret the data --- our model (from~\cite{Battagliaetal2012}) predicts slightly more tSZ power than the UPP+$20$\% HSE bias model used in~\cite{Planck2013ymap}, and thus requires a lower value of $\sigma_8$ to fit the data.  It is also worth noting that the errors bars in Eq.~(\ref{eq.Ommsig8constraintClyy}) are slightly smaller than those from~\cite{Planck2013ymap}, despite our addition of the tSZ trispectrum to the covariance matrix.  This result is due to the fact that the likelihood in~\cite{Planck2013ymap} included two additional foreground parameters (for point sources and clustered CIB), which were marginalized over in the determination of cosmological constraints.  There is also possibly a small difference due to the fact that the trispectrum provides another cosmology-dependent quantity in the likelihood function in addition to the power spectrum itself (see~\cite{Eifleretal2009} for a discussion of this effect in the context of the weak lensing power spectrum).

\begin{figure}
\centering
\includegraphics[width=\textwidth]{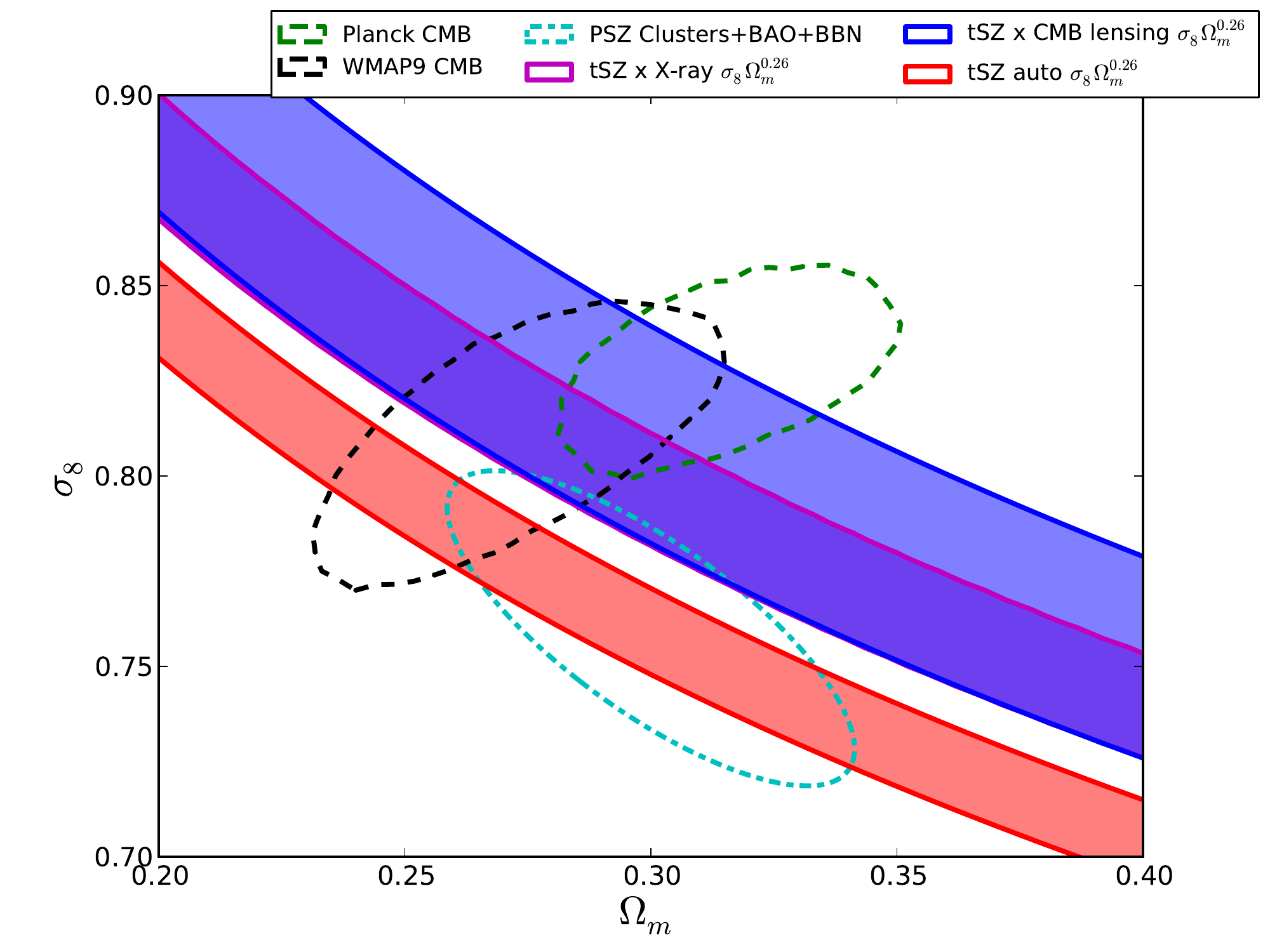}
\caption{ Constraints on $\sigma_8$ and $\Omega_m$ from the tSZ -- CMB lensing cross-power spectrum measurement presented in this work (blue shaded), the tSZ auto-power spectrum measurement from~\cite{Planck2013ymap} re-analyzed in this work (red shaded), the Planck+WMAP polarization CMB analysis~\cite{Planck2013params} (green dashed), WMAP9 CMB analysis~\cite{Hinshawetal2013} (black dashed), number counts of Planck tSZ clusters~\cite{Planck2013counts} (cyan dashed), and a recent cross-correlation of tSZ signal from Planck and WMAP with an X-ray ``$\delta$''-map based on ROSAT data~\cite{Hajianetal2013} (magenta shaded).  All contours shown are $68$\% confidence intervals only (for visual clarity).  See the text for further discussion.
\label{fig.Ommsig8likeplot}}
\end{figure}

Comparison of the constraints presented in Eqs.~(\ref{eq.Ommsig8constraint}) and~(\ref{eq.Ommsig8constraintClyy}) clearly indicates that the tSZ auto-power spectrum measurement from~\cite{Planck2013ymap} favors a lower value of this parameter combination than our measurement of the tSZ -- CMB lensing cross-power spectrum.  The most plausible explanation for this result involves systematics, in particular the subtraction of residual power due to leakage of CIB and IR and radio point sources into the $y$-map in~\cite{Planck2013ymap}.  As can be seen by eye in Fig.~\ref{fig.Clyymeas}, the unsubtracted (blue) points agree quite well with our fiducial theoretical model over a fairly wide multipole range ($50 \lsim \ell \lsim 300$), although contamination is clearly present at higher multipoles.  The residual power from non-tSZ contaminants is assessed with simulations in~\cite{Planck2013ymap}, and an additional $50$\% error is included in the final results due to uncertainties in the subtraction of this power.  If the subtraction procedure has removed too much power, clearly $\sigma_8 \left(\Omega_m / 0.282 \right)^{0.26}$ will be biased low.  A detailed characterization of all the foregrounds is needed to carefully assess this issue, but a rough estimate (assuming that the tSZ auto-power spectrum amplitude scales as $\sigma_8^8$) indicates that a value of $\sigma_8 \approx 0.80$ would be obtained if the foreground power subtraction has been over-estimated by $\approx 20$--$25$\%.  Further study of the CIB, Galactic dust, and point source power spectra will be useful to clarify the accuracy of the subtraction procedure.  Along the same lines, it is also possible that our result from the tSZ -- CMB lensing cross-power spectrum is biased slightly high due to residual CIB emission in the $y$-map; however, the thorough systematics analysis in Section~\ref{sec:errors} gives us confidence that we have treated this issue with sufficient precision.

\begin{figure}
\centering
\includegraphics[width=\textwidth]{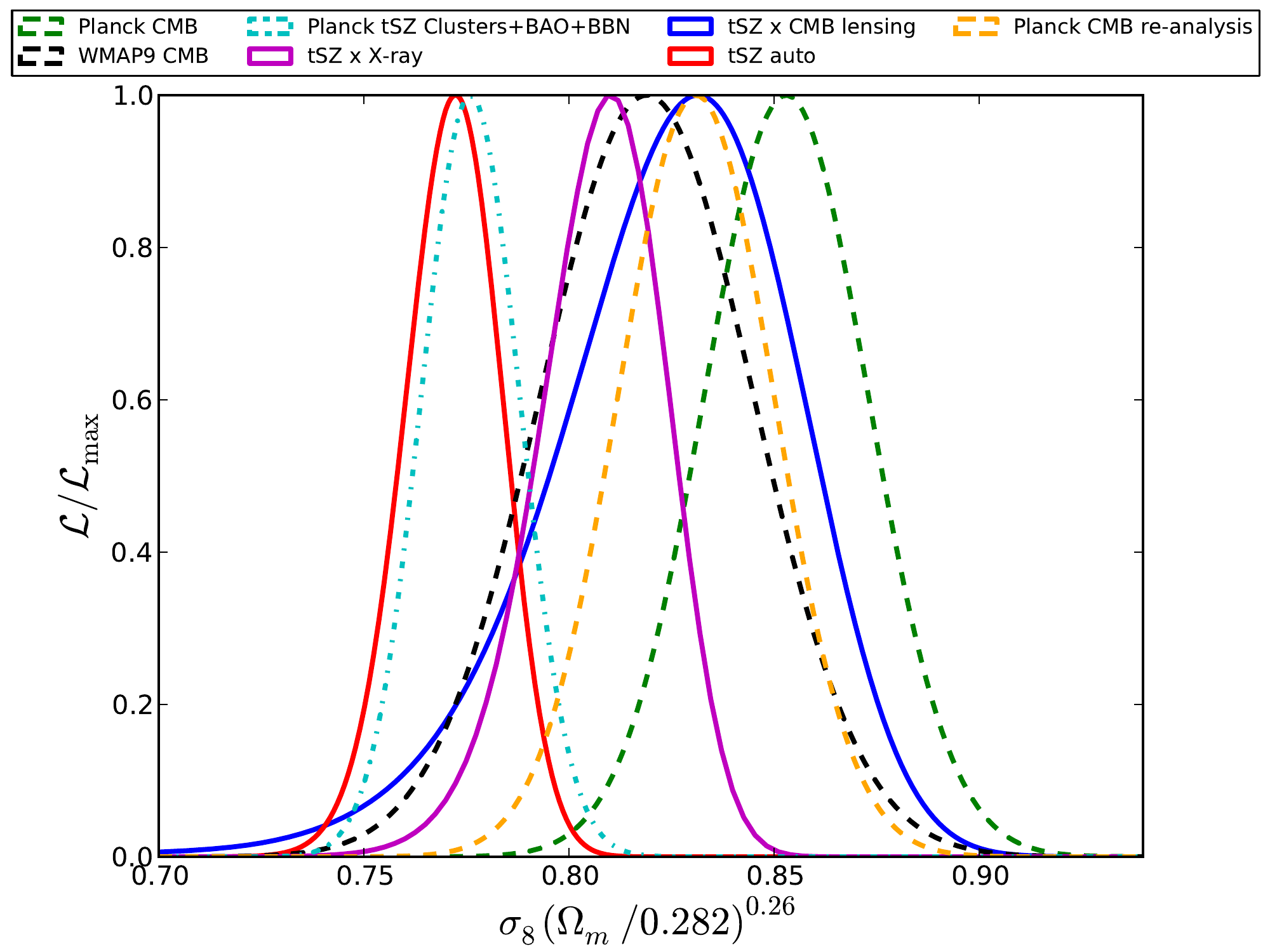}
\caption{ Normalized one-dimensional likelihoods for the degenerate parameter combination $\sigma_8 ( \Omega_m / 0.282 )^{0.26}$ that is best constrained by the tSZ auto- and cross-power spectra measurements.  The probes shown are identical to those in Fig.~\ref{fig.Ommsig8likeplot}, except for an additional result from a recent re-analysis of the Planck CMB + WMAP polarization data~\cite{Spergeletal2013}.  See the text for further discussion.
\label{fig.Ommsig8deg1Dlikeplot}}
\end{figure}

The other possible explanation for the discrepancy between Eqs.~(\ref{eq.Ommsig8constraint}) and~(\ref{eq.Ommsig8constraintClyy}) lies in the physics of the ICM.  However, this explanation requires an unexpected mass dependence for the relative importance of non-thermal pressure support or feedback effects in the ICM (i.e., the HSE mass bias).  Recall that the tSZ -- CMB lensing cross-power spectrum is sourced by lower mass, higher redshift objects than the tSZ auto-power spectrum (see Section~\ref{sec:theory}).  The low value of Eq.~(\ref{eq.Ommsig8constraintClyy}) compared to Eq.~(\ref{eq.Ommsig8constraint}) implies that a larger value of the HSE mass bias is required to reconcile $\sigma_8$ and $\Omega_m$ inferred from the tSZ auto-power spectrum measurement from~\cite{Planck2013ymap} with the WMAP9 or Planck values of these parameters than is required for the measurement of the tSZ -- CMB lensing cross-power spectrum presented in this paper.  Eq.~(\ref{eq.HSEbiasconstraint}) directly demonstrates that our results do not require an extreme value of the HSE mass bias in order to remain consistent with a WMAP9 cosmology.  However, given the mass- and redshift-dependences of the tSZ auto- and tSZ -- CMB lensing cross-spectra, Eqs.~(\ref{eq.HSEbiasconstraint}) and~(\ref{eq.HSEbiasconstraintClyy}) imply that the HSE mass bias (or the influence of non-thermal ICM physics) is larger for more massive, lower redshift groups and clusters.  This mass dependence is the opposite of the general expectation from simulations or simple theoretical arguments~(e.g., \cite{Shawetal2010}), which posit that the most massive clusters in the universe are dominated by their gravitational potential energy, with energetic feedback and non-thermal effects playing a smaller role than in less massive systems (but see~\cite{Rudd-Nagai2009} for an effect that does have this type of mass dependence).  Thus, it seems more plausible that the discrepancy between Eqs.~(\ref{eq.Ommsig8constraint}) and~(\ref{eq.Ommsig8constraintClyy}) arises from systematic effects such as those discussed in the previous paragraph.

We present these constraints on $\sigma_8$ and $\Omega_m$ and compare them to results from various cosmological probes in Figs.~\ref{fig.Ommsig8likeplot} and~\ref{fig.Ommsig8deg1Dlikeplot}.  We show results from the following analyses:
\begin{itemize}
\item Planck+WMAP polarization+high-$\ell$ (ACT+SPT) CMB analysis~\cite{Planck2013params}: $\sigma_8 = 0.829 \pm 0.012$ and $\Omega_m = 0.315^{+0.016}_{-0.018}$, or, $\sigma_8 (\Omega_m / 0.27)^{0.3} = 0.87 \pm 0.02$;
\item WMAP9 CMB-only analysis (note that this is CMB-only, and hence the best-fit parameters are slightly different than our fiducial model, which is WMAP9+eCMB+BAO+$H_0$)~\cite{Hinshawetal2013}: $\sigma_8 = 0.821 \pm 0.023$ and $\Omega_m = 0.279 \pm 0.025$;
\item Planck tSZ cluster number counts + BAO + BBN~\cite{Planck2013counts}: $\sigma_8 ( \Omega_m / 0.27)^{0.3} = 0.782 \pm 0.010$, or, $\sigma_8 = 0.77 \pm 0.02$ and $\Omega_m = 0.29 \pm 0.02$; 
\item Cross-correlation of tSZ signal from Planck and WMAP with an X-ray ``$\delta$''-map based on ROSAT data~\cite{Hajianetal2013}: $\sigma_8 (\Omega_m / 0.30)^{0.26} = 0.80 \pm 0.02$, i.e., $\sigma_8 (\Omega_m / 0.282)^{0.26} = 0.81 \pm 0.02$;
\item Re-analysis of Planck+WMAP polarization CMB~\cite{Spergeletal2013} (Fig.~\ref{fig.Ommsig8deg1Dlikeplot} only): $\sigma_8 ( \Omega_m / 0.30)^{0.26} = 0.818 \pm 0.019$, i.e., $\sigma_8 ( \Omega_m / 0.282)^{0.26} = 0.831 \pm 0.019$.
\end{itemize}

The tSZ auto-power spectrum results shown are those derived in our re-analysis of the measurements from~\cite{Planck2013ymap}, as detailed above.  Also, the contours shown for the tSZ auto- and cross-power spectra in Fig.~\ref{fig.Ommsig8likeplot} assume that only a single parameter ($\sigma_8 \left(\Omega_m / 0.282 \right)^{0.26}$) is fit to the data.  Note that all contours shown in Fig.~\ref{fig.Ommsig8likeplot} are $68$\% confidence intervals only.  Fig.~\ref{fig.Ommsig8likeplot} indicates that there is general concordance between the various probes, with the exception of the tSZ auto-power spectrum and tSZ cluster count constraints.  The discordance of these two probes is more visually apparent in Fig.~\ref{fig.Ommsig8deg1Dlikeplot}, which shows the one-dimensional constraints on $\sigma_8 \left(\Omega_m / 0.282 \right)^{0.26}$ from each probe.  The low amplitude of the tSZ auto-power spectrum is discussed in the previous two paragraphs.  A thorough investigation of the low $\sigma_8$ values inferred from the tSZ cluster counts is beyond the scope of this paper, but it is possible that the ICM physics discussed in the previous paragraph is again responsible, or that systematic effects due to e.g. the complicated selection function arising from the inhomogeneous noise in the Planck maps are responsible.  Theoretical uncertainties in the exponential tail of the mass function could also be an issue, as the clusters used in the counts analysis are much more massive (and hence rarer) than those which dominate the various tSZ statistics shown in Figs.~\ref{fig.Ommsig8likeplot} and~\ref{fig.Ommsig8deg1Dlikeplot}.

Finally, it is worth mentioning that at some level the value of $\sigma_8$ inferred from low-redshift tSZ measurements must disagree with that inferred from primordial CMB measurements, due to the non-zero masses of neutrinos.  Neutrino oscillation experiments imply a minimum mass for the most massive of the three neutrino species of $\approx 0.05$ eV; thus the energy density in neutrinos is at least $\approx 1$\% of the energy density in baryons~\cite{McKeown-Vogel2004}.  While this contribution seems small, it is enough to reduce the number of $z > 1, 10^{15} \, M_{\odot}/h$ clusters by $\approx 25$\% and yield a $\approx 1.5$\% decrease in the value of $\sigma_8$ inferred from low-redshift large-scale structure measurements compared to that inferred from $z=1100$ CMB measurements~\cite{Ichiki-Takada2012}.  For example, if the true CMB-inferred value is $\sigma_8 = 0.817$ (WMAP9), then the true tSZ-inferred value should be $\sigma_8 \approx 0.805$.  While this effect does not completely explain the current discrepancy between the results shown in Figs.~\ref{fig.Ommsig8likeplot} and~\ref{fig.Ommsig8deg1Dlikeplot}, it is worth keeping in mind as the statistical precision of these measurements continues to improve.

\subsection{Simultaneous Constraints on Cosmology and the ICM}
\label{sec:interpjoint}

In the previous two sections, we have constrained ICM pressure profile models while fixing the background cosmology, and constrained cosmological parameters while holding the ICM model fixed.  We now demonstrate the feasibility of simultaneously constraining both the ICM and cosmological parameters by combining our measurement of the tSZ -- CMB lensing cross-power spectrum with the tSZ auto-power spectrum measurement from~\cite{Planck2013ymap}.  Heuristically, since the cross-power spectrum amplitude scales roughly as $\sigma_8^6 \Omega_m^{1.5} P_0$, while the auto-power spectrum amplitude scales roughly as $\sigma_8^8 \Omega_m^{2.1} P_0^2$, it is clear that by combining the different probes we can begin to break the cosmology--ICM degeneracy that currently hinders cosmological constraints from tSZ measurements.  The parameter $P_0$ here represents the overall normalization of the pressure--mass relation, as discussed prior to Eq.~(\ref{eq.P0constraint}).  It is important to note that this approach to simultaneously constraining the ICM and cosmological parameters is only valid for an ICM model without scatter (we do not consider scatter in our prescription for the pressure profile as a function of halo mass and redshift; we simply implement the fitting function from the ``AGN feedback'' simulations of~\cite{Battagliaetal2012}).  If scatter were included, the situation would be slightly more complicated, as the tSZ auto-spectrum would probe $\langle P_0^2 \rangle = P_0^2 + \sigma_{P_0}^2$, where $\sigma_{P_0}^2$ is the scatter in the normalization of the pressure--mass relation, whereas the cross-spectrum would probe $\langle P_0 \rangle = P_0$.  For our purposes, $\sigma_{P_0}^2 = 0$, and hence both power spectra probe $P_0$.  Including scatter would not invalidate this framework, but would simply require fitting for an additional parameter in the analysis\footnote{We thank M.~Becker and E.~Rozo for emphasizing these points to us.}.  This is likely a useful direction to pursue in future tSZ statistical analyses.

\begin{figure}
\centering
\includegraphics[width=\textwidth]{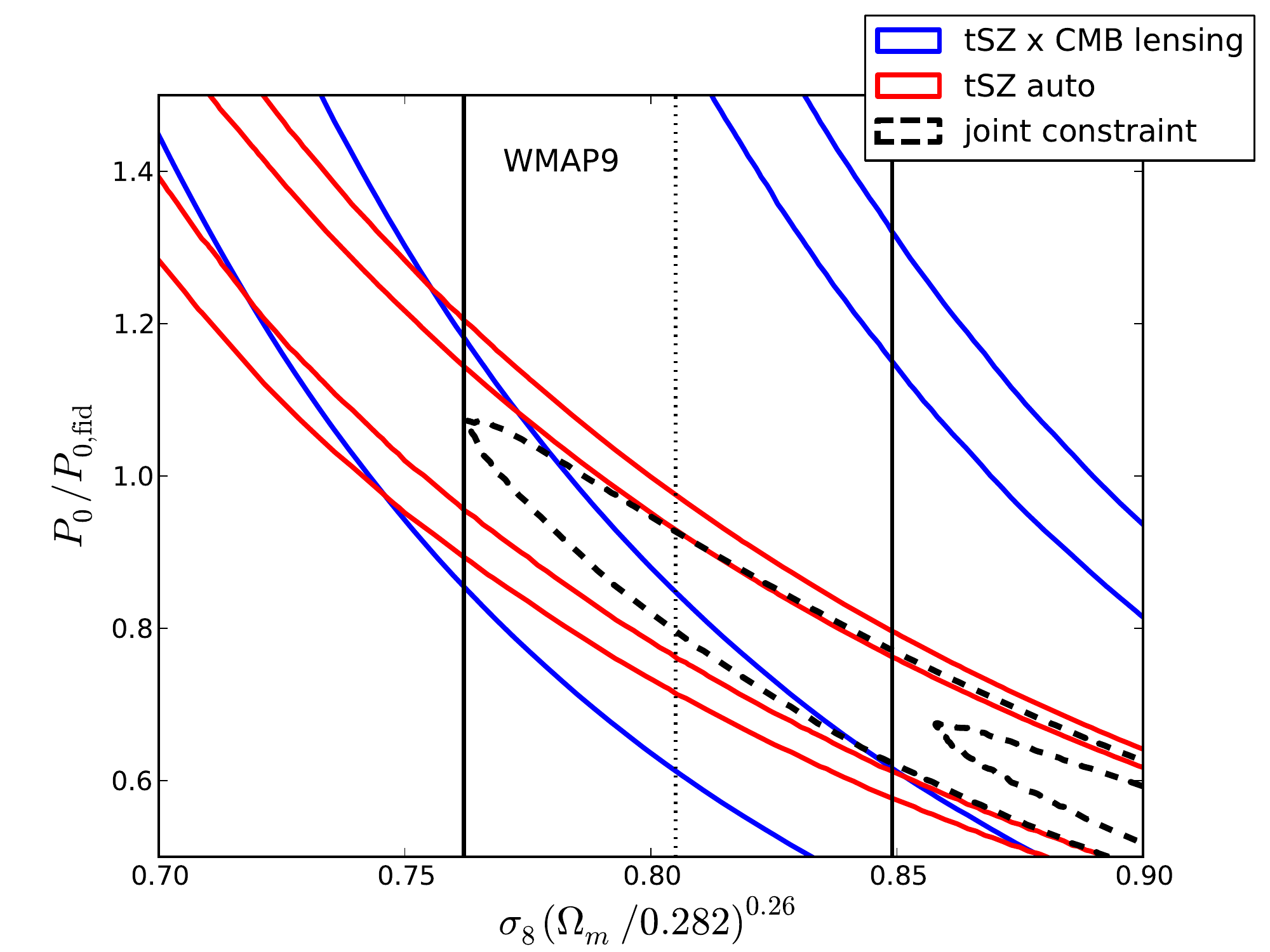}
\caption{ Constraints on $\sigma_8 (\Omega_m/0.282)^{0.26}$ and $P_0$ (the amplitude of the pressure--mass relation normalized to its fiducial value in our ICM model~\cite{Battagliaetal2012}) from the tSZ -- CMB lensing cross-power spectrum measurement presented in this work (blue), the tSZ auto-power spectrum measurement from~\cite{Planck2013ymap} re-analyzed in this work (red), and the combination of the two probes (black dashed).  The different dependences of the two probes on these parameters allow the degeneracy between the ICM and cosmology to be broken (albeit weakly in this initial application) using tSZ measurements alone.  The contours show the $68$\% and $95$\% confidence intervals for the tSZ probes; in addition, the vertical lines show the $68$\% confidence interval for the WMAP9 constraint on this cosmological parameter combination, which is obviously independent of $P_0$.  Standard values for the pressure--mass normalization ($P_0 \approx 0.8$--$1.0$) are compatible with all of the probes.
\label{fig.Ommsig8degP0likeplot}}
\end{figure}

We implement a two-parameter likelihood for the joint constraint, consisting of $P_0$ and $\sigma_8 (\Omega_m / 0.282)^{0.26}$.  Note that all other parameters in the ICM model are held fixed; only the overall normalization is varied.  We also neglect any covariance between the tSZ -- CMB lensing cross-power spectrum and the tSZ auto-power spectrum.  Since the two signals are sourced by somewhat different populations of clusters (see Figs.~\ref{fig.dCldzell100}--\ref{fig.Clmasscontribs}) and the noise in the auto-power measurement is dominated by foreground uncertainties, this seems reasonable in the current analysis, but for future studies with higher SNR this covariance should be accurately quantified.  The results of the joint analysis are shown in Fig.~\ref{fig.Ommsig8degP0likeplot}.  Using either the tSZ -- CMB lensing cross-power spectrum or tSZ auto-power spectrum alone, it is clear that the parameters are completely degenerate.  However, by combining the two probes, the allowed region dramatically shrinks --- the degeneracy is broken.  At the current signal-to-noise levels for these measurements, the degeneracy-breaking power is fairly weak, but this situation will improve with forthcoming data from Planck, ACTPol, and SPTPol.  As shown in Fig.~\ref{fig.Ommsig8degP0likeplot}, the current combination of the two measurements leads to a preference for very low values of $P_0$ and high values of $\sigma_8 (\Omega_m / 0.282)^{0.26}$.  This result is driven by the low amplitude of the measured tSZ auto-power spectrum from~\cite{Planck2013ymap}, which was discussed in detail in the previous section.  Given current uncertainties in the origin of this low amplitude (e.g., systematics due to residual foreground power subtraction), we refrain from quoting precise parameter constraints from this joint analysis.  Regardless, the $95$\% confidence interval in Fig.~\ref{fig.Ommsig8degP0likeplot} includes regions of parameter space consistent with both standard ICM models ($P_0 \approx 0.8$--$1.0$) and standard cosmological parameters from other probes ($\sigma_8 (\Omega_m / 0.282)^{0.26} \approx 0.8$).  For clear visual comparison, Fig.~\ref{fig.Ommsig8degP0likeplot} shows the $68$\% confidence interval for $\sigma_8 (\Omega_m / 0.282)^{0.26}$ from the WMAP9 CMB-only analysis, which overlaps significantly with the allowed region from the combination of the tSZ probes.  Fig.~\ref{fig.Ommsig8degP0likeplot} thus represents a clear proof-of-principle for this technique.  By simultaneously analyzing multiple tSZ statistics, the ICM--cosmology degeneracy can be broken and robust constraints on both can be obtained~\cite{Hill-Sherwin2013,Hilletal2013}.


\section{Discussion and Outlook}
\label{sec:discussion}

In this paper we have presented the first detection of the correlation between the tSZ and CMB lensing fields.  Based on the theoretical calculations presented in Section~\ref{sec:theory}, this cross-correlation signal is sensitive to the ICM physics in some of the highest redshift and least massive groups and clusters ever probed.  The measurement shows clear signatures of both the one- and two-halo terms, probing the correlation between ionized gas and dark matter over scales ranging from $\approx 0.1 \, \mathrm{Mpc}/h$ at $z \approx 0.05$ to $\approx 50 \, \mathrm{Mpc}/h$ at $z \approx 2$.  Interpreting the measurement with halo model calculations, we do not find evidence for the presence of diffuse, unbound gas that lies outside of collapsed halo regions (the ``missing baryons'' --- see discussion in Section~\ref{sec:interpICM}).  Further interpretation of these measurements using cosmological hydrodynamics simulations will be extremely useful.

An additional area in clear need of future improvement is the characterization of foreground contamination in the Compton-$y$ map reconstruction and the estimation of the tSZ auto-power spectrum, with the tSZ -- CIB correlation a particular point of concern.  We make the $y$-map constructed in this work publicly available, but stress that it contains signal from a number of non-tSZ sources, including Galactic dust, CIB, and unresolved point sources.  However, it may prove a useful resource for further cross-correlation studies, provided that contamination from these other sources can be properly understood and mitigated (as done in this analysis for the CIB contamination).

Our measurement of the tSZ -- CMB lensing cross-power spectrum is consistent with extrapolations of existing ICM physics models in the literature.  In particular, we find that our fiducial model (the ``AGN feedback'' model from~\cite{Battagliaetal2012}) is consistent with the data, assuming a WMAP9 background cosmology.  We do not find evidence for extreme values of the HSE mass bias in the UPP model of~\cite{Arnaudetal2010}, obtaining $0.60 < (1-b) < 1.38$ at the $99.7$\% confidence level.  Working in the context of our fiducial gas physics model, we constrain the cosmological parameters $\sigma_8$ and $\Omega_m$, obtaining $\sigma_8 (\Omega_m / 0.282)^{0.26} = 0.824 \pm 0.029$ at $68$\% C.L.  In addition, we re-analyze the tSZ auto-power spectrum measurement from~\cite{Planck2013ymap}, and combine the results with our cross-power measurement to break the degeneracy between ICM physics and cosmological parameters using tSZ statistics alone.  Within the $95$\% confidence interval, standard ICM models and WMAP9 or Planck CMB-derived cosmological parameters are consistent with the results (Fig.~\ref{fig.Ommsig8degP0likeplot}).  Higher SNR measurements will soon greatly improve on these constraints, and may allow robust conclusions regarding $\sigma_8$ and massive neutrinos to be drawn from the tSZ signal, overcoming systematic uncertainties arising from the ICM physics.  However, currently the tSZ data alone are driven to rather extreme regions of parameter space by the low amplitude of the measured tSZ auto-power spectrum.  As discussed in Section~\ref{sec:interpcosm}, this low amplitude could be the result of over-subtraction of residual foreground power in~\cite{Planck2013ymap}, or could be due to ICM physics currently missing in the theoretical calculations.  In this context, it is worth noting that the ACT and SPT constraints on the tSZ power at $\ell \approx 3000$ are also low compared to our fiducial model, although these constraints are obtained in a different approach based on fitting multifrequency measurements of the high-$\ell$ CMB power spectrum.  However, at these very small angular scales, the tSZ power spectrum is more sensitive to the details of the gas distribution in lower-mass halos (e.g., gas blown out of groups by AGN feedback), and thus the low amplitude of the ACT and SPT values may indeed be due to ICM physics considerations.  Upcoming data from ACTPol~\cite{Niemacketal2010}, SPTPol~\cite{Austermannetal2012}, and Planck will shed further light on these issues.

Overall, our measurement of the tSZ -- CMB lensing cross-correlation is a direct confirmation that hot, ionized gas traces dark matter throughout the universe over a wide range of physical scales.

\begin{acknowledgments}
We thank Graeme Addison for providing the theoretical model of the CIB power spectrum at 857 GHz, as well as for a number of useful exchanges.  We thank Duncan Hanson for guidance regarding the Planck CMB lensing potential map, Olivier Dor\'{e} for providing the CIB--CMB lensing cross-power spectrum modeling results from~\cite{Planck2013CIBxlens}, and Jacques Delabrouille for providing an early release of the Planck Sky Model code.  We thank Nick Battaglia and Amir Hajian for providing results from~\cite{Hajianetal2013} for comparisons in Section~\ref{sec:interp}, as well as many informative discussions.  We are also grateful to Daisuke Nagai, Enrico Pajer, Uros Seljak, Blake Sherwin, and Kendrick Smith for many helpful conversations.  While this manuscript was in preparation, we became aware of related work in~\cite{vanWaerbekeetal2013}, and subsequently exchanged correspondence with the authors of that study; we leave a detailed comparison of the results for future work.  JCH and DNS are supported by NASA Theory Grant NNX12AG72G and NSF AST-1311756.
\end{acknowledgments}


\begin{thebibliography}{3}
\bibitem[Hinshaw et al.(2013)]{Hinshawetal2013} Hinshaw, G., Larson, D., Komatsu, E., et al.\ 2013, \apjs, 208, 19 
\bibitem[Bennett et al.(2013)]{Bennettetal2013} Bennett, C.~L., Larson, D., Weiland, J.~L., et al.\ 2013, \apjs, 208, 20 
\bibitem[Planck Collaboration et al.(2013)]{Planck2013params} Planck Collaboration, Ade, P.~A.~R., Aghanim, N., et al.\ 2013, arXiv:1303.5076 
\bibitem[Sievers et al.(2013)]{Sieversetal2013} Sievers, J.~L., Hlozek, R.~A., Nolta, M.~R., et al.\ 2013, \jcap, 10, 60 
\bibitem[Story et al.(2012)]{Storyetal2013} Story, K.~T., Reichardt, C.~L., Hou, Z., et al.\ 2012, arXiv:1210.7231 
\bibitem[Blanchard \& Schneider(1987)]{Blanchard-Schneider1987} Blanchard, A., \& Schneider, J.\ 1987, \aap, 184, 1 
\bibitem[Zeldovich \& Sunyaev(1969)]{Zeldovich-Sunyaev1969} Zel'dovich, Y.~B., \& Sunyaev, R.~A.\ 1969, \apss, 4, 301 
\bibitem[Sunyaev \& Zeldovich(1970)]{Sunyaev-Zeldovich1970} Sunyaev, R.~A., \& Zeldovich, Y.~B.\ 1970, \apss, 7, 3 
\bibitem[Carlstrom et al.(2002)]{Carlstrometal2002} Carlstrom, J.~E., Holder, G.~P., \& Reese, E.~D.\ 2002, \araa, 40, 643 
\bibitem[Marriage et al.(2011)]{Marriageetal2011} Marriage, T.~A., Acquaviva, V., Ade, P.~A.~R., et al.\ 2011, \apj, 737, 61 
\bibitem[Vanderlinde et al.(2010)]{Vanderlindeetal2010} Vanderlinde, K., Crawford, T.~M., de Haan, T., et al.\ 2010, \apj, 722, 1180 
\bibitem[Planck Collaboration et al.(2013)]{Planck2013counts} Planck Collaboration, Ade, P.~A.~R., Aghanim, N., et al.\ 2013, arXiv:1303.5080
\bibitem[Hand et al.(2012)]{Handetal2012} Hand, N., Addison, G.~E., Aubourg, E., et al.\ 2012, Physical Review Letters, 109, 041101 
\bibitem[Smith et al.(2007)]{Smithetal2007} Smith, K.~M., Zahn, O., \& Dor{\'e}, O.\ 2007, \prd, 76, 043510 
\bibitem[Hirata et al.(2008)]{Hirataetal2008} Hirata, C.~M., Ho, S., Padmanabhan, N., Seljak, U., \& Bahcall, N.~A.\ 2008, \prd, 78, 043520 
\bibitem[Das et al.(2011)]{Dasetal2011} Das, S., Sherwin, B.~D., Aguirre, P., et al.\ 2011, Physical Review Letters, 107, 021301 
\bibitem[van Engelen et al.(2012)]{vanEngelenetal2012} van Engelen, A., Keisler, R., Zahn, O., et al.\ 2012, \apj, 756, 142 
\bibitem[Goldberg \& Spergel(1999)]{Goldberg-Spergel1999} Goldberg, D.~M., \& Spergel, D.~N.\ 1999, \prd, 59, 103002 
\bibitem[Cooray \& Hu(2000)]{Cooray-Hu2000} Cooray, A., \& Hu, W.\ 2000, \apj, 534, 533 
\bibitem[Cooray et al.(2000)]{Coorayetal2000} Cooray, A., Hu, W., \& Tegmark, M.\ 2000, \apj, 540, 1 
\bibitem[Cooray(2000)]{Cooray2000} Cooray, A.\ 2000, \prd, 62, 103506 
\bibitem[Peiris \& Spergel(2000)]{Peiris-Spergel2000} Peiris, H.~V., \& Spergel, D.~N.\ 2000, \apj, 540, 605 
\bibitem[Afshordi et al.(2004)]{Afshordietal2004} Afshordi, N., Loh, Y.-S., \& Strauss, M.~A.\ 2004, \prd, 69, 083524 
\bibitem[Planck Collaboration et al.(2013)]{Planck2013NG} Planck Collaboration, Ade, P.~A.~R., Aghanim, N., et al.\ 2013, arXiv:1303.5084
\bibitem[Planck Collaboration et al.(2013)]{Planck2013ISW} Planck Collaboration, Ade, P.~A.~R., Aghanim, N., et al.\ 2013, arXiv:1303.5079
\bibitem[Hasselfield et al.(2013)]{Hasselfieldetal2013} Hasselfield, M., Hilton, M., Marriage, T.~A., et al.\ 2013, \jcap, 7, 8
\bibitem[Reichardt et al.(2013)]{Reichardtetal2013} Reichardt, C.~L., Stalder, B., Bleem, L.~E., et al.\ 2013, \apj, 763, 127 
\bibitem[Nozawa et al.(2006)]{Nozawaetal2006} Nozawa, S., Itoh, N., Suda, Y., \& Ohhata, Y.\ 2006, Nuovo Cimento B Serie, 121, 487 
\bibitem[Battaglia et al.(2012)]{Battagliaetal2012} Battaglia, N., Bond, J.~R., Pfrommer, C., \& Sievers, J.~L.\ 2012, \apj, 758, 75 
\bibitem[Battaglia et al.(2010)]{Battagliaetal2010} Battaglia, N., Bond, J.~R., Pfrommer, C., Sievers, J.~L., \& Sijacki, D.\ 2010, \apj, 725, 91
\bibitem[Sijacki et al.(2007)]{Sijackietal2007} Sijacki, D., Springel, V., Di Matteo, T., \& Hernquist, L.\ 2007, \mnras, 380, 877 
\bibitem[McCarthy et al.(2010)]{McCarthyetal2010} McCarthy, I.~G., Schaye, J., Ponman, T.~J., et al.\ 2010, \mnras, 406, 822 
\bibitem[Borgani et al.(2004)]{Borganietal2004} Borgani, S., Murante, G., Springel, V., et al.\ 2004, \mnras, 348, 1078 
\bibitem[Piffaretti \& Valdarnini(2008)]{Piffaretti-Valdarnini2008} Piffaretti, R., \& Valdarnini, R.\ 2008, \aap, 491, 71 
\bibitem[Nagai et al.(2007)]{Nagaietal2007} Nagai, D., Kravtsov, A.~V., \& Vikhlinin, A.\ 2007, \apj, 668, 1 
\bibitem[Cooray \& Sheth(2002)]{Cooray-Sheth2002} Cooray, A., \& Sheth, R.\ 2002, \physrep, 372, 1 
\bibitem[Hill \& Pajer(2013)]{Hill-Pajer2013} Hill, J.~C., \& Pajer, E.\ 2013, \prd, 88, 063526 
\bibitem[Limber(1954)]{Limber1954} Limber, D.~N.\ 1954, \apj, 119, 655
\bibitem[Komatsu \& Seljak(2002)]{Komatsu-Seljak2002} Komatsu, E., \& Seljak, U.\ 2002, \mnras, 336, 1256 
\bibitem[Cooray(2001)]{Cooray2001} Cooray, A.\ 2001, \prd, 64, 063514 
\bibitem[Planck Collaboration et al.(2013)]{Planck2013overview} Planck Collaboration, Ade, P.~A.~R., Aghanim, N., et al.\ 2013, arXiv:1303.5062 
\bibitem[Planck Collaboration et al.(2013)]{Planck2013HFI} Planck Collaboration, Ade, P.~A.~R., Aghanim, N., et al.\ 2013, arXiv:1303.5067 
\bibitem[Planck Collaboration et al.(2013)]{Planck2013LFI} Planck Collaboration, Aghanim, N., Armitage-Caplan, C., et al.\ 2013, arXiv:1303.5063 
\bibitem[Planck Collaboration et al.(2013)]{Planck2013ymap} Planck Collaboration, Ade, P.~A.~R., Aghanim, N., et al.\ 2013, arXiv:1303.5081
\bibitem[Planck Collaboration et al.(2013)]{Planck2013sources} Planck Collaboration, Ade, P.~A.~R., Aghanim, N., et al.\ 2013, arXiv:1303.5088 
\bibitem[Planck Collaboration et al.(2013)]{Planck2013lensing} Planck Collaboration, Ade, P.~A.~R., Aghanim, N., et al.\ 2013, arXiv:1303.5077 
\bibitem[Planck Collaboration et al.(2013)]{Planck2013HFIbands} Planck Collaboration, Ade, P.~A.~R., Aghanim, N., et al.\ 2013, arXiv:1303.5070 
\bibitem[Planck Collaboration et al.(2013)]{Planck2013CIBxlens} Planck Collaboration, Ade, P.~A.~R., Aghanim, N., et al.\ 2013, arXiv:1303.5078 
\bibitem[Planck Collaboration et al.(2013)]{Planck2013CIB} Planck Collaboration, Ade, P.~A.~R., Aghanim, N., et al.\ 2013, arXiv:1309.0382 
\bibitem[Planck Collaboration et al.(2013)]{Planck2013SZcatalog} Planck Collaboration, Ade, P.~A.~R., Aghanim, N., et al.\ 2013, arXiv:1303.5089 
\bibitem[Eriksen et al.(2004)]{Eriksenetal2004} Eriksen, H.~K., Banday, A.~J., G{\'o}rski, K.~M., \& Lilje, P.~B.\ 2004, \apj, 612, 633 
\bibitem[Remazeilles et al.(2011)]{Remazeillesetal2011} Remazeilles, M., Delabrouille, J., \& Cardoso, J.-F.\ 2011, \mnras, 410, 2481 
\bibitem[Addison et al.(2012)]{Addisonetal2012} Addison, G.~E., Dunkley, J., \& Spergel, D.~N.\ 2012, \mnras, 427, 1741 
\bibitem[Mesinger et al.(2012)]{Mesingeretal2012} Mesinger, A., McQuinn, M., \& Spergel, D.~N.\ 2012, \mnras, 422, 1403 
\bibitem[Hincks et al.(2013)]{Hincksetal2013} Hincks, A.~D., Hajian, A., \& Addison, G.~E.\ 2013, \jcap, 5, 4 
\bibitem[Remazeilles et al.(2013)]{Remazeillesetal2013} Remazeilles, M., Aghanim, N., \& Douspis, M.\ 2013, \mnras, 430, 370
\bibitem[Crawford et al.(2013)]{Crawfordetal2013} Crawford, T.~M., Schaffer, K.~K., Bhattacharya, S., et al.\ 2013, arXiv:1303.3535 
\bibitem[Planck Collaboration et al.(2013)]{Planck2012Coma} Planck Collaboration, Ade, P.~A.~R., Aghanim, N., et al.\ 2013, \aap, 554, A140 
\bibitem[Wilson et al.(2012)]{Wilsonetal2012} Wilson, M.~J., Sherwin, B.~D., Hill, J.~C., et al.\ 2012, \prd, 86, 122005
\bibitem[Hill \& Sherwin(2013)]{Hill-Sherwin2013} Hill, J.~C., \& Sherwin, B.~D.\ 2013, \prd, 87, 023527
\bibitem[Hill et al. (2013)]{Hilletal2013} Hill, J.~C., Sherwin, B.~D., Smith, K.~M., et al.\ in prep.
\bibitem[Arnaud et al.(2010)]{Arnaudetal2010} Arnaud, M., Pratt, G.~W., Piffaretti, R., et al.\ 2010, A \& A, 517, A92 
\bibitem[Sehgal et al.(2010)]{Sehgaletal2010} Sehgal, N., Bode, P., Das, S., et al.\ 2010, \apj, 709, 920 
\bibitem[Dunkley et al.(2011)]{Dunkleyetal2011} Dunkley, J., Hlozek, R., Sievers, J., et al.\ 2011, \apj, 739, 52 
\bibitem[Sun et al.(2011)]{Sunetal2011} Sun, M., Sehgal, N., Voit, G.~M., et al.\ 2011, \apjl, 727, L49 
\bibitem[Planck Collaboration et al.(2013)]{Planck2012stack} Planck Collaboration, Ade, P.~A.~R., Aghanim, N., et al.\ 2013, \aap, 550, A131 
\bibitem[Hajian et al.(2013)]{Hajianetal2013} Hajian, A., Battaglia, N., Spergel, D.~N., et al.\ 2013, arXiv:1309.3282 
\bibitem[Shaw et al.(2010)]{Shawetal2010} Shaw, L.~D., Nagai, D., Bhattacharya, S., \& Lau, E.~T.\ 2010, \apj, 725, 1452
\bibitem[Lewis \& Challinor(2006)]{Lewis-Challinor2006} Lewis, A., \& Challinor, A.\ 2006, \physrep, 429, 1 
\bibitem[Navarro et al.(1997)]{NFW1997} Navarro, J.~F., Frenk, C.~S., \& White, S.~D.~M.\ 1997, \apj, 490, 493
\bibitem[Duffy et al.(2008)]{Duffyetal2008} Duffy, A.~R., Schaye, J., Kay, S.~T., \& Dalla Vecchia, C.\ 2008, \mnras, 390, L64
\bibitem[Bryan \& Norman(1998)]{Bryan-Norman1998} Bryan, G.~L., \& Norman, M.~L.\ 1998, \apj, 495, 80 
\bibitem[Tinker et al.(2008)]{Tinkeretal2008} Tinker, J., Kravtsov, A.~V., Klypin, A., et al.\ 2008, \apj, 688, 709 
\bibitem[Tinker et al.(2010)]{Tinkeretal2010} Tinker, J.~L., Robertson, B.~E., Kravtsov, A.~V., et al.\ 2010, \apj, 724, 878 
\bibitem[Komatsu \& Kitayama(1999)]{Komatsu-Kitayama1999} Komatsu, E., \& Kitayama, T.\ 1999, \apjl, 526, L1
\bibitem[Sherwin et al.(2012)]{Sherwinetal2012} Sherwin, B.~D., Das, S., Hajian, A., et al.\ 2012, \prd, 86, 083006 
\bibitem[Binney \& Tremaine(2008)]{Binney-Tremaine} Binney, J., \& Tremaine, S.\ 2008, Galactic Dynamics: Second Edition, by James Binney and Scott Tremaine.~ISBN 978-0-691-13026-2 (HB).~Published by Princeton University Press, Princeton, NJ USA, 2008.,  
\bibitem{Bhattacharyaetal2012} Bhattacharya, S., Nagai, D., Shaw, L., Crawford, T., \& Holder, G.~P.\ 2012, \apj, 760, 5 
\bibitem[Trac et al.(2011)]{Tracetal2011} Trac, H., Bode, P., \& Ostriker, J.~P.\ 2011, \apj, 727, 94 
\bibitem[Shaw et al.(2009)]{Shawetal2009} Shaw, L.~D., Zahn, O., Holder, G.~P., \& Dor{\'e}, O.\ 2009, \apj, 702, 368 
\bibitem[Addison et al.(2013)]{Addisonetal2013} Addison, G.~E., Dunkley, J., \& Bond, J.~R.\ 2013, \mnras, 436, 1896  
\bibitem[Hirata \& Seljak(2003)]{Hirata-Seljak2003} Hirata, C.~M., \& Seljak, U.\ 2003, \prd, 67, 043001 
\bibitem[Okamoto \& Hu(2003)]{Okamoto-Hu2003} Okamoto, T., \& Hu, W.\ 2003, \prd, 67, 083002 
\bibitem[Song et al.(2003)]{Songetal2003} Song, Y.-S., Cooray, A., Knox, L., \& Zaldarriaga, M.\ 2003, \apj, 590, 664 
\bibitem[Sachs \& Wolfe(1967)]{Sachs-Wolfe1967} Sachs, R.~K., \& Wolfe, A.~M.\ 1967, \apj, 147, 73 
\bibitem[Osborne et al.(2013)]{Osborneetal2013} Osborne, S.~J., Hanson, D., \& Dor{\'e}, O.\ 2013, arXiv:1310.7547 
\bibitem[van Engelen et al.(2013)]{vanEngelenetal2013} van Engelen, A., Bhattacharya, S., Sehgal, N., et al.\ 2013, arXiv:1310.7023 
\bibitem[Eifler et al.(2009)]{Eifleretal2009} Eifler, T., Schneider, P., \& Hartlap, J.\ 2009, \aap, 502, 721 
\bibitem[Rudd \& Nagai(2009)]{Rudd-Nagai2009} Rudd, D.~H., \& Nagai, D.\ 2009, \apjl, 701, L16 
\bibitem[Battaglia et al.(2013)]{Battagliaetal2013} Battaglia, N., Bond, J.~R., Pfrommer, C., \& Sievers, J.~L.\ 2013, \apj, 777, 123 
\bibitem[Ichiki \& Takada(2012)]{Ichiki-Takada2012} Ichiki, K., \& Takada, M.\ 2012, \prd, 85, 063521 
\bibitem[McKeown \& Vogel(2004)]{McKeown-Vogel2004} McKeown, R.~D., \& Vogel, P.\ 2004, \physrep, 394, 315 
\bibitem[Spergel et al.(2013)]{Spergeletal2013} Spergel, D., Flauger, R., \& Hlozek, R.\ 2013, arXiv:1312.3313
\bibitem[Niemack et al.(2010)]{Niemacketal2010} Niemack, M.~D., Ade, P.~A.~R., Aguirre, J., et al.\ 2010, \procspie, 7741
\bibitem[Austermann et al.(2012)]{Austermannetal2012} Austermann, J.~E., Aird, K.~A., Beall, J.~A., et al.\ 2012, \procspie, 8452
\bibitem[Van Waerbeke et al.(2013)]{vanWaerbekeetal2013} Van Waerbeke, L., Hinshaw, G., \& Murray, N.\ 2013, arXiv:1310.5721 


\end{thebibliography}
\end{document}